\documentclass[usegraphicx,usenatbib]{mn2e}
\pdfoutput=1
\usepackage[totalwidth=480pt,totalheight=680pt]{geometry}
\usepackage{amssymb}
\usepackage{color}

\usepackage{mathrsfs}

\newcommand{\lya}{\ifmmode\mathrm{Ly}\alpha\else{}Ly$\alpha$\fi}
\newcommand{\igm}{\ifmmode\mathrm{IGM}\else{}IGM\fi}
\newcommand{\bao}{\ifmmode\mathrm{BAO}\else{}BAO\fi}
\newcommand{\hi}{\ifmmode\mathrm{HI}\else{}HI\fi}
\newcommand{\hii}{\ifmmode\mathrm{HII}\else{}HII\fi}
\newcommand{\hei}{\ifmmode\mathrm{HeI}\else{}HeI\fi}
\newcommand{\heii}{\ifmmode\mathrm{HeII}\else{}HeII\fi}
\newcommand{\cmb}{\ifmmode\mathrm{CMB}\else{}CMB\fi}
\newcommand{\qso}{\ifmmode\mathrm{QSO}\else{}QSO\fi}

\title[Cosmology from clustering of \lya{} galaxies]
{Cosmology from clustering of $\bmath{\lya{}}$ galaxies: breaking non-gravitational $\bmath{\lya{}}$ radiative transfer degeneracies using the bispectrum}

\author[B. Greig et al.] {Bradley~Greig,$^{1,2}$\thanks{E-mail:~bgreig@student.unimelb.edu.au~(BG)}, Eiichiro Komatsu,$^{3,4,5}$\thanks{~komatsu@mpa-garching.mpg.de~(EK)} \& J. Stuart B.~Wyithe;$^{1,2}$\thanks{~swyithe@unimelb.edu.au~(JSBW)} \\ $^1$School of Physics, University of Melbourne, Parkville, Victoria 3010, Australia\\
$^2$ARC Centre of Excellence for All-sky Astrophysics (CAASTRO)\\
$^3$Texas Cosmology Center and the Department of Astronomy, The University of Texas at Austin, 1 University Station, C1400, \\
\, Austin, TX 78712, USA\\
$^4$Kavli Institute for the Physics and Mathematics of the Universe, Todai Institutes for Advanced Study, \\
\, the University of Tokyo, Kashiwa, Japan 277-8583 (Kavli IPMU, WPI)\\
$^5$Max-Planck-Institut f\"{u}r Astrophysik, Karl-Schwarzschild Str. 1, 85741 Garching, Germany}

\begin{document}
\maketitle 
\begin{abstract}
Large surveys for \lya{} emitting (LAE) galaxies have been proposed as a
 new method for measuring clustering of the galaxy population at high
 redshift with the goal of determining cosmological parameters. However,
 \lya{} radiative transfer effects may modify the observed clustering of
 LAE galaxies in a way that mimics gravitational effects, potentially
 reducing the precision of cosmological constraints. For
 example, the effect of the linear redshift-space distortion on the
 power spectrum of LAE galaxies is potentially degenerate with \lya{}
 radiative transfer effects owing to the dependence of observed flux on
 intergalactic medium velocity gradients. In this paper, we show that
 the three-point function (bispectrum) can distinguish between
 gravitational and non-gravitational effects, and thus breaks these
 degeneracies, making it possible to recover cosmological parameters from
 LAE galaxy surveys. 
 Constraints on the
 angular diameter distance and the Hubble expansion rate can
also be improved
 by combining power spectrum and bispectrum measurements. 
\end{abstract} 

\section[Introduction]{Introduction}

For the past three decades galaxy redshift
surveys have served as the traditional method for constraining
cosmological parameters such as the matter density of the
universe and the equation of state of dark energy, by measuring the clustering of galaxies.  
These have been restricted to $z<1$ due to
the increasingly fainter galaxy magnitudes and larger required cosmic
volumes, which render spectroscopy of large numbers of
photometrically selected early type galaxies plausible only at such
low redshifts. 
Recently the WiggleZ collaboration has pushed galaxy
clustering work to $z\sim1$ using 
emission lines from star-forming galaxies \citep{blake/etal:2011c,blake/etal:2011b,blake/etal:2011a,blake/etal:2012}.

Ly$\alpha$ emitting (LAE) galaxies are detectable out to high redshift
\citep{Iye:2006p13032,Kashikawa:2006p13047,Lehnert:2010p13102,Ouchi:2010p13150},
due to their strong line emission. Indeed over the
previous few years, the number of detected LAE sources has
steadily grown and the sample sizes of LAE galaxies have reached sufficient
size for clustering studies
\citep{Gawiser:2007p13156,Kovac:2007p13181,Orsi:2008p13553,Guaita:2010p13210,Ouchi:2010p13150}. 

While the existing samples of LAE galaxies are still too small for
cosmological purposes, the rate of detection of these LAE galaxies will
significantly improve with the upcoming Hobby-Eberly Telescope Dark
Energy Experiment (HETDEX, \citealt{Hill:2004p12593,Hill:2008p12595}),
whose aim is to spectroscopically measure the redshifts of 800$\,$000
LAE galaxies in the redshift range 1.9 $\leq$ $z$
$\leq$ 3.5 \citep{Hill:2004p12593,Hill:2008p12595} with the total sky
coverage of 420 square degrees and the total volume coverage of $10~{\rm
Gpc}^3$. This survey is specifically designed to use the
clustering of 
LAE galaxies to make the
precise measurement of the distance scales, both the angular diameter
distance ($D_{A}$) and the Hubble rate ($H$), 
as a function of $z$ out to $z\sim 3$.

In order for us to use the clustering of LAE galaxies to
measure the distance scales, we must understand how the clustering of
LAE galaxies is related to the underlying matter distribution.
Simulations by \citet{Zheng:2010p8285,Zheng:2011p8289} and
\citet{Laursen:2011p8323} have investigated the radiative transfer
effects on the \lya{} emission of LAE galaxies both within the
circumgalactic environment around the halo and from the resonant
scattering of diffuse neutral hydrogen in the intergalactic medium
(\igm{}). Of particular interest is the clustering of the LAE galaxies:
\citet{Zheng:2011p8289} find that line-of-sight gradients in the
peculiar velocity of LAE galaxies could lead to an observed reduction in
the line-of-sight clustering amplitude of the galaxies, counteracting
the strength of the typical Kaiser effect \citep{Kaiser:1987p8786}
caused by gravitation. Conversely, the clustering of LAE galaxies transverse to the line-of-sight is found to be significantly boosted by the \lya{} radiative transfer effects. In addition to the local effects of peculiar velocity gradients, the transmission of the \lya{} emission line of LAE galaxies through the diffuse IGM could also be affected by fluctuations in the UV ionizing background and to changes in the neutral hydrogen fraction associated with changes in the density around the local environment. 

To understand these effects, \citet{Zheng:2011p8289} and
\citet{Wyithe:2011p12569} have derived analytic models to describe the
observed modifications of the power spectrum of LAE
galaxies. Both derive quantities that directly relate to 
the non-gravitational effects expected from the \lya{} radiative
effects. \citet{Wyithe:2011p12569} use this model to study
the expected recovery of both cosmological and \lya{} radiative
transfer parameters from a survey corresponding to HETDEX. They find that some
cosmological parameters derived only from the power
spectrum are degenerate with the \lya{} radiative transfer effects, and
that this has direct consequences for the accuracy with which
cosmological parameters can be recovered from the LAE galaxy
power spectrum. Prior
knowledge of the magnitude of the radiative transfer effects can improve
the recovery of the cosmological constraints. 

In this paper we further investigate the effects of
non-gravitational LAE clustering on the recovery of cosmological
parameters. We extend and improve the linear theory work of
\citet{Wyithe:2011p12569} by including the three-point
correlation function
(bispectrum) and by combining with the power spectrum, to break the
first order degeneracies of the \lya{} radiative transfer effects and
cosmological parameters. To calculate the bispectrum,
we use a next-to-leading order Eulerian perturbation theory approach
(\citealt{Bernardeau:2002p8437} and references within), and derive
expressions valid into the mildly non-linear regime. We also derive higher-order expressions for the \lya{} radiative transfer effects,  including
the higher-order effects of redshift-space distortions. We 
then study how well a joint analysis of the power spectrum and the
bispectrum can break the cosmological and
radiative transfer degeneracies.
We provide the expected constraints on cosmological
parameters through the application of Fisher matrices, with specific
reference to the HETDEX survey. 

This paper is set out as follows. In Section \ref{sec:problem} we
outline the degeneracies between cosmological and \lya{} radiative
transfer parameters, and in Section \ref{sec:PS} perform a Fisher matrix
analysis of the LAE galaxy power spectrum. In Section \ref{sec:method}
we outline and describe existing Eulerian perturbation theory
expressions and provide the derivation of higher-order corrections for
the \lya{} radiative transfer effects, in order to construct both a
bispectrum and a reduced bispectrum model. In Sections \ref{sec:RBS} and
\ref{sec:higherorder} we perform Fisher matrix analyses of the reduced
bispectrum alone, and a combined power spectrum and bispectrum in order
to provide cosmological parameter estimates. We finish with
our summary and final remarks in Section \ref{sec:conclusion}. In our
numerical calculations we consider the standard set of cosmological
parameters \citep{Komatsu:2011p12557}, with $\Omega_{m} = 0.275$,
$\Omega_{\Lambda} = 0.725$, $\Omega_{b} =
0.0458$, $n_s = 0.968$, $h = 0.702$ and $\sigma_{8} = 0.816$.

\section[Clustering of Lyman-alpha emitters]{Clustering of $\bmath{\lya{}}$ emitters}\label{sec:problem}
In this section we summarize the linear theory clustering of LAE galaxies. For a galaxy redshift survey, one can write a simple expression relating the power spectrum of galaxies to the underlying matter distribution. However, for LAE galaxies, \citet{Zheng:2011p8289} and \citet{Wyithe:2011p12569} show that \lya{} radiative transfer effects modify this relationship. 
 
\subsection{Galaxy power spectrum}

In linear theory, the power spectrum of galaxies in redshift space is given by
\begin{equation} \label{eq:gal}
P_{\rm gal}(k)=(b_1+f\mu^2)^2 P_L(k),
\end{equation}
where $b_{1}$ is the linear galaxy bias, since galaxies are biased
tracers of the matter density field \citep{Kaiser:1984p8759}, $f \equiv{\rm d
ln}\,D(a)/{\rm d ln}\,a$ is the growth rate of structure,
$D(a)$ is the linear growth factor, $P_{L}(k)$ is the power
spectrum of the linear density fluctuations, and $\mu$ is
the cosine of the angle between the line-of-sight vector
$\hat{\bmath{z}}$ and the wavevector $\bmath{k}$, i.e., $\mu \equiv
\bmath{k}\cdot\hat{\bmath{z}}/k$. 

Of particular interest for cosmology is the recovery of the growth rate
of structure, $f$, which can be parametrized as $f =
\Omega_{m}^{\gamma}$. The value that $\gamma$ takes can distinguish
between cosmological models described by general relativity or other
gravitational 
descriptions \citep{Linder:2005p13557}.
However, the galaxy bias, $b_{1}$, and the growth rate of structure, $f$, are
degenerate in the above model. Hence, what is actually measurable from a
galaxy redshift survey is the linear redshift-space distortion
parameter, $\beta = f/b_1$. With this parameter, the above
expression becomes
\begin{equation} \label{eq:galbeta}
P_{\rm gal}(k)=b^{2}_{1}(1+\beta\mu^2)^2 P_L(k).
\end{equation}

\subsection{LAE power spectrum} \label{sec:1stOrderLya}

\lya{} radiative transfer effects potentially modify the power spectrum
of LAE galaxies away from the standard galaxy description (Equation
\ref{eq:gal}). This arises because the observed number density of \lya{}
galaxies at a fixed observed flux depends on how many \lya{} photons escape to observers. Thus the observed density of galaxies can be modified by the local environment nearby to the \lya{} galaxies. 

The \lya{} optical depth through the \igm{} depends on the local density ($\rho$), the ionizing background strength ($\Gamma$), and also on the strength of the local peculiar velocity gradient along the line-of-sight (${\rm d}v_{z}/{\rm d}r$), 
\begin{eqnarray}
\tau \propto \frac{\rho^{2}}{\Gamma T^{0.7} \frac{{\rm d}v_{z}}{{\rm d}r}} \propto \frac{\rho^{2-0.7(\gamma-1)}}{\Gamma \frac{{\rm d}v_{z}}{{\rm d}r}},
\end{eqnarray}
where $\gamma$ is the polytropic index, used to relate the temperature
to the underlying density field as $T \propto \rho^{\gamma-1}$ with
$\gamma = 1.4$ \citep{Hui:1997p339}. 

In this section we shall
describe the effect of \lya{}
transmission fluctuations on the observed clustering signal of LAEs, following \citet{Wyithe:2011p12569}. Modifications to the
intrinsic \lya{} luminosity by the \igm{} induce a change in number
counts of LAE galaxies.  
The number density of LAE sources, $n_{\lya{}}$, that are observed above some observational flux threshold, $F_{0}$, can be expressed relative to the average, $\bar{n}_{\lya{}}(>L_{0},\rho_{0},\Gamma_{0},\delta(\bmath{x}))$, as
\begin{equation} \label{eq:numdens}
n_{\rm Ly\alpha}(>F_{0}) = \bar{n}_{\rm Ly\alpha}(>L_{0},\rho_{0},\Gamma_{0},\delta(\bmath{x}))\times[1+\delta_{g}(\bmath{x})],
\end{equation}
where $\delta(\bmath{x})$ is the large-scale matter-density
perturbation, and $\delta_{g}(\bmath{x})$ is the perturbation in the
galaxy number density. 
Now the average number density of LAE galaxies depends on the
fluctuations due to the non-gravitational \lya{} radiative transfer
effects; namely, the local density in the LAE environment, $\rho$,
ionizing background, $\Gamma$, and peculiar velocity gradients, ${\rm
d}v_{z}/{\rm d}r$. 

We first obtain the expression for the mean number density
of observed LAE galaxies. Taylor expanding about the three radiative transfer
effects, we obtain
\begin{eqnarray} \label{eq:Taylor}
\bar{n}_{\rm Ly\alpha}(>L_{0},\rho_{0},\Gamma_{0},\delta(\bmath{x})) = \bar{n}^{(0)}_{\rm Ly\alpha}\left(1 + \bar{n}^{(1)}_{\lya{}}\right),
\end{eqnarray}
where $\bar{n}^{(0)}_{\rm Ly\alpha}$ is just the mean number of LAE galaxies, and $\bar{n}^{(1)}_{\lya{}}$ is the first-order Taylor-expanded expression evaluated around their mean quantity, 
\begin{eqnarray} \label{eq:Taylorex1}
\bar{n}^{(1)}_{\lya{}} & = &\frac{1}{\bar{n}^{(0)}_{\rm Ly\alpha}}(\Gamma - \Gamma_{0})\frac{\partial\bar{n}_{\rm Ly\alpha}}{\partial\Gamma}\line(0,1){15}_{\line(0,1){15}F_{0},\Gamma_{0}} + (\rho - \rho_{0})\frac{\partial\bar{n}_{\rm Ly\alpha}}{\partial\rho}\line(0,1){15}_{\line(0,1){15}F_{0},\rho_{0}} \nonumber \\
& & + \left(\frac{{\rm d} v_{z}}{{\rm d}(a r_{\rm com})}- H\right)\frac{\partial\bar{n}_{\rm Ly\alpha}}{\partial\frac{{\rm d} v_{z}}{{\rm d}(a r_{\rm com})}}\line(0,1){15}_{\line(0,1){15}F_{0},\rho_{0}}.
\end{eqnarray}
Here, $H$ is the Hubble rate, and the line-of-sight velocity
gradient is taken with respect to the comoving distance, $r_{\rm com}$.
We rewrite this expression for $\bar{n}^{(1)}_{\lya{}}$ as
\begin{eqnarray} \label{eq:1stOrder}
\bar{n}^{(1)}_{\lya{}} & = & \delta_{\Gamma}(\bmath{x})C_{\Gamma} + \delta_{\rho}(\bmath{x}) C_{\rho} + \delta_{v}(\bmath{x})C_{v}.
\end{eqnarray}
The constants $C_{\Gamma}$, $C_{\rho}$, and $C_{v}$, 
are defined in Appendix \ref{app:Lyaeffects}, and
capture the distinct physical effects caused by changes to the local
environment corresponding to either changes to the ionizing background,
the density, or the velocity gradient along the line-of-sight,
respectively. The other quantities are defined as
$\delta_\Gamma\equiv (\Gamma-\Gamma_0)/\Gamma_0$, $\delta_\rho\equiv
(\rho-\rho_0)/\rho_0$, and $\delta_v\equiv (Ha)^{-1}dv_z/dr_{\rm com}$.

Next, the number density of LAE galaxies observed above a flux limit can be written as,
\begin{eqnarray} \label{eq:intermediate}
n_{\rm Ly\alpha}(>F_{0}) = \bar{n}^{(0)}_{\rm Ly\alpha}[1+\delta_{g}(\bmath{x})]\left[1+\bar{n}^{(1)}_{\lya{}}\right],
\end{eqnarray}
where we have substituted Equation \ref{eq:Taylor} into Equation
\ref{eq:numdens}. 
Rewriting this expression as fluctuations in the number of LAE galaxies relative to the mean number of galaxies expected without \lya{} radiative transfer effects, $\bar{n}^{(0)}_{\rm Ly\alpha}$, we obtain
\begin{equation} \label{eq:Lyafluct}
\delta_{\lya{}}(\bmath{x}) = \frac{n_{\rm Ly\alpha}(>F_{0})}{\bar{n}^{(0)}_{\rm Ly\alpha}} - 1  = [1+\delta_{g}(\bmath{x})]\left[1+\bar{n}^{(1)}_{\lya{}}\right] - 1.
\end{equation}
Expanding Equation \ref{eq:Lyafluct} and taking the Fourier transform,
one finds, to the first order, 
\begin{equation}
\delta_{\rm Ly\alpha,s}(\bmath{k})  = \left[b_{1} \left(1 + C_{\Gamma}\right) + C_{\rho} + f\mu^{2}(1- C_{v})\right]\delta(\bmath{k}),
\end{equation}
where we have used the linear-theory predictions:
$\delta_{\Gamma}(\bmath{k}) = 
b_{1}C_{\Gamma}\delta(\bmath{k})$, $\delta_{\rho}(\bmath{k}) =
\delta(\bmath{k})$, and $\delta_{v}(\bmath{k}) =
-f\mu^{2}C_{v}\delta(\bmath{k})$. Here, we have implicitly performed the
linear redshift-space transformation (see Appendix \ref{app:full}).

In this work we define the redshift-space LAE galaxy power spectrum as
\begin{eqnarray}
\langle \delta_{\lya{},s}(\bmath{k}_{1})\delta_{\lya{},s}(\bmath{k}_{2}) \rangle = (2\pi)^{3}P_{\lya{},s}(k)\delta^{D}(\bmath{k}_{1} + \bmath{k}_{2}),
\end{eqnarray}
where $P_{\rm Ly\alpha,s}(k)$ is given by
\begin{equation} \label{eq:1storderLyaPS}
P_{\rm Ly\alpha,s}(k)  =  \left[b_{1} \left(1 + C_{\Gamma}\right) + C_{\rho} + f\mu^{2}(1- C_{v})\right]^{2}P_{L}(k),
\end{equation}
and $P_{L}(k)$ is the linear real-space matter power spectrum. Relative
to Equation \ref{eq:gal}, the addition of \lya{} radiative transfer
effects can lead to changes in the amplitude of the measured power spectrum.\footnote{In Equation \ref{eq:1storderLyaPS}, we do not consider the scale dependence of the ionizing background fluctuations, which is the major difference between our expression and the expression in \citet{Wyithe:2011p12569}. The ionizing background fluctuations are expected to be scale dependant, important on large scales (set by the mean free path of the ionizing photons) and becoming negligible on small scales. However, including the scale dependence associated with the ionizing background fluctuations increases the model complexity, providing additional model degeneracies. We feel that this simplification is justified since the transmission models investigated by \citet{Wyithe:2011p12569} find the magnitude of ionizing background fluctuations ($C_{\Gamma}$) to be small compared to the other two radiative transfer effects. Hence, while ignoring the scale dependence is a simplification, the overall impact of removing this scale dependence should be minor.} 

Equation \ref{eq:1storderLyaPS} contains the main contributing terms of
\citet{Zheng:2011p8289}. However, we do not include the transverse
line-of-sight velocity-gradient or the density-gradient (as provided by
\citealt{Zheng:2011p8289}). As shown in \citet{Zheng:2011p8289}, the
effect of the density-gradient adds additional scale-dependant terms to
the expression for the clustering of LAE galaxies on small scales. In
this work, we are working at much larger scales, and so can ignore this
scale-dependence and allow the density-gradient terms to be absorbed
into the existing parameters of Equation \ref{eq:1storderLyaPS}. 

The inclusion of \lya{} radiative transfer effects introduces
degeneracies between the cosmological parameters and \lya{} radiative
transfer parameters. In particular, from Equation
\ref{eq:1storderLyaPS}, we note the degeneracy between the growth rate
of structure, $f$, and the line-of-sight peculiar velocity radiative
transfer effect, $C_{v}$. Additionally, the galaxy bias, $b_1$, is
degenerate with the local environment density, $C_{\rho}$, and the
fluctuations in the ionizing background, $C_{\Gamma}$.

To simplify the expression, consider the following redefinition of Equation \ref{eq:1storderLyaPS},
\begin{eqnarray} \label{eq:PStilde}
P_{\rm Ly\alpha,s}(k) & = & \tilde{b}^{2}_{1}\left[1 + \tilde{\beta}\mu^{2}(1- C_{v})\right]^{2}P_{L}(k),
\end{eqnarray}
where
\begin{eqnarray} \label{eq:b1tilde}
\tilde{b}_{1} \equiv b_{1} + C_{\rho} + b_{1}C_{\Gamma}.
\end{eqnarray}
This includes the large-scale effects of density and ionizing background
which modify the observed galaxy clustering, and $\tilde{\beta}$ which
is a modified linear redshift-space distortion parameter corresponding
to the modified galaxy bias, $\tilde{\beta} \equiv f/\tilde{b}_{1}$.  

Now, the problem is clear: while Equation~\ref{eq:PStilde}
has the same structure as Equation~\ref{eq:galbeta}, the meaning of each
parameter is different. The correspondence is $b_1\to \tilde{b}_1$ and
$\beta\to \tilde{\beta}(1-C_v)$, which shows the parameter degeneracy.
In the next 
section we illustrate the resulting effect of radiative transfer
parameters on the potential cosmological constraints. 

\section[Cosmological constraints based on the linear LAE galaxy power spectrum]{Cosmological constraints based on the linear LAE galaxy power spectrum} \label{sec:PS}

To generate the expected constraints on cosmological
parameters for a given survey configuration, we calculate
the Fisher matrix which, for the galaxy power spectrum, can be written as
\citep[e.g.,][]{Seo:2003p11154}
\begin{eqnarray}
F_{ij} = \int^{k_{{\rm max}}}_{0}\frac{k^2 dk}{2\pi^{2}}\int^{1}_{0}\frac{\partial {\rm{ln}} P_{\lya{},s}}{\partial \theta_{i}}\frac{\partial {\rm{ln}} P_{\lya{},s}}{\partial \theta_{j}}w(k,\mu)d\mu,
\end{eqnarray}
where $w(k,\mu)$ is the weight given by
\begin{eqnarray}
w(k,\mu) \equiv \frac{1}{2}\left[\frac{n_{g}P_{\lya{},s}(k,\mu)}{1+n_{g}P_{\lya{},s}(k,\mu)}\right]^{2}V_{survey}.
\end{eqnarray}
Here, $\theta_{i}$ is the parameter set of our $i$th dimensional model,
$n_{g}$ is the number density of LAE sources, and $V_{survey}$ is the
volume of the redshift survey. 

We focus our attention on a survey like HETDEX, for which we assume the
linear galaxy bias to be $b_{1} = 2.2$. We generate constraints assuming
measurement of one redshift bin at the midpoint of the HETDEX redshift
range, $z_{{\rm min}} = 1.9$ and $z_{{\rm max}} = 3.5$. At this redshift we have $f
= 0.972$ for the growth rate of structure. We assume HETDEX
will detect 800$\,$000 LAE galaxies in a total survey area of 420
sq. deg. We restrict our analysis to the weakly non-linear regime,
selecting a maximum wavenumber, $k_{{\rm max}} = 0.3 \,h\,$Mpc$^{-1}$.

To generate the 1-$\sigma$ constraints for our cosmological parameters,
we construct the one-dimensional maximum likelihood distribution from
the Fisher matrix assuming a Gaussian distribution. The likelihood for
the $i$th model parameter is 
\begin{eqnarray} 
\mathscr{L}(\bmath{x}_{i}) = {\rm exp}\left[-\frac{1}{2}\bar{x}^{2}_{i} \left( F_{ii} - \sum^{n-1}_{j,k\neq i}F_{ij}(\bar{F}_{jk})^{-1}F_{ki}\right)\right],
\end{eqnarray}
where $\bar{x}_{i} \equiv (x_{i} - x_{o})$ is defined to be the
cosmological parameter value, $x_{i}$, subtracted by its fiducial value,
$x_{o}$. $F$ is the full Fisher matrix of the $n$ parameter model, and
$\bar{F}$ the reduced Fisher matrix of the $n-1$ parameter model with
the $i$th row and column removed. 

To generate the two-dimensional joint constraints on any two parameters, we use
\begin{eqnarray} \label{eq:2Dlikelihood}
\mathscr{L}(\bmath{x}_{i},\bmath{x}_{j}) &=& {\rm exp}\left\{-\frac{1}{2}\left[ \bar{x}^{2}_{i} \left( F_{ii} - \sum^{n-2}_{k,l\neq i}F_{ik}(\bar{F}_{kl})^{-1}F_{li} \right) \right. \right. \nonumber \\
& & \left. \left. +  \bar{x}^{2}_{j}\left(F_{jj} - \sum^{n-2}_{k,l\neq j}F_{jk}(\bar{F}_{kl})^{-1}F_{lj}\right) \right. \right. \nonumber \\
& & \left. \left. + 2 \bar{x}_{i}\bar{x}_{j}\left(F_{ij} - \sum^{n-2}_{k,l\neq i,j}F_{ik}(\bar{F}_{kl})^{-1}F_{lj}\right)\right]\right\}, \nonumber \\
\end{eqnarray}
which contains the cross term which determines the correlation between
the two parameters being considered. Here, $F$ is the full Fisher matrix of the $n$ parameter model, and $\bar{F}$ is the reduced Fisher matrix of the $n-2$ parameter model, with the $i$th and $j$th rows and columns removed. 

To investigate the degeneracies due to \lya{} radiative transfer
parameters, we consider recovery of cosmological parameters from three
power spectra: 
\begin{itemize}
\item[1.] The galaxy power spectrum given by Equation~\ref{eq:galbeta}, 
\item[2.] A fiducial LAE power spectrum given by 
Equation~\ref{eq:PStilde} with the fiducial values of radiative transfer
parameters set to vanish, i.e., $C_\Gamma=C_\rho=C_v=0$ (but these
radiative transfer parameters are marginalized over), and 
\item[3.] A LAE power spectrum given by 
Equation~\ref{eq:PStilde} with the fiducial values of radiative transfer
parameters set to some indicative values.
\end{itemize}

\begin{table}
\begin{tabular}{@{}lccc}
\hline
Parameter & Marginalization & PS  \\
&  & 1-$\sigma$ (per cent)\\
\hline
$\beta$ & ${\rm ln}(A)$ & 0.0091 (2.06)& \\
$\beta$ & ${\rm ln}(A),{\rm ln}(D_{A}),{\rm ln}(H)$ & 0.0213 (4.82)& \\
${\rm ln}(D_{A})$ & ${\rm ln}(A),\beta,{\rm ln}(H)$ & 0.0110 (1.10)\\
${\rm ln}(H)$ & ${\rm ln}(A),\beta,{\rm ln}(D_{A})$ & 0.0132 (1.32)\\
\hline
\end{tabular}
\caption{The 1-$\sigma$ constraints for the linear redshift-space
 distortion parameter, $\beta$, the angular diameter distance, ${\rm
 ln}(D_{A})$, and the Hubble rate, ${\rm ln}(H)$, for a 
 galaxy redshift survey with HETDEX-like survey
 parameters. The other model parameters are 
 marginalized over, but no \lya{} radiative transfer effects are
 included, i.e., the power spectrum is given by Equation \ref{eq:galbeta}.}
\label{tab:Beta}
\end{table}

\begin{figure*} 
	\begin{center}
		\includegraphics[trim = 2.5cm 2.5cm 2.2cm 2cm, scale = 0.27]{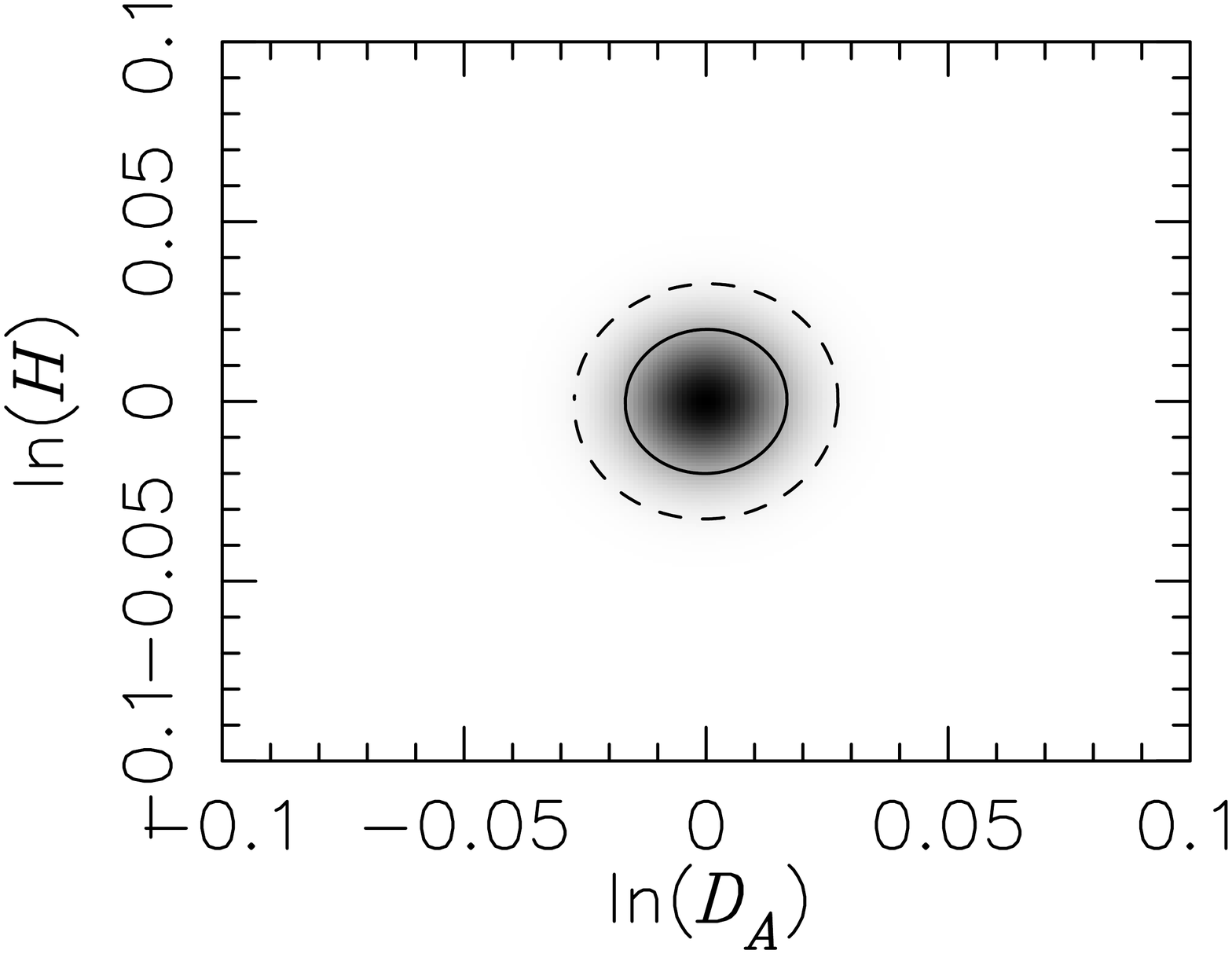}
		\includegraphics[trim = 6.5cm 2.5cm 2.2cm 2cm, scale = 0.27]{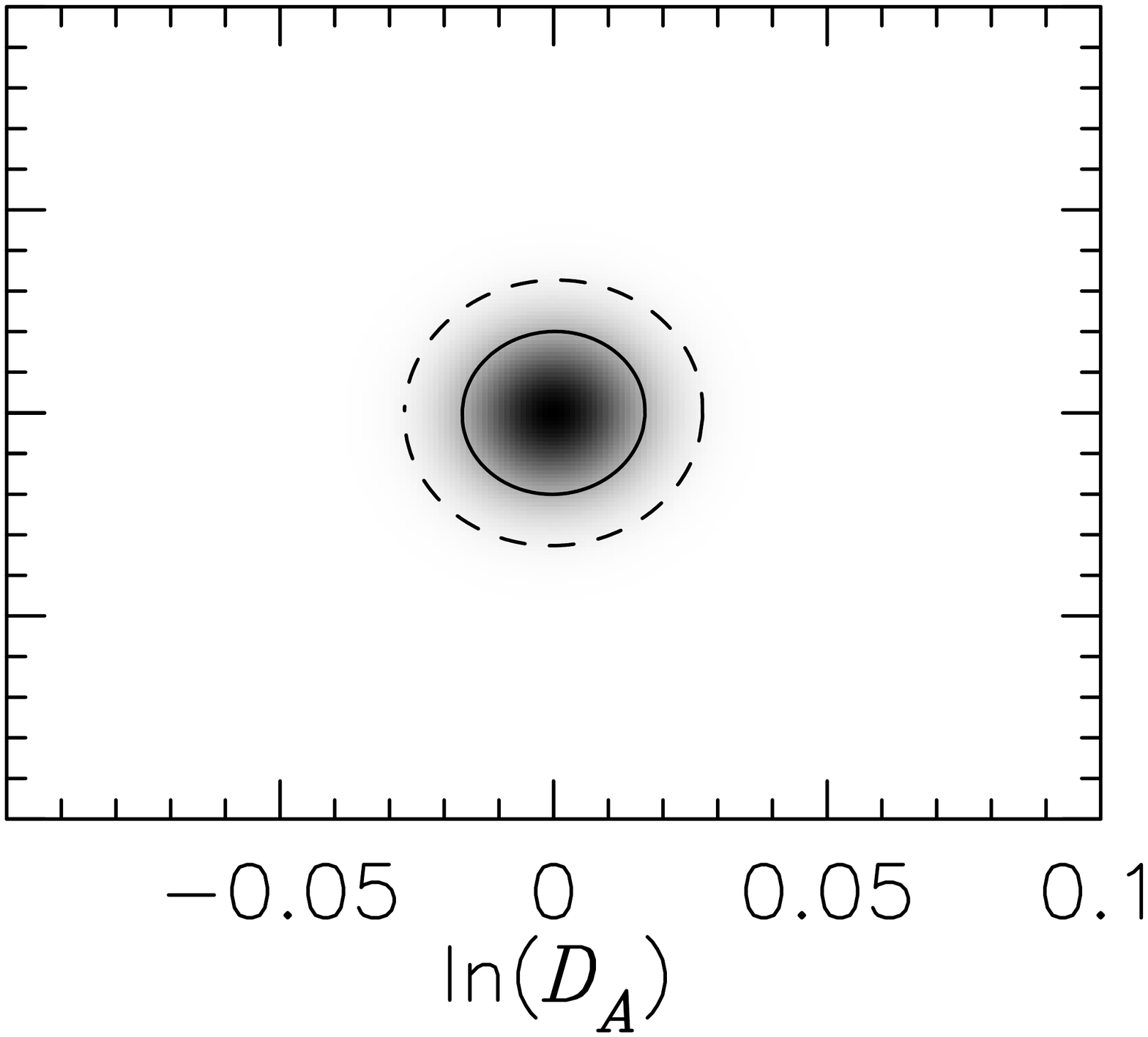}
		\includegraphics[trim = 6.5cm 2.5cm 2.2cm 2cm, scale = 0.27]{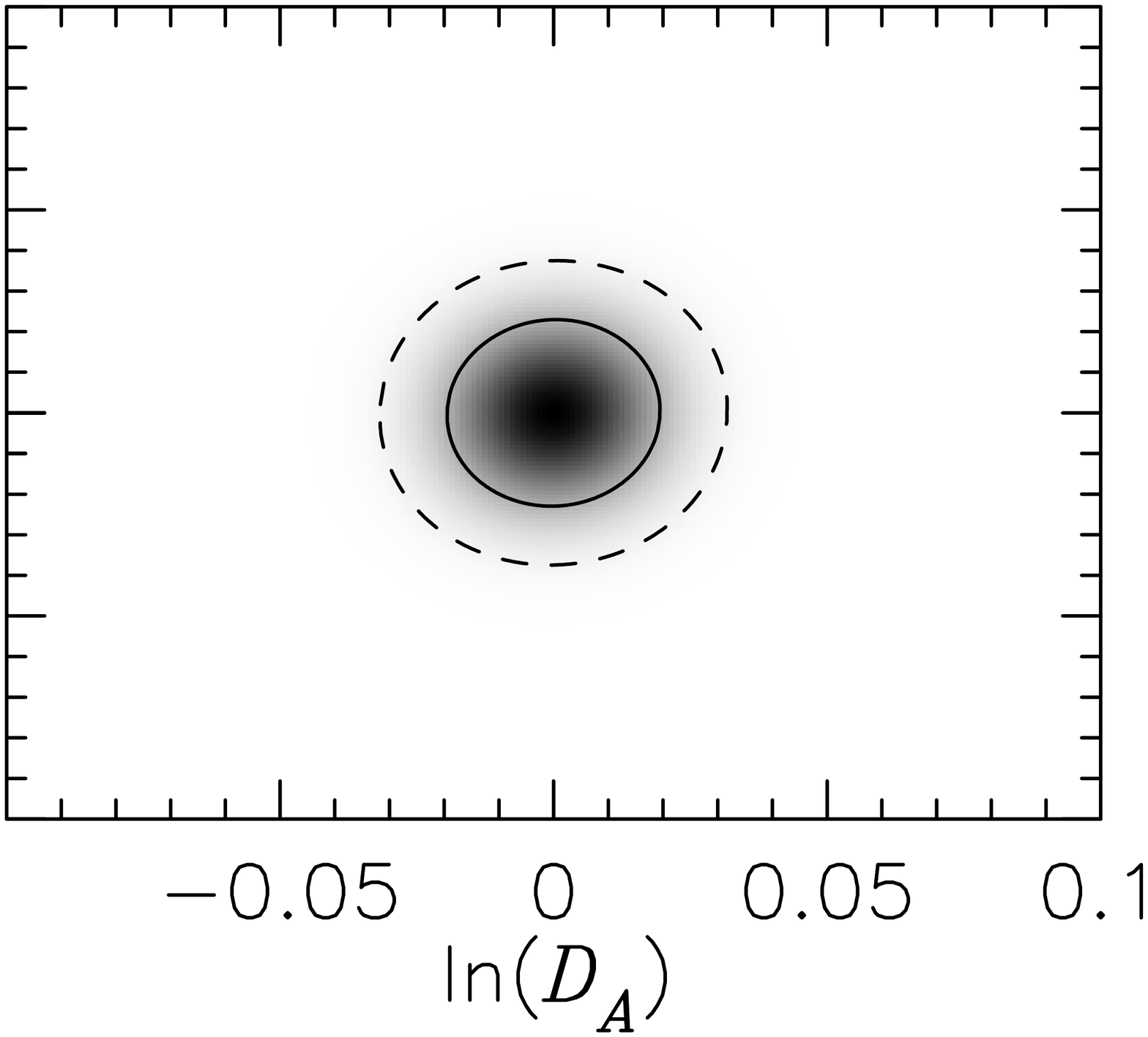}
	\end{center}
\caption{Two-dimensional marginalized joint distribution for the two
 cosmological distance scales: the angular diameter distance ($D_{A}$)
 and the Hubble rate ($H$). (\textit{Left}) a typical galaxy redshift
 survey (no \lya{} effects; no marginalization over $C_{v}$),
 (\textit{middle}) a fiducial LAE  galaxy redshift survey
 (the fiducial values of the \lya{} radiative transfer
 parameters set to vanish; marginalized over $C_{v}$) and
 (\textit{right}) a LAE galaxy redshift survey including the first-order
 \lya{} radiative transfer effects given by $C_{\Gamma} = 0.05$,
 $C_{\rho} = -0.39$, and $C_{v} = 0.11$. The error ellipse
 is slightly bigger for this case because the effective bias of LAE
 galaxies, $\tilde{b}_1=1.9$, is about 15\% smaller than the fiducial
 value, $b_1=2.2$, reducing the amplitude of the power spectrum relative
 to the shot noise.
 The solid and
 dashed curves show the 1- and 2-$\sigma$ constraints
 generated from the likelihood distribution, respectively. Scale selected to aid
 comparison with Figure \ref{fig:DA_Hcomplete}.} 
\label{fig:DA_H}
\end{figure*}

\begin{table*}
\begin{tabular}{@{}lcccccc}
\hline
Parameter & Marginalization & No Priors on $C_v$ & Perfect knowledge & $\sigma_{C_{v}} = 0.01$ & $\sigma_{C_{v}} = 0.1$ & $\sigma_{C_{v}} = 0.5$ \\
&  & 1-$\sigma$ (per cent) & 1-$\sigma$  (per cent) & 1-$\sigma$  (per cent) & 1-$\sigma$  (per cent) & 1-$\sigma$  (per cent) \\
\hline

$\tilde{\beta}$ & ${\rm ln}(A)$,$C_{v}$ & - & 0.0091 (2.06) & 0.0101 (2.29) & 0.0451 (10.21) & 0.2211 (50.04)\\
$\tilde{\beta}$ & ${\rm ln}(A),{\rm ln}(D_{A}),{\rm ln}(H)$,$C_{v}$ & - & 0.0213 (4.82) & 0.0218 (4.93) & 0.0491 (11.11) & 0.2220 (50.24)\\
${\rm ln}(D_{A})$ & ${\rm ln}(A),\tilde{\beta},{\rm ln}(H)$,$C_{v}$ & 0.0110 (1.10) & 0.0110 (1.10)& 0.0110 (1.10)& 0.0110 (1.10) & 0.0110 (1.10)\\
${\rm ln}(H)$ & ${\rm ln}(A),\tilde{\beta},{\rm ln}(D_{A})$,$C_{v}$ & 0.0132 (1.32)& 0.0132 (1.32) & 0.0132 (1.32)& 0.0132 (1.32)& 0.0132 (1.32)\\
\hline
\end{tabular}
\caption{The 1-$\sigma$ constraints for $\tilde{\beta}$, $\ln(D_{A})$, and
 $\ln(H)$ for our fiducial LAE galaxy redshift survey (the fiducial
 values of the \lya{} radiative transfer parameters set to vanish) marginalized 
 over remaining model parameters shown in the second column. We compare
 varying priors added to the  
 radiative transfer parameter, $C_{v}$, which suffers from a large
 degeneracy with $\tilde{\beta}$. The power spectrum model is given by Equation \ref{eq:PStilde}.}
\label{tab:betapriorsfiducial}
\end{table*}

\subsection[Galaxy power spectrum]{Galaxy power spectrum}

For the galaxy power spectrum given by Equation~\ref{eq:galbeta}, the galaxy
bias is completely degenerate with the amplitude of the power spectrum
($\sigma_{8}$). Furthermore, one cannot directly measure the growth rate
of structure, $f$, but only the parameter $\beta$. We show later that by
considering the bispectrum, one can directly probe $f$. 

The cosmological parameters we determine from this model are therefore
the overall amplitude [${\rm ln}(A)$], the linear redshift-space distortion parameter [$\beta$], and the two distance measurements given by the angular diameter distance [${\rm ln}(D_{A})$] and the Hubble rate [${\rm ln}(H)$]. The information on the galaxy bias is factored into the linear redshift-space distortion parameter and we redefine the amplitude to include galaxy bias. The constrains for $\beta$ as well as ${\rm ln}(D_{A})$ and ${\rm ln}(H)$ represent the best case scenario for how accurately we can recover the cosmological parameters from a galaxy power spectrum analysis. 

In Table \ref{tab:Beta}, we provide the 1-$\sigma$ constraints on
$\beta$, and the two distance scales, $D_{A}$ and $H$. For a galaxy
redshift survey with HETDEX-like survey parameters,  the
expected uncertainty on the linear redshift-space
distortion, $\beta$, is $0.021$ ($4.8$ per
cent), on the angular diameter distance it is $1.1$
per cent, and on the Hubble rate it is $1.3$ per
cent. Our distance constraints for the above model are consistent with
the results of \citet{Shoji:2009p8219}.  

Of particular interest for cosmological analyses is the two-dimensional joint constraints on the two distance measures, ${\rm ln}(D_{A})$ and ${\rm ln}(H)$. In the left panel of Figure \ref{fig:DA_H}, we show the 1-$\sigma$ and 2-$\sigma$ joint constraints on ${\rm ln}(D_{A})$ and ${\rm ln}(H)$, which give the baseline for comparison with the recovery of the cosmological distance parameters for $D_{A}$ and $H$ for the remainder of this work.

\begin{figure*} 
	\begin{center}
		\includegraphics[trim = 4cm 3cm 2.2cm 2cm, scale = 0.39]{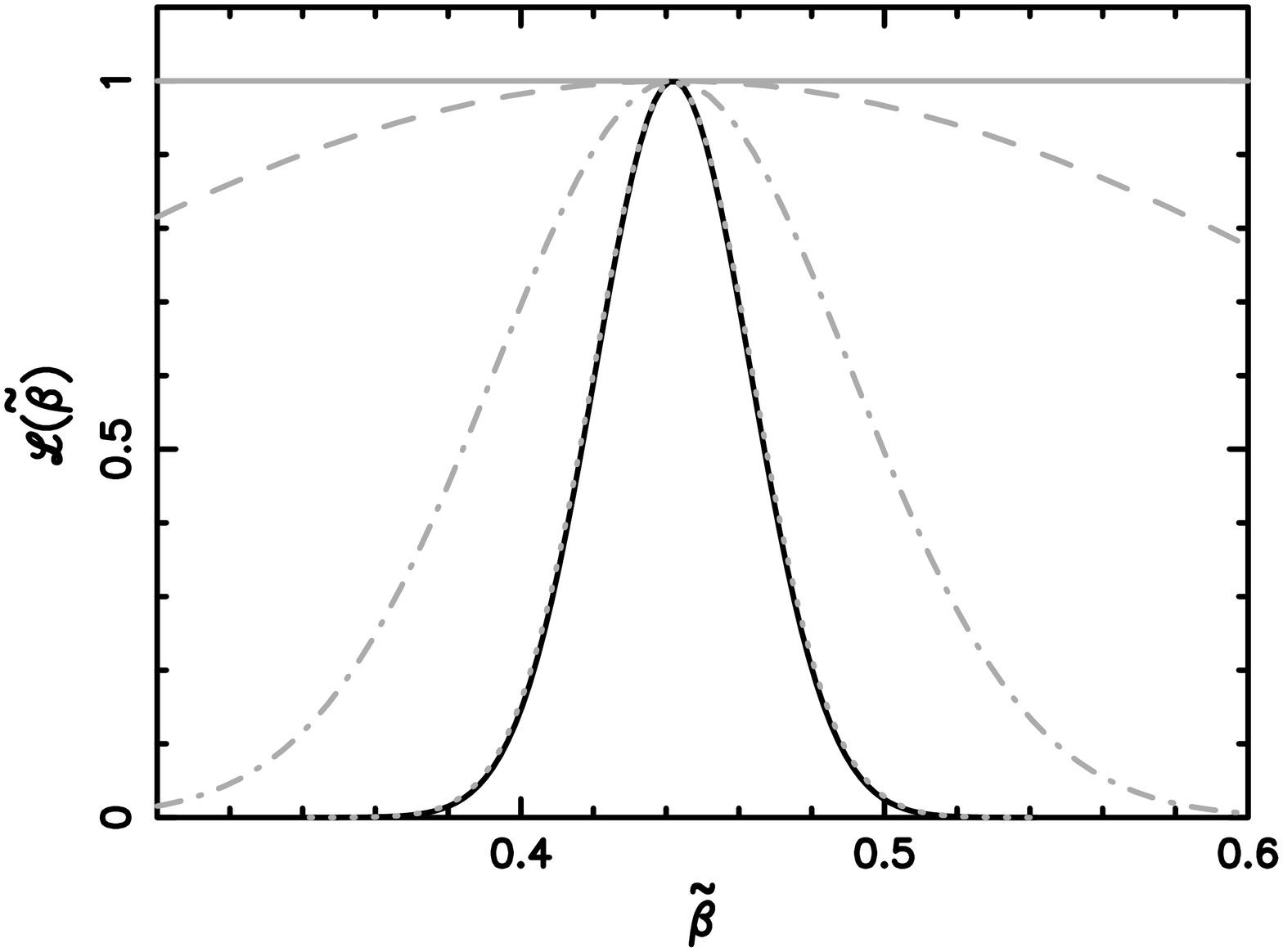}
		\includegraphics[trim = 6cm 3cm 2.2cm 2cm, scale = 0.39]{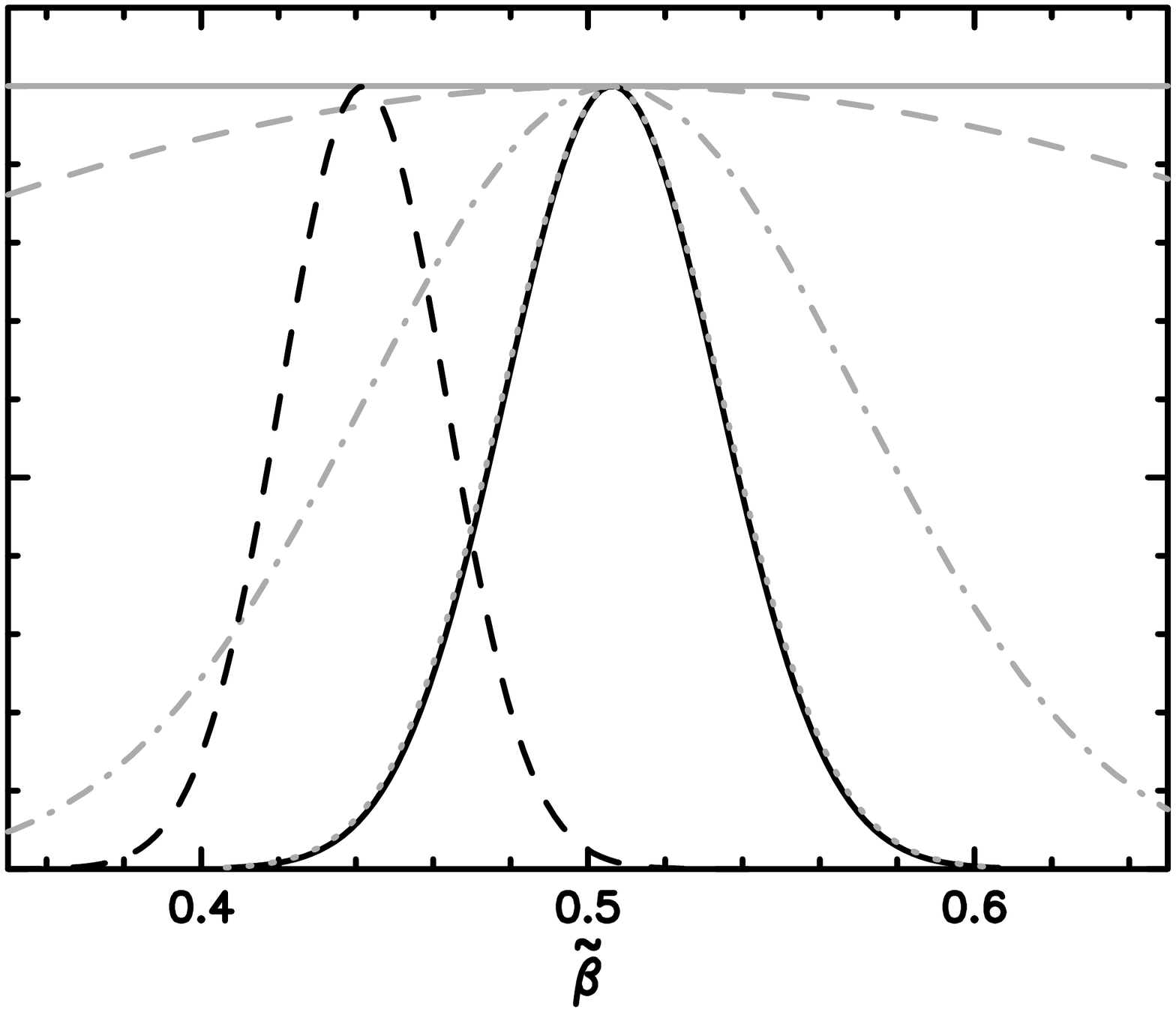}
	\end{center}
\caption{One-dimensional marginalized likelihood distributions for the
 linear redshift-space distortion parameter ($\tilde{\beta}$) generated from;
 (\textit{left}) the fiducial LAE galaxy power spectrum (no \lya{}
 effects, but marginalized over $C_{v}$), and (\textit{right}) a LAE
 galaxy power spectrum including \lya{} radiative transfer effects
 ($C_{\Gamma} = 0.05$, $C_{\rho} = -0.39$, and $C_{v} = 0.11$). The
 resultant offset in $\tilde{\beta}$ in the LAE galaxy power spectrum
 (\textit{right panel}) is due to the modified bias because of the
 included \lya{} radiative transfer effects. The various curves denote
 different priors added to $C_{v}$ and are as follows, \textit{black solid:}
 perfect knowledge ($\sigma_{C_{v}} = 0.0001$), \textit{grey dotted:}
 $\sigma_{C_{v}} = 0.01$, \textit{grey dot-dashed:} $\sigma_{C_{v}} = 0.1$,
 \textit{grey dashed:} $\sigma_{C_{v}} = 0.5$, and \textit{grey solid:} no priors added. In the right panel, the \textit{black dashed} offset curve is the comparison to the case corresponding to
 perfect knowledge on $C_{v}$ from the fiducial LAE galaxy power
 spectrum (\textit{black solid curve, left panel}). } 
\label{fig:betalikelihoods}
\end{figure*}

\begin{table*}
\begin{tabular}{@{}lccccccccc}
\hline
Parameter & Marginalization & No Priors on $C_v$ & Perfect knowledge & $\sigma_{C_{v}} = 0.01$ & $\sigma_{C_{v}} = 0.1$ & $\sigma_{C_{v}} = 0.5$ \\
&  & 1-$\sigma$ (per cent) & 1-$\sigma$ (per cent) & 1-$\sigma$ (per cent) & 1-$\sigma$ (per cent) & 1-$\sigma$ (per cent) \\
\hline

$\tilde{\beta}$ & ${\rm ln}(A)$,$C_{v}$ & - & 0.0121 (2.39)& 0.0134 (2.65)& 0.0582 (11.49)& 0.2847 (56.23)\\
$\tilde{\beta}$ & ${\rm ln}(A),{\rm ln}(D_{A}),{\rm ln}(H)$,$C_{v}$ & - & 0.0278 (5.49) & 0.0283 (5.59) & 0.0633 (12.50)& 0.2858 (56.45)\\
${\rm ln}(D_{A})$ & ${\rm ln}(A),\tilde{\beta},{\rm ln}(H)$,$C_{v}$ & 0.0128 (1.28)&0.0128  (1.28)&0.0128  (1.28)&0.0128  (1.28)&0.0128  (1.28)&\\
${\rm ln}(H)$ & ${\rm ln}(A),\tilde{\beta},{\rm ln}(D_{A})$,$C_{v}$ & 0.0151 (1.51)&0.0151 (1.51) &0.0151 (1.51) &0.0151 (1.51) &0.0151 (1.51) &\\
\hline
\end{tabular}
\caption{Same as Table~\ref{tab:Beta}, but for the LAE galaxy redshift survey including first order \lya{} radiative transfer effects, $C_{\Gamma} = 0.05$, $C_{\rho} = -0.39$ and $C_{v} = 0.11$.}
\label{tab:betapriors}
\end{table*}

\subsection{Fiducial LAE galaxy power spectrum}

We model the LAE galaxy power spectrum using Equation \ref{eq:PStilde}. In our fiducial case, we set all \lya{} radiative transfer coefficients to zero, i.e., $\tilde{b}_{1} = b_{1}$, $\tilde{\beta} = \beta$, and $C_{v} = 0$. Although we set the \lya{} radiative transfer effects to zero, we still marginalize over the possible existence of $C_{v}$ in this case.

The amplitude of the power spectrum is completely degenerate with the
modified galaxy bias, and so we can redefine the amplitude to include
$\tilde{b}_{1}$. The LAE galaxy power spectrum includes  $C_{v}$, the
\lya{} radiative transfer effect associated with the line-of-sight
peculiar velocity gradient. Hence, for the LAE galaxy power spectrum,
the model contains 5 parameters, $\ln(A)$, $\tilde{\beta}$, $C_{v}$,
$\ln(D_{A})$, and $\ln(H)$. The linear redshift-space distortion
parameter, $\tilde{\beta}$, and the radiative transfer effect, $C_{v}$,
are completely degenerate; however, with the addition of priors on
$C_{v}$, one can break the degeneracy and improve the constraints on the
linear distortion parameter, $\tilde{\beta}$ \citep{Wyithe:2011p12569}. 

In Table \ref{tab:betapriorsfiducial}, we provide the resulting
1-$\sigma$ constraints on the linear redshift-space distortion
parameter, $\tilde{\beta}$, as well as the distance constraints,
marginalized over the remaining model parameters including $C_{v}$. The
columns from left to right in Table \ref{tab:betapriorsfiducial}
consider priors added to $C_{v}$; essentially perfect knowledge of
$C_{v}$, $\sigma_{C_{v}} = 0.0001$\footnote{Throughout this work we
define the `perfect knowledge on $C_{v}$' as a prior set to $\sigma_{C_{v}} =
0.0001$, which is to ensure that our Fisher matrix elements remain
finite, yet still mimic the behaviour for perfectly understood
parameters.}, $\sigma_{C_{v}} = 0.01$, $\sigma_{C_{v}} = 0.1$, and
$\sigma_{C_{v}} = 0.5$. The inclusion of $C_{v}$ into the model
significantly impacts the recovery of the linear redshift-space
distortion parameter, $\tilde{\beta}$, whereas the distance constraints
remain unaffected by the marginalization over the radiative transfer
effects. This differs from \citet{Wyithe:2011p12569} where inclusion of
scale-dependent ionizing background fluctuations lead to reduced
distance constraints. 

With sufficiently tight priors on $C_{v}$, the constraints on
$\tilde{\beta}$ approach the results given in the previous
section, as expected. If we have a poor understanding of $C_{v}$,
however, the ability to recover $\tilde{\beta}$ drops by an order of
magnitude. See the left panel of Figure
\ref{fig:betalikelihoods} for a graphical representation of the effect
of the priors.

On the other hand, the distance constraints are unaffected
by the degeneracy between $\tilde{\beta}$ and $C_{v}$. (Compare the middle
panel of Figure~\ref{fig:DA_H} with the left panel.) This is because
$\beta$ and $C_v$ enter into the power spectrum in the same way: as far
as the distance scales are concerned, it
makes no difference whether one marginalizes over $\beta$ in
Equation~\ref{eq:galbeta} or $\tilde{\beta}(1-C_v)$ in Equation~\ref{eq:PStilde}.

\subsection{LAE galaxy power spectrum}

\citet{Wyithe:2011p12569} show that the magnitude of the \lya{} radiative transfer parameters varies significantly depending on the LAE model considered. In particular the magnitude of the effect is significantly larger in the absence of a galactic outflow, so that the absorption is dominated by infalling \igm{}. For illustration we consider this `infall' model with an escape fraction of 10 per cent, as this is the model with the largest magnitude \lya{} radiative transfer effects. Hence, when we include the radiative transfer effects into our model, we set $C_{\Gamma} = 0.05$, $C_{\rho} = -0.39$, and $C_{v} = 0.11$ in Equation \ref{eq:PStilde}.

In Table \ref{tab:betapriors}, we provide estimates for the recovery of
cosmological parameters when we include the radiative transfer
effects. The linear bias is modified from its fiducial value of $b_{1} =
2.2$ to $\tilde{b}_{1} = 1.9$, acting to reduce the observed clustering
of LAE galaxies. As the level of the shot noise is the same,
a reduced effective bias implies a lower signal-to-noise for measuring
the power spectrum of LAE galaxies. As a result, the expected
constraints on the angular diameter distance and the Hubble rate are
worse than the previous two cases. One can see this clearly in the right
panel of Figure~\ref{fig:DA_H}.

The same is true for $\tilde{\beta}$: due to a smaller
effective bias, the fractional precision by which we can determine
$\tilde{\beta}$ is slightly worse than the previous cases. (One can see
this by comparing the rows of `$\beta$' in
Table~\ref{tab:betapriorsfiducial} and \ref{tab:betapriors}.) Note also
that, as the fiducial value of $\tilde{b}_1$ is different due to the
radiative transfer parameters, the fiducial value of
$\tilde{\beta}=f/\tilde{b}_1$ is also different.

In Figure~\ref{fig:betalikelihoods}, we show the one-dimensional
likelihood distributions for $\tilde{\beta}$ for both the fiducial LAE
galaxy model (left panel) and the LAE galaxy model (with \lya{} effects,
right panel). As already described in the previous section, the recovery
of $\tilde{\beta}$ from these models is highly sensitive to the priors
on $C_v$,
with marginal improvement on the priors breaking the degeneracy between
$\tilde{\beta}$ and $C_{v}$. In the right panel, we compare the
likelihood distribution for the case of perfect knowledge of $C_{v}$ for
the fiducial LAE galaxy model (the dashed line) to the LAE galaxy model
(the solid line), showing the 
degree of offset that the \lya{} radiative transfer parameters have on
the fiducial value of $\tilde{\beta}$. 

Thus inclusion of the \lya{} radiative transfer parameters impacts the
recovery of cosmological constraints, most notably the growth rate of
structure $f$ through the recovery of the linear redshift-space
distortion parameter $\tilde{\beta}$. Unless we have a good
prior knowledge on the value of $C_v$, it seems hopeless to determine
$\tilde{\beta}$ with any precision. Fortunately, one can break the
degeneracy between $\tilde{\beta}$ and $C_v$ by including the
three-point function (bispectrum), as we shall show next.

\section[Bispectrum and non-linear clustering of LAE galaxies]{Bispectrum and non-linear clustering of LAE galaxies} \label{sec:method}

If primordial perturbations are Gaussian,  linear density fields are
also Gaussian, in which case the bispectrum of linear density fields
vanishes. The bispectrum is defined as
\begin{equation}
\langle \delta(\bmath{k}_{1})\delta(\bmath{k}_{2})\delta(\bmath{k}_{3}) \rangle =
(2\pi)^{3}B(\bmath{k}_{1},\bmath{k}_{2},\bmath{k}_{3}) \delta^{D}(\bmath{k}_{1} + \bmath{k}_{2} + \bmath{k}_{3}).
\end{equation}

However, non-linear gravitational evolution of density
fields, and non-linear gravitational and non-gravitational
evolution of galaxy bias make the observed galaxy density fields
non-Gaussian. As a result, the observed bispectrum does not vanish,
providing information regarding non-linear evolution of density fields. 

In the previous section, we have considered the linear-theory power spectrum model for the LAE galaxy population. Structure formation is inherently a non-linear process, and by considering the true non-linear galaxy power spectrum one would expect to increase the constraining power. However, the non-linear power spectrum alone will not achieve this, due to the additional parameters required to fully describe it. Hence the simple linear LAE galaxy power spectrum is preferred instead of the increased model complexity provided by the non-linear LAE galaxy power spectrum. On the other hand additional information on large scales may be contained in the non-Gaussianity associated with structure formation.

We use this information to break the degeneracy between cosmological
parameters and \lya{} radiative transfer parameters. There are three
effects: (1) gravitational evolution of matter density fields, (2)
gravitational and non-gravitational evolution of galaxy formation
(captured by galaxy bias), and (3) non-gravitational \lya{}
radiative transfer effects. 

\subsection[Eulerian perturbation theory]{Eulerian perturbation theory}

First, we summarize the non-linear gravitational evolution of density
fields. Specifically, we apply standard Eulerian perturbation theory
(\citealt{Bernardeau:2002p8437} and references within) which, at larger
redshifts, has been shown to describe the power spectrum
measured from N-body simulations accurately \citep{Jeong:2006p5318}. 

The
next-to-leading order corrections to the matter density field, $\delta$,
as well as to the velocity-divergence field, $\eta$, are generated from the following expressions:
\begin{eqnarray} \label{eq:PTdens}
\delta(\bmath{k},z) &= &\sum^{\infty}_{n=1} D^{n}(z)\int \frac{\rm d^3\bmath{q}_{1}}{(2\pi)^3} \int \frac{\rm d^3\bmath{q}_{n-1}}{(2\pi)^3} \nonumber \\
& & \times \int d^3\bmath{q}_{n}\delta^{D}(\bmath{k}-\sum^{n}_{i=1} \bmath{q}_{i})F^{(s)}_{n} (\bmath{q}_{1},\bmath{q}_{2},...,\bmath{q}_{n}) \nonumber \\
& & \times \delta_{1}(\bmath{q}_{1})\delta_{1}(\bmath{q}_{2})...\delta_{1}(\bmath{q}_{n})
\end{eqnarray}
\begin{eqnarray} \label{eq:PTvel}
\eta(\bmath{k},z) &=& \sum^{\infty}_{n=1} D^{n}(z)\int \frac{\rm d^3\bmath{q}_{1}}{(2\pi)^3} \int \frac{\rm d^3\bmath{q}_{n-1}}{(2\pi)^3} \nonumber \\
& & \times \int d^3\bmath{q}_{n}\delta^{D}(\bmath{k}-\sum^{n}_{i=1} \bmath{q}_{i}) G^{(s)}_{n} (\bmath{q}_{1},\bmath{q}_{2},...,\bmath{q}_{n}) \nonumber \\
& & \times \delta_{1}(\bmath{q}_{1})\delta_{1}(\bmath{q}_{2})...\delta_{1}(\bmath{q}_{n}),
\end{eqnarray}
where $D(z)$ is the linear growth factor describing the evolution of the
linear density field, $\delta_{1}(\bmath{q}_{i})$, which is a Gaussian random
field, and  $F^{(s)}_{n}$ and $G^{(s)}_{n}$ are symmetrized kernel expressions generated from recursive relations \citep{Jain:1994p8584}. We deal only with the next-to-leading order expressions, for which the kernels are well known:
\begin{equation}
F^{(s)}_{2}(\bmath{q}_{1},\bmath{q}_{2})  =  \frac{5}{7} + \frac{2}{7}\frac{(\bmath{q}_{1}\cdot\bmath{q}_{2})^{2}}{q^{2}_{1}q^{2}_{2}} + \frac{\bmath{q}_{1}\cdot\bmath{q}_{2}}{2}\left(\frac{1}{q^{2}_{1}} + \frac{1}{q^{2}_{2}}\right),
\end{equation}
\begin{equation}
G^{(s)}_{2}(\bmath{q}_{1},\bmath{q}_{2})  =  \frac{3}{7} + \frac{4}{7}\frac{(\bmath{q}_{1}\cdot\bmath{q}_{2})^{2}}{q^{2}_{1}q^{2}_{2}} + \frac{\bmath{q}_{1}\cdot\bmath{q}_{2}}{2}\left(\frac{1}{q^{2}_{1}} + \frac{1}{q^{2}_{2}}\right).
\end{equation}

\subsection[Galaxy bias]{Galaxy bias}

Galaxies are biased tracers of the underlying dark matter density field
\citep{Kaiser:1984p8759}.  Pushing into the weakly non-linear regime, we
anticipate contributions from both the linear and non-linear mapping of
galaxies to the dark matter field. 

The bias of galaxies differs from population to population, and their
exact value depends on the underlying galaxy formation
processes. Typically we expect a scale-dependant bias relating the
clustering of the galaxies to the underlying matter density on small
scales, but on large scales we expect the bias to be scale-independent. 

To estimate the clustering of galaxies, we Taylor-expand the
fluctuations in the number density of galaxies, $\delta_{g}(\bmath{x})$, in terms of the underlying matter density field fluctuations (\citealt{Fry:1993p8713,McDonald:2006p8376}):
\begin{eqnarray} \label{eq:galaxybias}
\delta_{g}(\bmath{x}) = \epsilon(\bmath{x}) + b_{1}\delta(\bmath{x}) + \frac{1}{2}b_{2}\delta(\bmath{x})^{2} + ...\,,
\end{eqnarray}
where $\delta(\bmath{x})$ is the non-linear matter density field, and $b_{1}$ and
$b_{2}$ are the linear and non-linear bias parameters, respectively. The
$\epsilon(\bmath{x})$ term is a stochasticity parameter describing the
non-deterministic relationship between galaxies and the underlying
matter distribution \citep{Yoshikawa:2001p8880}. 
We shall assume that
$\epsilon$ is a Gaussian field which is not correlated with $\delta$,
i.e., $\langle\epsilon^3\rangle=0$ and
$\langle\epsilon\delta\rangle=0$. Under this assumption, $\epsilon$ does
not contribute to the bispectrum, and thus we shall ignore stochasticity
throughout this paper.


\subsection[LAE kernel expressions]{LAE kernel expressions}

In Section \ref{sec:1stOrderLya}, we outline the derivation for the
first-order \lya{} radiative transfer effects. To derive higher-order
expressions for the \lya{} radiative transfer effects, we 
Taylor-expand 
$n_{\rm Ly\alpha}(>F_{0})$ about the three non-gravitational \lya{}
radiative transfer effects, in analogy to the
galaxy bias derivation \citep{Fry:1993p8713}.
We obtain 
\begin{eqnarray} \label{eq:Taylor2nd}
\bar{n}_{\rm Ly\alpha}(>L_{0},\rho_{0},\Gamma_{0},\delta(\bmath{x})) = \bar{n}^{(0)}_{\rm Ly\alpha}\left(1 + \bar{n}^{(1)}_{\lya{}} + \bar{n}^{(2)}_{\lya{}} \right),
\end{eqnarray}
where $\bar{n}^{(1)}_{\lya{}}$ and $\bar{n}^{(2)}_{\lya{}}$ are the
first- and second-order Taylor-expanded expressions, respectively,
evaluated around the 
mean quantity $\bar{n}^{(0)}_{\rm Ly\alpha}$. The term
$\bar{n}^{(1)}_{\lya{}}$ is given in Equation \ref{eq:Taylorex1}, and
$\bar{n}^{(2)}_{\lya{}}$ is given by 
\begin{eqnarray} \label{eq:Taylorex2}
\bar{n}^{(2)}_{\lya{}} & = & \frac{1}{2}\frac{1}{\bar{n}^{(0)}_{\rm Ly\alpha}}\left[(\Gamma - \Gamma_{0})\frac{\partial}{\partial\Gamma}\line(0,1){15}_{\line(0,1){15}F_{0},\Gamma_{0}} + (\rho - \rho_{0})\frac{\partial}{\partial\rho}\line(0,1){15}_{\line(0,1){15}F_{0},\rho_{0}} \right. \nonumber \\
& & \left.+\left(\frac{{\rm d} v_{z}}{{\rm d}(a r_{\rm com})}- H\right)\frac{\partial}{\partial\frac{{\rm d} v_{z}}{{\rm d}(a r_{\rm com})}}\line(0,1){15}_{\line(0,1){15}F_{0},\rho_{0}}\right]^{2}\bar{n}_{\rm Ly\alpha}.\nonumber \\
\end{eqnarray}
The number density of LAE galaxies above a flux limit can then be written as
\begin{eqnarray} \label{eq:intermediate2ndOrder}
n_{\rm Ly\alpha}(>F_{0}) = \bar{n}^{(0)}_{\rm Ly\alpha}[1+\delta_{g}(\bmath{x})]\left[1+\bar{n}^{(1)}_{\lya{}} + \bar{n}^{(2)}_{\lya{}}\right],
\end{eqnarray}
where we have substituted Equation \ref{eq:Taylor2nd} into Equation
\ref{eq:numdens}. 

To proceed further, we firstly expand the expressions in Equations \ref{eq:Taylorex1} and \ref{eq:Taylorex2}, and then recast the above derivatives as explicit constants with respect to their radiative transfer effect. Once this has been performed, we can rewrite Equations \ref{eq:Taylorex1} and \ref{eq:Taylorex2} as
\begin{eqnarray} \label{eq:1stOrder}
\bar{n}^{(1)}_{\lya{}} & = & \delta_{\Gamma}(\bmath{x})C_{\Gamma} + \delta_{\rho}(\bmath{x}) C_{\rho} + \delta_{v}(\bmath{x})C_{v}
\end{eqnarray}
and,
\begin{eqnarray} \label{eq:2ndOrder}
\bar{n}^{(2)}_{\lya{}} & = &  \frac{1}{2}\left[C_{\Gamma \Gamma}\delta_{\Gamma}^{2}(\bmath{x}) + C_{\rho \rho}\delta_{\rho}^{2}(\bmath{x}) + C_{v v}\delta_{v}^{2}(\bmath{x})\right]  \nonumber \\
& & + C_{\Gamma \rho}\delta_{\Gamma}(\bmath{x})\delta_{\rho}(\bmath{x}) + C_{\Gamma v}\delta_{\Gamma}(\bmath{x})\delta_{v}(\bmath{x}) \nonumber \\
& & + C_{\rho v}\delta_{\rho}(\bmath{x}) \delta_{v}(\bmath{x}).
\end{eqnarray}

Equations \ref{eq:1stOrder} and \ref{eq:2ndOrder} contain the first- and
second-order \lya{} radiative transfer coefficients. 
In Appendix \ref{app:Lyaeffects} we derive the explicit expressions for
the first-order \lya{} radiative transfer coefficients, and use the same
basic ideas to also derive the second-order coefficients.

To the second order, the fluctuations in the number density of LAE galaxies relative to the mean are given by
\begin{equation} \label{eq:Lyafluct2ndOrder}
\delta_{\lya{}}(\bmath{x}) = [1+\delta_{g}(\bmath{x})]\left[1+\bar{n}^{(1)}_{\lya{}}+\bar{n}^{(2)}_{\lya{}}\right] - 1.
\end{equation}
To generate the \lya{} radiative transfer kernels that describe the
 modification to the galaxy power spectrum and the higher-order
 corrections, we expand Equation \ref{eq:Lyafluct2ndOrder}
up to the second order in fluctuations, and take the
 Fourier transform of the corresponding expression (Appendix
 \ref{app:full}). In the resulting expression, the fluctuations of the
 local density, $\delta_{\rho}(\bmath{x})$, are given by Equation
 \ref{eq:PTdens}. The fluctuations in the line-of-sight peculiar velocity field in Fourier space can be expressed as $\delta_{v}(\bmath{k}) = - f\mu^{2}\eta(\bmath{k})$, where $\eta(\bmath{k})$ is the fluctuation in the velocity field given by Equation \ref{eq:PTvel}. The expression due to fluctuations in the UV background can be somewhat more complicated and is outlined in Appendix \ref{app:full}. 

In Appendix \ref{app:full} we also derive the redshift-space expressions
for the \lya{} radiative transfer effects. Finally, in analogy with
Equations \ref{eq:PTdens} and \ref{eq:PTvel}, we write the
fluctuations in the number density of LAE galaxies in real space as
\begin{eqnarray} \label{eq:realspace}
\delta_{\lya{}}(\bmath{k},z) &= &\sum^{\infty}_{n=1} D^{n}(z)\int \frac{\rm d^3\bmath{q}_{1}}{(2\pi)^3} \int \frac{\rm d^3\bmath{q}_{n-1}}{(2\pi)^3}\nonumber \\
& & \times \int d^3\bmath{q}_{n}\delta^{D}(\bmath{k}-\sum^{n}_{i=1} \bmath{q}_{i})Z^{(s)}_{n} (\bmath{q}_{1},\bmath{q}_{2},...,\bmath{q}_{n})  \nonumber \\
& & \times\delta_{1}(\bmath{q}_{1})\delta_{1}(\bmath{q}_{2})...\delta_{1}(\bmath{q}_{n}),
\end{eqnarray}
and those in redshift-space as
\begin{eqnarray} \label{eq:redshiftspace}
\delta_{\lya{},s}(\bmath{k},z) &= &\sum^{\infty}_{n=1} D^{n}(z)\int \frac{\rm d^3\bmath{q}_{1}}{(2\pi)^3} \int \frac{\rm d^3\bmath{q}_{n-1}}{(2\pi)^3}  \nonumber \\
& & \times \int d^3\bmath{q}_{n}\delta^{D}(\bmath{k}-\sum^{n}_{i=1} \bmath{q}_{i})K^{(s)}_{n} (\bmath{q}_{1},\bmath{q}_{2},...,\bmath{q}_{n})  \nonumber \\
& & \times\delta_{1}(\bmath{q}_{1})\delta_{1}(\bmath{q}_{2})...\delta_{1}(\bmath{q}_{n}),
\end{eqnarray}
where the \lya{} radiative transfer kernels, $Z^{(s)}_{n}$ and
 $K^{(s)}_{n}$, are given in Appendix \ref{app:full}.

\subsection[LAE bispectrum and reduced
  bispectrum]{LAE bispectrum and reduced bispectrum}

The second-order terms in
Equation~\ref{eq:redshiftspace} yield a non-vanishing bispectrum of the
fluctuations in the number density of LAE galaxies in redshift space:
\begin{eqnarray} \label{eq:bispectrum}
B_{\lya{},s}(\bmath{k}_{1},\bmath{k}_{2},\bmath{k}_{3}) &=& 2\left[K^{(s)}_{1}(\bmath{k}_{1})K^{(s)}_{1}(\bmath{k}_{2})K^{(s)}_{2}(\bmath{k}_{1},\bmath{k}_{2}) \right. \nonumber \\
& & \left. \times P_{L}(k_{1})P_{L}(k_{2}) + ({\rm 2\,\,cyc.})\right],
\end{eqnarray}
where $K^{(s)}_{1}(\bmath{k})$ and
$K^{(s)}_{2}(\bmath{k}_{1},\bmath{k}_{2})$ are found in Appendix \ref{app:full}.

The bispectrum in redshift space depends on six
variables: three wavenumbers, $k_{1}$, $k_{2}$, and $k_{3}$, giving the
sides of a triangle; and the cosines of the angles that these three
vectors make with the line-of-sight direction, $\mu_{1}$, $\mu_{2}$, and
$\mu_{3}$. However, due to the triangular condition, these six variables
are not all independent. Instead, the bispectrum can be written as a
function of five independent 
variables \citep{Scoccimarro:1999p9850,Smith:2008p8996}: three
parameters ($k_{1}$, $k_{2}$, and the angle between them, ${\rm
cos}(\theta_{12})$) define the shape of the triangle; and the
remaining two parameters ($\mu_{1}$ and $\phi$) define the orientation
of the triangles with respect to the line-of-sight. 

By convention, we align the first wavevector, $k_{1}$, to the
line-of-sight direction, $\hat{\bmath{z}}$, about which the triangle can
be rotated through the azimuthal direction ($\hat{\phi}$). The
cosines of the angles that $k_2$ and $k_3$ make with the line-of-sight
direction are given by
\begin{eqnarray}
\mu_{2} &=& \mu_{1}{\rm cos}(\theta_{12}) - \sqrt{1-\mu^{2}_{1}}{\rm sin}(\theta_{12}){\rm cos}(\phi) \\
\mu_{3} &=& - \frac{k_{1}}{k_{3}}\mu_{1} - \frac{k_{2}}{k_{3}}\mu_{2}.
\end{eqnarray}
Here the last equality comes from the triangular condition, $\bmath{k}_{1} + \bmath{k}_{2} + \bmath{k}_{3} = \bmath{0}$.

The `reduced' bispectrum is given by the ratio of the
bispectrum to 
the products of the power spectra:
\begin{eqnarray}
\label{eq:reducedbispectrumdef}
Q_{{\rm Ly\alpha,s}}(\bmath{k}_{1},\bmath{k}_{2},\bmath{k}_{3}) \equiv \frac{B_{{\rm Ly\alpha,s}}(\bmath{k}_{1},\bmath{k}_{2},\bmath{k}_{3})}{P_{\lya{},s}(\bmath{k}_{1})P_{\lya{},s}(\bmath{k}_{2}) + \rm{ 2 \,cyc.}}. 
\end{eqnarray}
This quantity is insensitive to the overall amplitude of
the power spectrum, as the second-order expression for the
bispectrum given in Equation~\ref{eq:bispectrum} is proportional to the
products of the power spectra. This properly removes the degeneracy
between the galaxy bias parameters and the amplitude of the matter power
spectrum. 

\subsection{Fisher matrix}

Before generating the expected cosmological constraints
using the Fisher matrix, let us first summarize the model
parameters characterizing the higher-order (non-linear)
terms.

It is important to note that, unlike the previous work which
simply multiplies the real-space bispectrum by the linear redshift
distortion factors
\citep{Scoccimarro:1999p9850,Sefusatti:2006p752,Sefusatti:2007p9477}, we
include the {\it full} wavenumber dependence of the redshift-space distortion
up to the second order. By including the full second-order
redshift-space distortion, we gain additional information
which helps
to further break the degeneracies between the
cosmological information and the radiative transfer effects, especially
those associated with the velocity gradient.  

The second-order redshift-space \lya{} kernel, after removing the scale
dependence of the ionizing background effect, is given by (see Appendix
\ref{app:full}) 
\begin{eqnarray} \label{eq:2ndredkernel}
K^{(s)}_{2} (\bmath{k}_{1},\bmath{k}_{2}) & = & \frac{1}{2}\tilde{b}_{2} - \frac{1}{2}f(\mu^{2}_{1} + \mu^{2}_{2})\tilde{C} + \tilde{b}_{1}F^{(s)}_{2}(\bmath{k}_{1},\bmath{k}_{2}) \nonumber \\
& & + f\mu_{12}^{2}(1-C_{v})G^{(s)}_{2}(\bmath{k}_{1},\bmath{k}_{2}) + \frac{1}{2}f^{2}\mu^{2}_{1}\mu^{2}_{2}C_{v v} \nonumber \\
& & + \frac{1}{2}(k_{12}\mu_{12} f)\left\{\frac{k_{1z}}{k^{2}_{1}}\left[\tilde{b}_{1} - f\mu_{2}^{2}C_{v}\right] + \right. \nonumber \\
& & \left. \frac{k_{2z}}{k^{2}_{2}}\left[\tilde{b}_{1} - f\mu_{1}^{2}C_{v}\right]\right\} \nonumber \\ 
& & + \frac{1}{2}(k_{12}\mu_{12}
 f)^{2}\left[\frac{k_{1z}k_{2z}}{k^{2}_{1}k^{2}_{2}}\right] \nonumber\\
 & = & \frac{1}{2}\tilde{b}_{2} - \frac{1}{2}f(\mu^{2}_{1} + \mu^{2}_{2})\tilde{C} + \tilde{b}_{1}F^{(s)}_{2}(\bmath{k}_{1},\bmath{k}_{2}) \nonumber \\
& & + f\mu_{12}^{2}(1-C_{v})G^{(s)}_{2}(\bmath{k}_{1},\bmath{k}_{2}) + \frac{1}{2}f^{2}\mu^{2}_{1}\mu^{2}_{2}C_{v v} \nonumber \\
& & + \frac{1}{2}\tilde{b}_{1} (k_{12}\mu_{12} f)\left[\frac{k_{1z}}{k^{2}_{1}}+\frac{k_{2z}}{k^{2}_{2}}\right]\nonumber \\ 
& & + \frac{1}{2}(k_{12}\mu_{12}
 f)^{2}(1-C_v)\left[\frac{k_{1z}k_{2z}}{k^{2}_{1}k^{2}_{2}}\right],
\end{eqnarray}
where we define
\begin{eqnarray}
\tilde{b}_{2} & \equiv & b^{2}_{1}\left(C_{\Gamma \Gamma} +
			      2C_{\Gamma}\right)+ C_{\rho \rho} + b_{2}(1 +
C_{\Gamma}) + 2b_{1}(C_{\Gamma\rho} + C_{\rho}),\nonumber \\
\\
\tilde{C} & \equiv & b_{1}C_{v} + C_{\rho v} + b_{1}C_{\Gamma v},\\
k_{ij}&\equiv& |\bmath{k}_i+\bmath{k}_j|, \\
\mu_{ij}&\equiv&
 \frac{(\bmath{k}_i+\bmath{k}_j)\cdot\hat{\bmath{z}}}{|\bmath{k}_i+\bmath{k}_j|},\\
k_{iz}&\equiv& \bmath{k}_i\cdot\hat{\bmath z}.
\end{eqnarray}
Here, $\tilde{b}_{2}$ is the effective non-linear galaxy bias modified by
various first- and second-order radiative transfer effects, and
$\tilde{b}_{1}$ is defined in Equation \ref{eq:b1tilde}. 

In our model, we choose to keep the second-order effect due to the
peculiar velocity gradient ($C_{vv}$) separate, as this has the
potential to be degenerate with the growth rate of structure,
$f$. Additionally, $\tilde{C}$, which contains the linear-order effects with respect to the peculiar velocity gradient, can also become degenerate with $f$. We note that setting the \lya{} radiative transfer effects to zero reduces Equation \ref{eq:2ndredkernel} to the typical second-order redshift-space galaxy kernel, as required.

To generate the expected constraints on the cosmological parameters, we calculate the Fisher matrix for both the bispectrum and the reduced bispectrum. The Fisher matrix for the bispectrum is
\begin{eqnarray}
F_{ij} = \sum_{k_{1},k_{2},k_{3}\leq k_{max}} \frac{1}{\sigma_{B}^{2}}\frac{\partial B_{g}(k,\mu)}{\partial \theta_{i}}\frac{\partial B_{g}(k,\mu)}{\partial \theta_{j}},
\end{eqnarray}
and for the reduced bispectrum is 
\begin{eqnarray}
F_{ij} = \sum_{k_{1},k_{2},k_{3}\leq k_{max}} \frac{1}{\sigma_{Q}^{2}}\frac{\partial Q_{g}(k,\mu)}{\partial \theta_{i}}\frac{\partial Q_{g}(k,\mu)}{\partial \theta_{j}},
\end{eqnarray}
where the Fisher matrices are summed over all possible triangular
configurations. 

The variances for the bispectrum and the reduced bispectrum are given,
respectively, by
\begin{eqnarray}
\sigma_{B}^{2} = \frac{s_{B}V_{survey}}{N_{t}}P_{tot}(k_{1})P_{tot}(k_{2})P_{tot}(k_{3})
\end{eqnarray}
and
\begin{eqnarray}
\sigma_{Q}^{2} = \frac{s_{B}V_{survey}}{N_{t}}\frac{P_{tot}(k_{1})P_{tot}(k_{2})P_{tot}(k_{3})}{[P_{\lya{},s}(k_{1})P_{\lya{},s}(k_{2}) + \rm{ 2\,cyc.}]^{2}}.
\end{eqnarray}
Here $s_{B}$ is the symmetric factor describing symmetry of the side
lengths of a given bispectrum triangle ($s_{B} = 6,2,1$ for equilateral, isosceles, and general triangles, respectively) and $P_{tot}(k)$ is the sum of the power spectrum component and the Poisson shot noise:
\begin{eqnarray}
P_{tot}(k) = P_{\lya{},s}(k) + \frac{1}{n_{g}},
\end{eqnarray}
where $n_{g}$ is the number density of LAE galaxies. The quantity $N_{t}$ is
the total number of available triangles: 
\begin{eqnarray} \label{eq:num_triangles}
N_{t} = \frac{V_{B}}{k_{F}^{3}},
\end{eqnarray}
where $k_{F}$ is the fundamental frequency and
\begin{equation}
       V_{B} = 2\pi d\mu d\phi k_{1}k_{2}k_{3}(\Delta k)^{3} \times \left\{ \begin{array}{cl} \;\! 1 \qquad {\rm if} \, k_{i} \neq k_{j} + k_{k},\\ \frac{1}{2} \qquad {\rm if} \, k_{i} = k_{j} + k_{k}.
\end{array} \right.
\end{equation}
This states that, for `collapsed' (or `co-linear') triangles defined
by $k_{i} = k_{j} + k_{k}$, the bispectrum volume is reduced by a factor
of two. (For the derivation of $V_{B}$, see Appendix \ref{app:VB}). In
the simplest spherically averaged scenario this reduces to 
\begin{equation}
       V_{B} = 32\pi^{2}k_{1}k_{2}k_{3}(\Delta k)^{3} \times \left\{ \begin{array}{cl} \;\! 1 \qquad {\rm if} \, k_{i} \neq k_{j} + k_{k},\\ \frac{1}{2} \qquad {\rm if} \, k_{i} = k_{j} + k_{k}.
\end{array} \right.
\end{equation}

\section{Constraints from the reduced bispectrum alone}
\label{sec:RBS} 

In Section \ref{sec:PS}, we show that the growth rate of structure, $f$ (or
$\tilde{\beta}$), is completely degenerate with the \lya{}
radiative transfer effect due to the velocity gradient,
$C_v$, as long as we rely only on the power spectrum.

However, the bispectrum provides additional constraining power that can
be used to break the degeneracy between the growth rate of structure,
$f$, the linear galaxy bias, $\tilde{b}_{1}$, and
$C_v$. This is because, unlike the power spectrum which
just tells us the amplitude of the fluctuations at a given scale, the
bispectrum tells us also {\it how} the structure forms. For example, one
needs information in the bispectrum in order to reproduce the `cosmic
web,' the filamentary structures in the universe. The power spectrum
cannot distinguish between the distribution with random phases and that
with the filamentary structures, as it is sensitive only to the
amplitude of the fluctuations. As a result, the bispectrum can
distinguish between the structures caused by gravitational and
non-gravitational effects.

In this section we firstly generate the expected
cosmological constraints from the reduced bispectrum. As mentioned
previously, the reduced bispectrum is insensitive to the
amplitude of the matter power spectrum. We again consider the same two models; a fiducial model where we set the radiative transfer coefficients to be zero but marginalize over them, and a model where we use explicit values for the \lya{} radiative transfer effects in our redshift-space expressions.

Since the recovery of the growth rate of structure $f$ is most affected
by the radiative transfer effects, we investigate the two-dimensional
joint likelihood distributions for $f$ with each of the other model
parameters (marginalized over all the remaining model parameters). 
For the remainder of this work, we set the non-linear galaxy bias to be
$b_{2}=1.5$.  

\subsection{Fiducial LAE reduced bispectrum}
\label{sec:fiducialRB}

\begin{figure*} 
	\begin{center}
		\includegraphics[trim = 8cm 0cm 0.2cm 2cm, scale = 0.165]{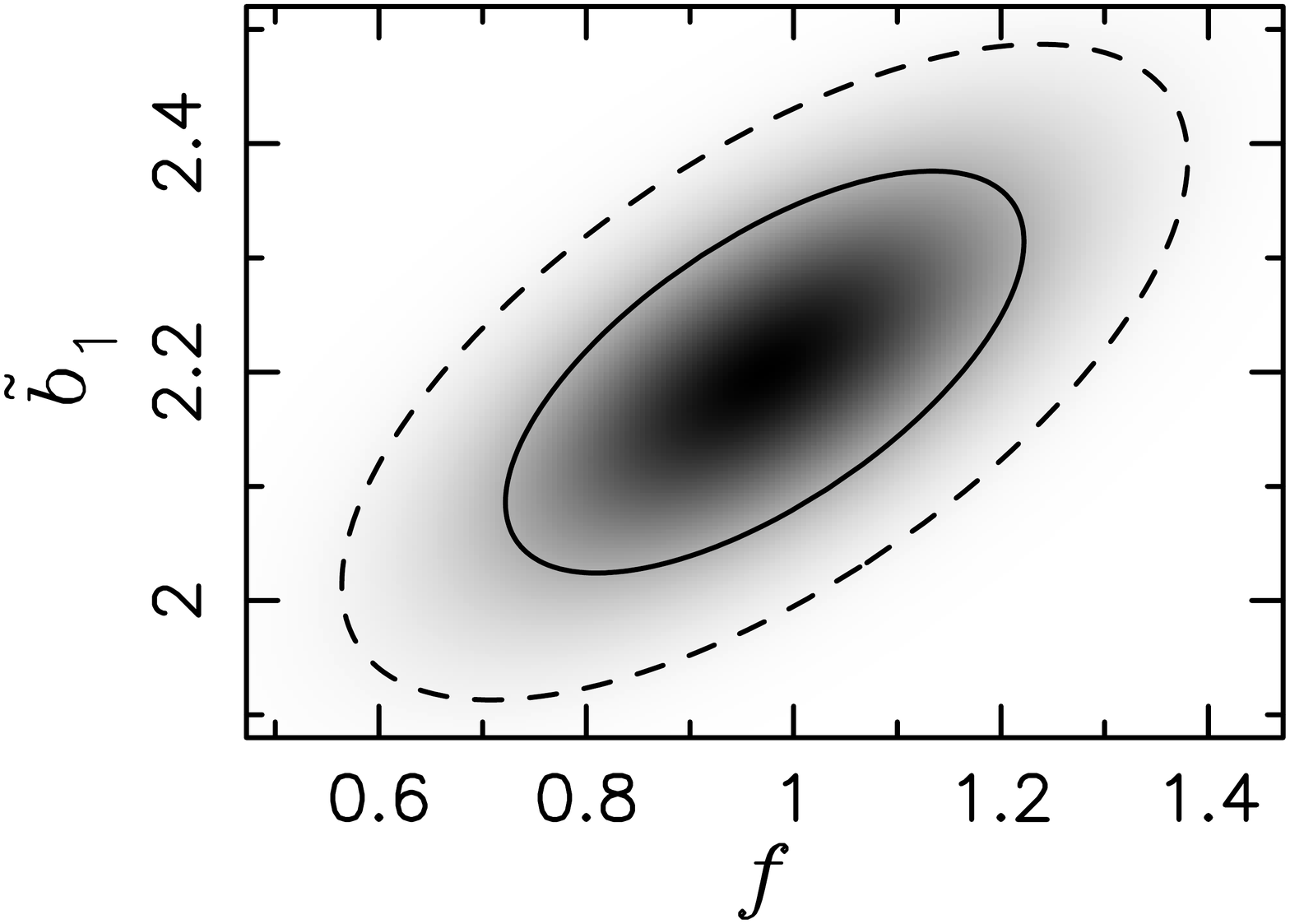}
		\includegraphics[trim = 1.5cm 0cm 0.2cm 2cm, scale = 0.165]{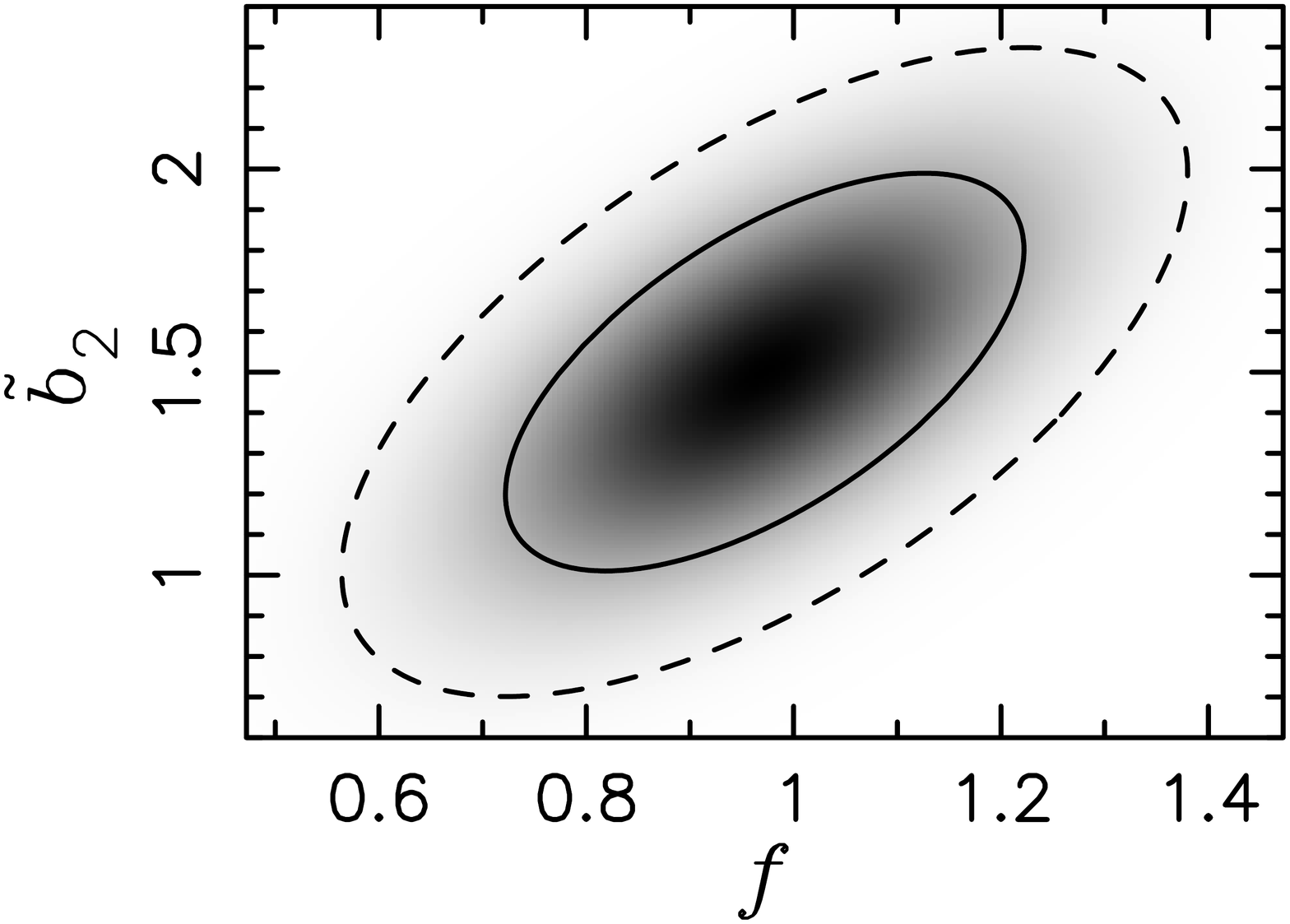}
		\includegraphics[trim = 1.5cm 0cm 0.2cm 2cm, scale = 0.165]{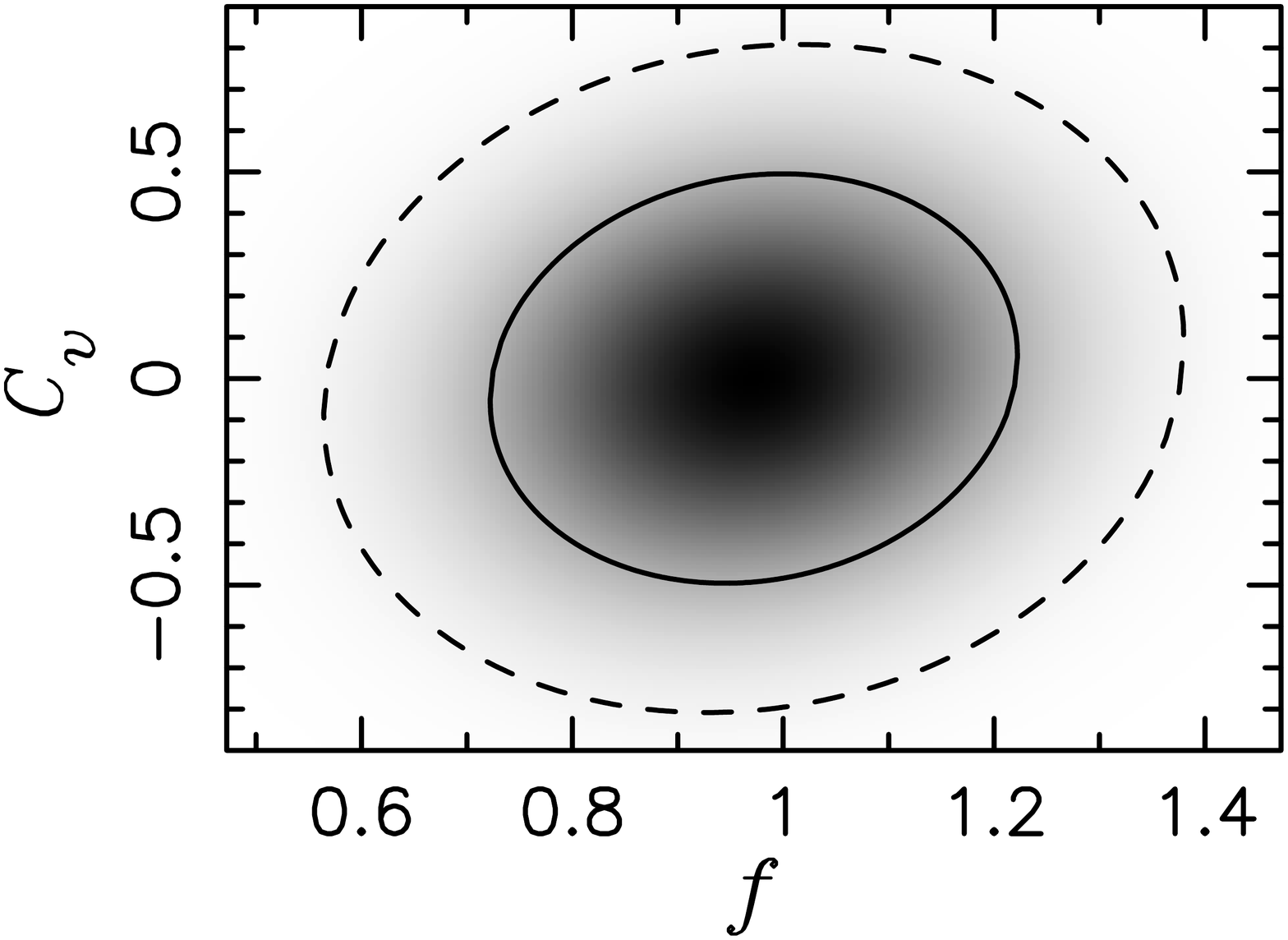}
		\includegraphics[trim = 1.5cm 0cm 6.2cm 2cm, scale = 0.165]{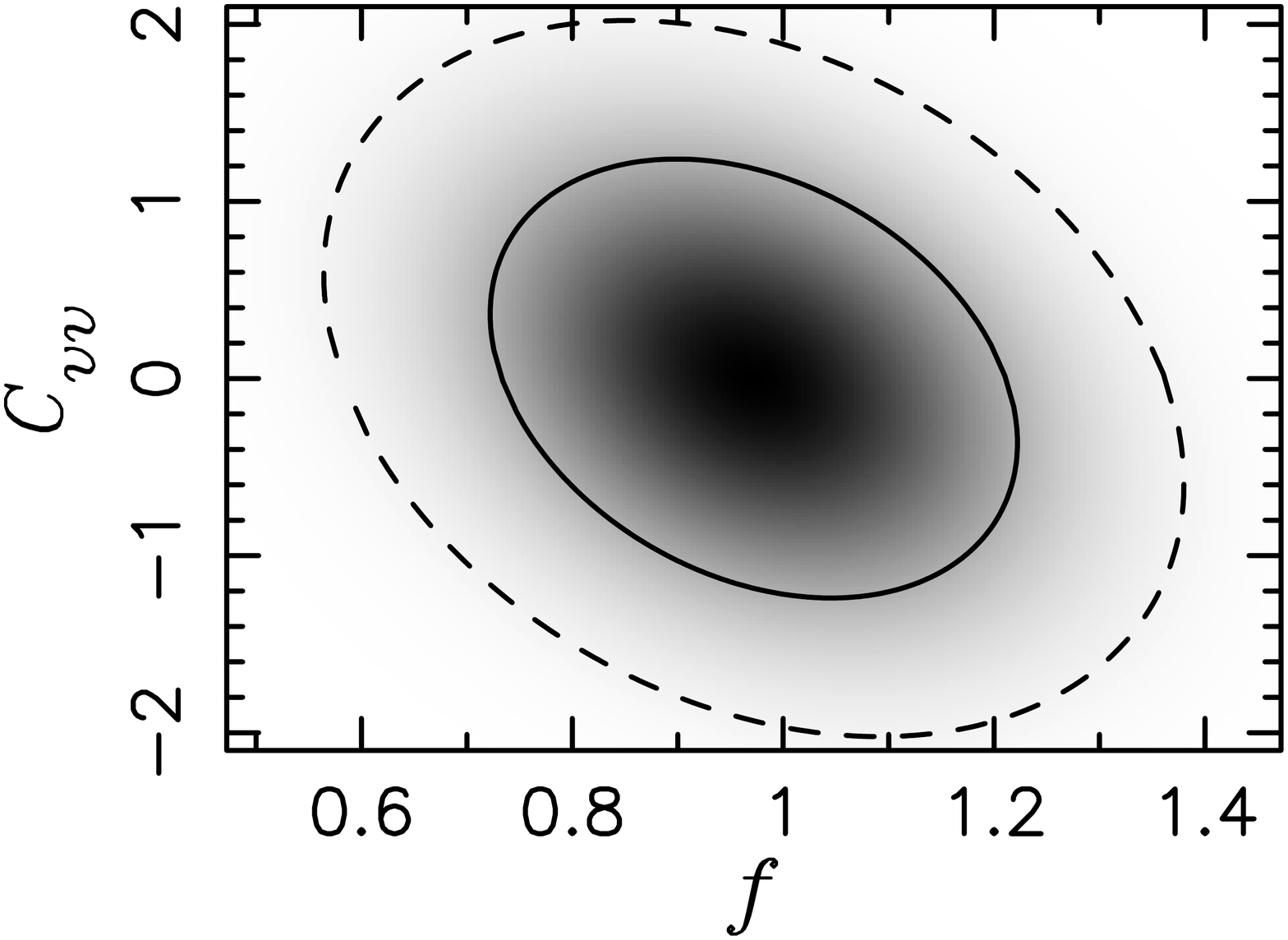}
		\includegraphics[trim = 8.5cm 3cm 0.2cm 1cm, scale = 0.165]{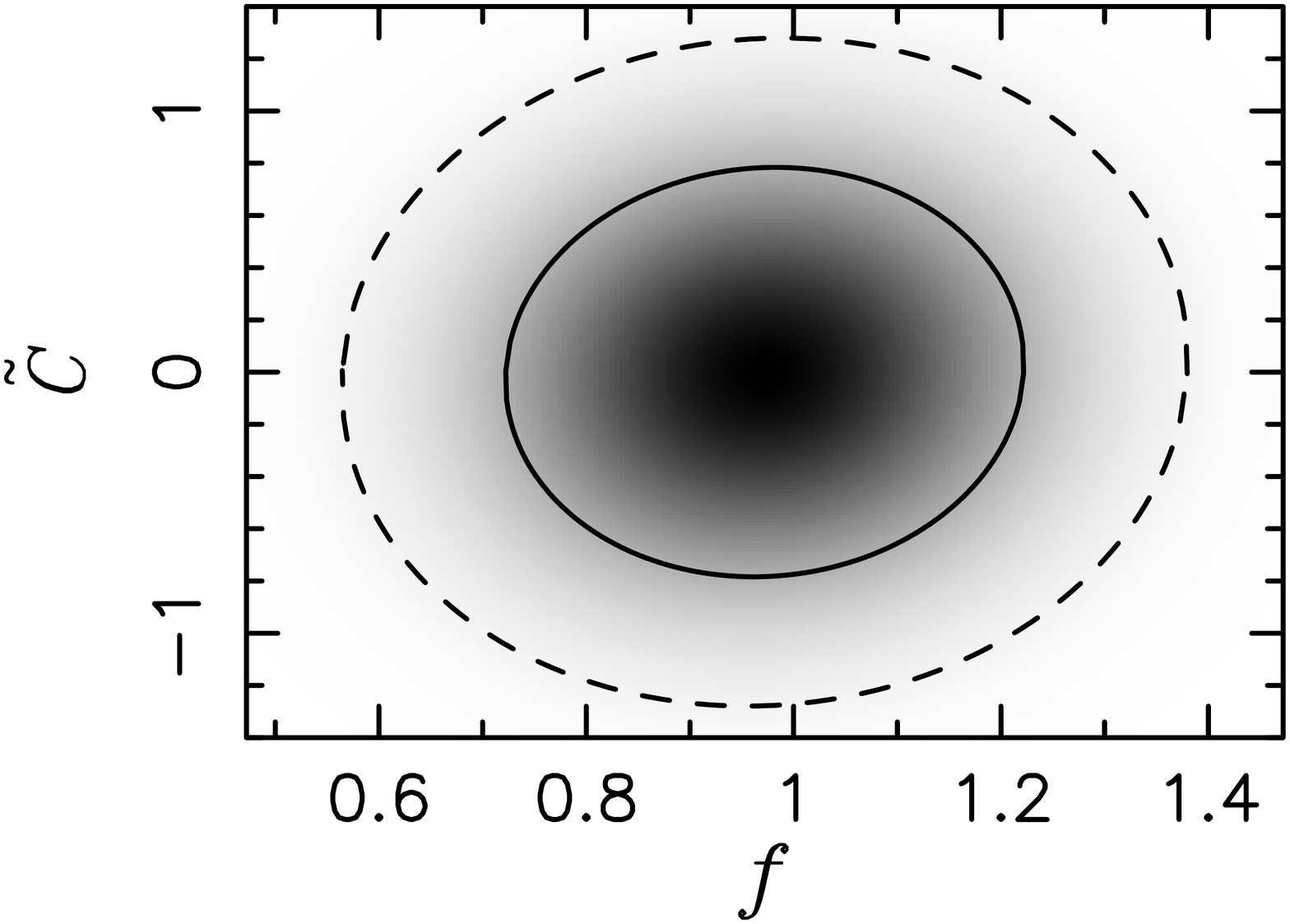}
		\includegraphics[trim = 1.5cm 3cm 0.2cm 1cm, scale = 0.165]{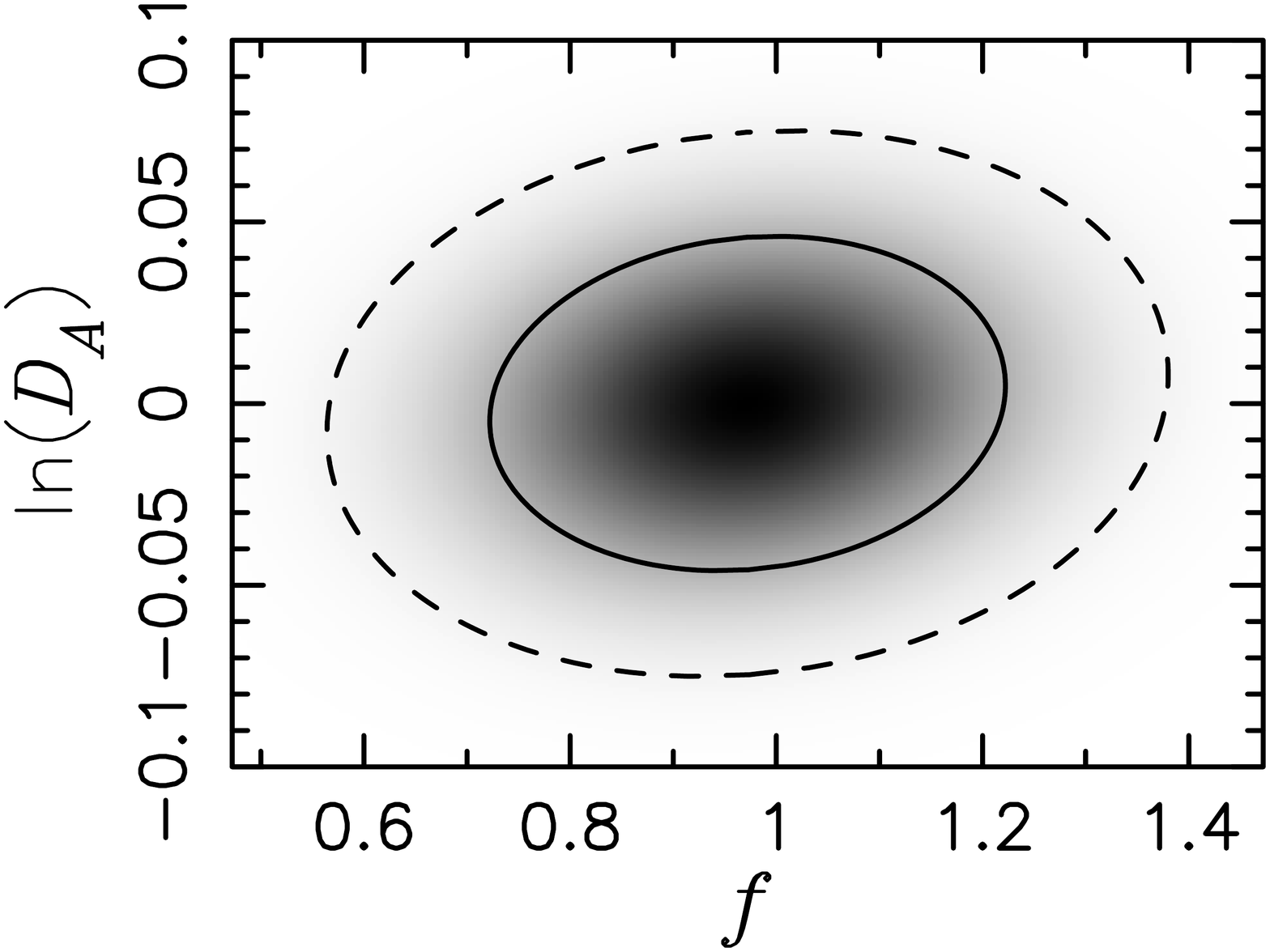}
		\includegraphics[trim = 1.5cm 3cm 6.2cm 1cm, scale = 0.165]{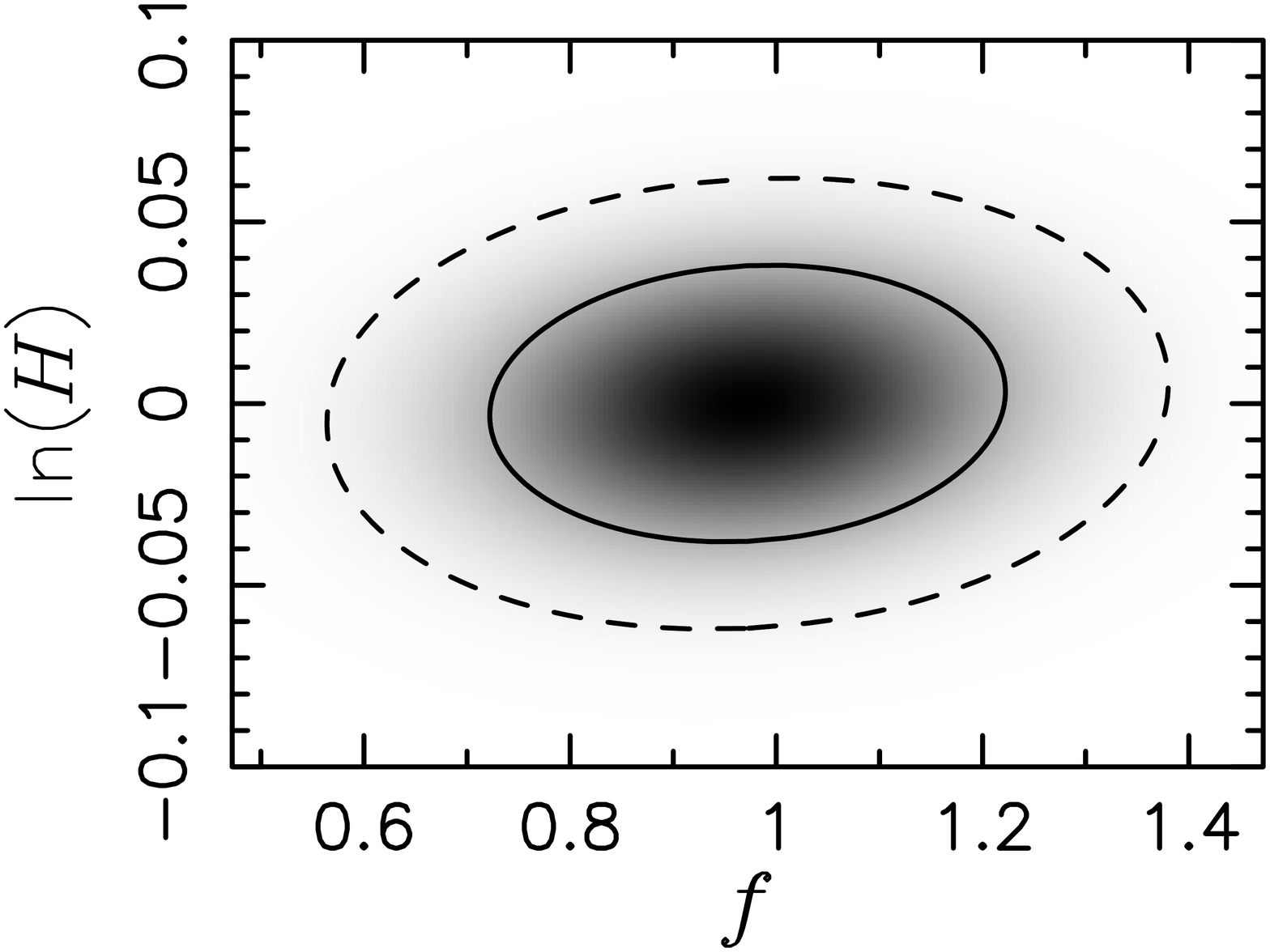}
	\end{center}
\caption{Two-dimensional joint marginalized likelihood distributions
 computed from the fiducial LAE galaxy {\it reduced} bispectrum
 alone (no \lya{} radiative transfer effects, but including
 marginalization over $C_{v}$, $C_{vv}$, and
 $\tilde{C}$). We show the correlations between the growth rate of
 structure, $f$, and various parameters including (clockwise from top
 left): the linear bias, $\tilde{b}_{1}$, non-linear bias,
 $\tilde{b}_{2}$, linear peculiar velocity \lya{} effect, $C_{v}$,
 non-linear peculiar velocity \lya{} effect, $C_{vv}$, the non-linear
 combination of other radiative transfer effect, $\tilde{C}$, angular
 diameter distance, $\ln(D_{A})$, and the Hubble rate $\ln(H)$. The
 solid and dashed curves show the 1- and 2-$\sigma$
 joint marginalized constraints, respectively.}
\label{fig:2DRBS}
\end{figure*}

\begin{figure*}  
	\begin{center}
		\includegraphics[trim = 9cm 2.5cm 2.2cm 2cm, scale = 0.27]{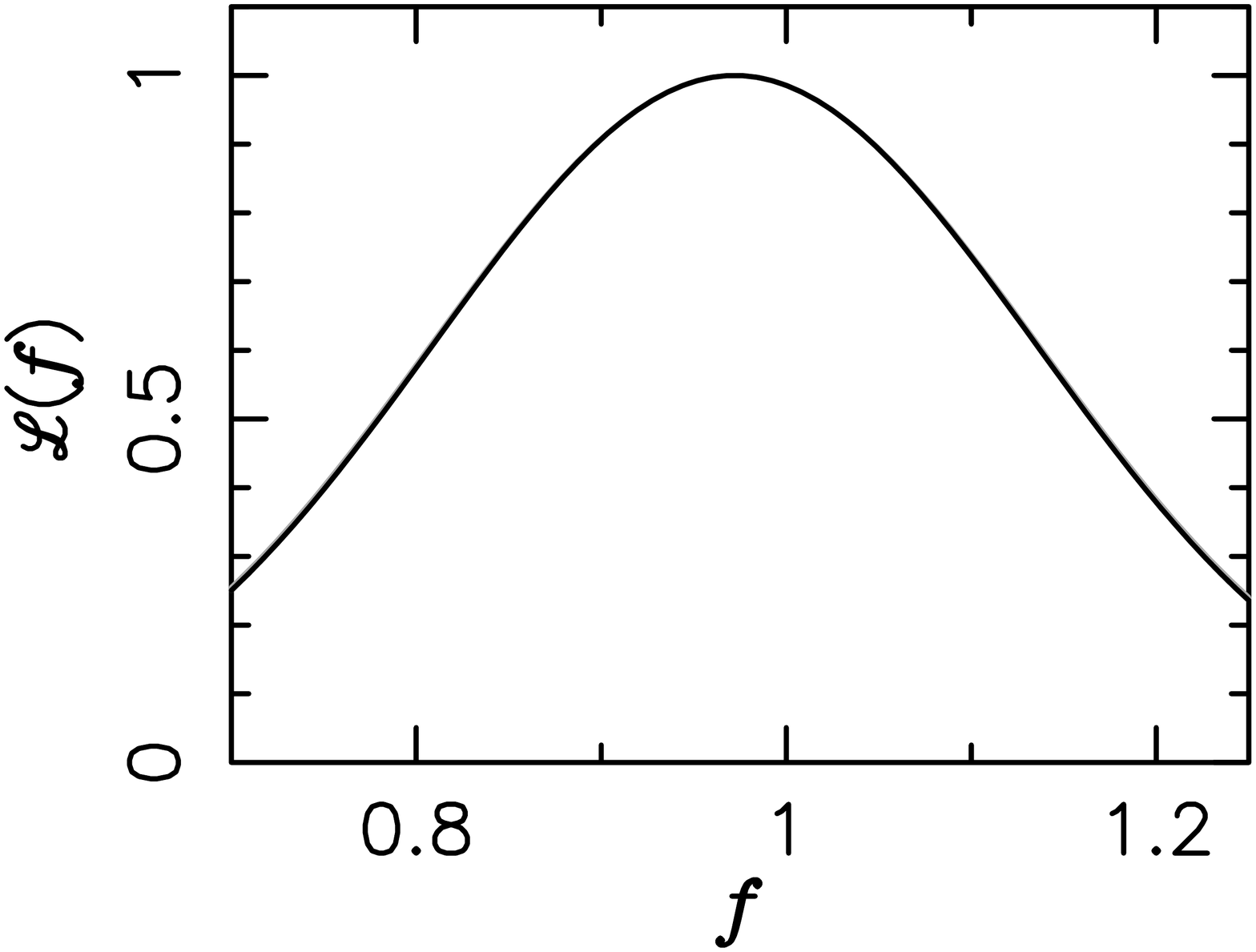}
		\includegraphics[trim = 6.1cm 2.5cm 2.2cm 2cm, scale = 0.27]{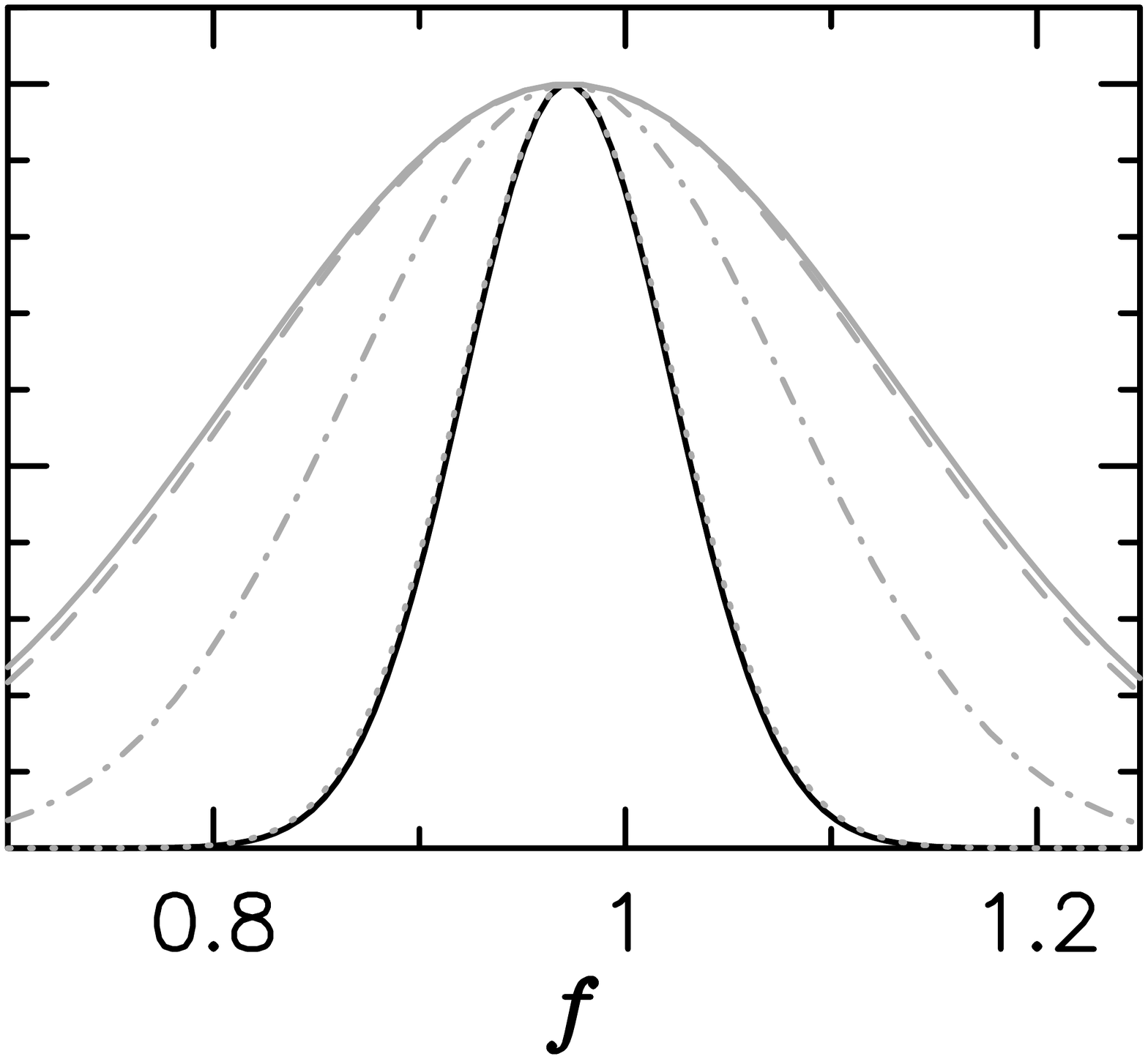}
		\includegraphics[trim = 6.1cm 2.5cm 8.2cm 2cm, scale = 0.27]{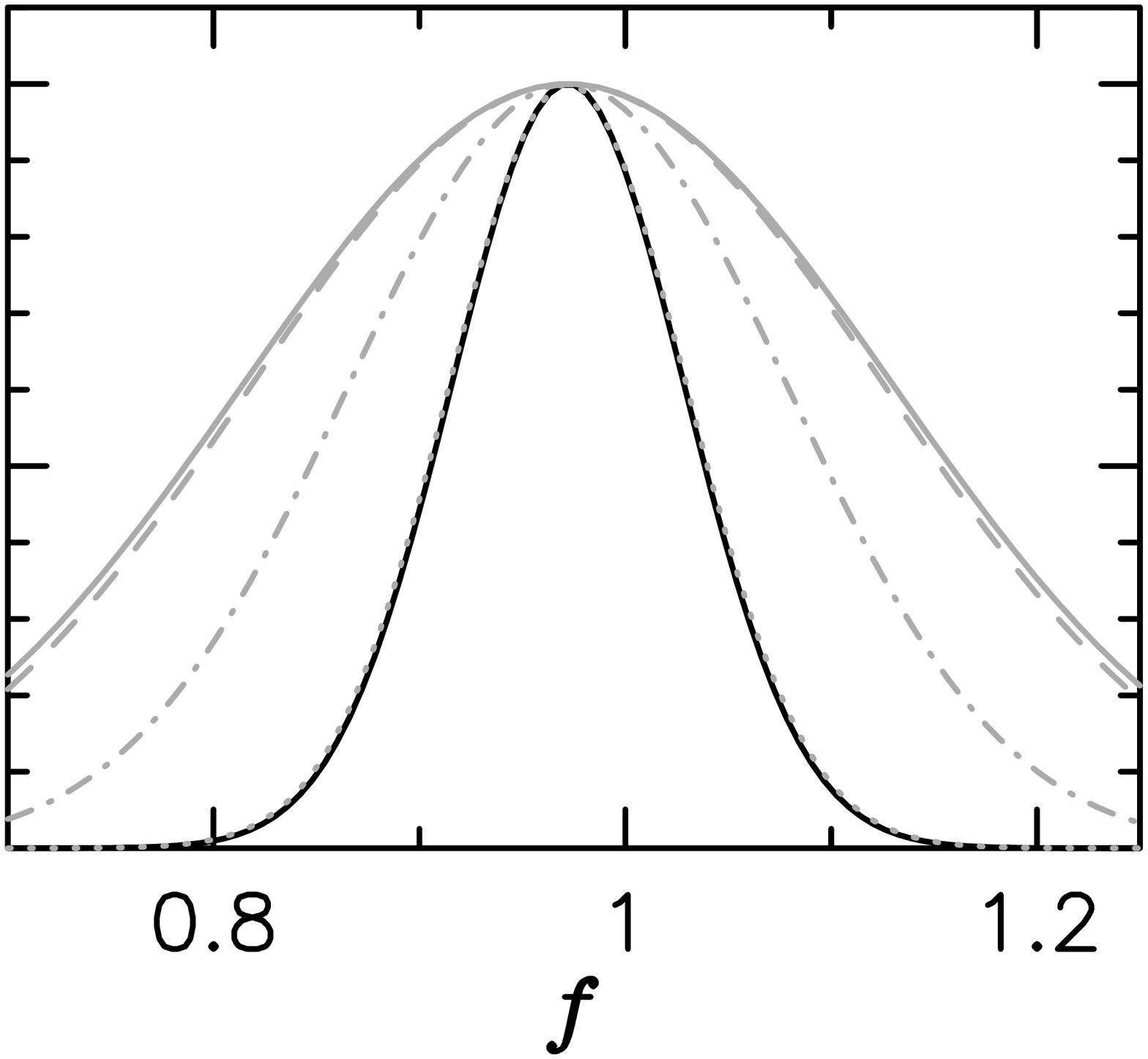}
	\end{center}
\caption{One-dimensional marginalized likelihood distributions for the growth rate of structure, $f$, for the fiducial case (no \lya{} radiative effects added, but including marginalization over $C_{v}$, $C_{vv}$ and $\tilde{C}$) generated from; (\textit{left}) the LAE galaxy reduced bispectrum only, (\textit{centre}) the LAE galaxy power spectrum combined with the LAE galaxy bispectrum, and (\textit{right}) the LAE galaxy power spectrum combined with the LAE galaxy reduced bispectrum. The various curves denote different priors added to $C_{v}$; \textit{black solid:} Perfect knowledge of $C_{v}$, \textit{grey dotted:} $\sigma_{C_{v}} = 0.01$, \textit{grey dot-dashed:} $\sigma_{C_{v}} = 0.1$, \textit{grey dashed:} $\sigma_{C_{v}} = 0.5$, and \textit{grey solid:} no priors added. }
\label{fig:fprior_fiducial}
\end{figure*}

\begin{table*}
\begin{tabular}{@{}lccccc}
\hline

Priors on $C_{v}$ & Parameter & Model & R BS & PS + BS & PS + R BS \\
& &  & 1-$\sigma$ (per cent) & 1-$\sigma$ (per cent) & 1-$\sigma$ (per cent)\\
\hline

No Priors & $f$ & Fiducial  & 0.1645 (16.92)& 0.1602 (16.48)& 0.1579 (16.24)\\
Perfect knowledge & $f$ & Fiducial  & 0.1635 (16.82) & 0.0507 (5.21)& 0.0565 (5.81)\\
0.01 & $f$ & Fiducial  & 0.1635 (16.82)& 0.0520 (5.35)& 0.0576 (5.93)\\
0.1 & $f$ & Fiducial  & 0.1636 (16.83)& 0.1055 (10.85)& 0.1063 (10.93)\\
0.5 & $f$ & Fiducial  & 0.1642 (16.89)& 0.1555 (16.00)& 0.1535 (15.79)\\
\hline
No Priors & ln($D_{A}$) & Fiducial  & 0.0303 (3.03)& 0.0076 (0.76)& 0.0103 (1.03)\\
0.01 & ln($D_{A}$) & Fiducial  & 0.0301 (3.01)& 0.0075 (0.75) & 0.0101 (1.01)\\
0.1 & ln($D_{A}$) & Fiducial  & 0.0302 (3.02)& 0.0075 (0.75) &  0.0102 (1.02)\\
\hline
No Priors & ln($H$) & Fiducial  & 0.0251 (2.51)& 0.0084 (0.84)& 0.0115 (1.15)\\
0.01 & ln($H$) & Fiducial  & 0.0249 (2.49)& 0.0083 (0.83)& 0.0112 (1.12)\\
0.1 & ln($H$) & Fiducial  &  0.0249 (2.49)& 0.0083 (0.83)&  0.0113 (1.13)\\
\hline
\end{tabular}
\caption{We show the 1-$\sigma$
 constraints expected from the reduced bispectrum (R BS), the power
 spectrum combined with the bispectrum (PS + BS), and the power spectrum
 combined with the reduced bispectrum (PS + R BS). No \lya{} radiative
 transfer effects are included, but the likelihood is marginalized
 over $C_{v}$, $C_{vv}$, and $\tilde{C}$. 
 The first five rows
 show the 1-$\sigma$ constraints on $f$ for various priors on $C_{v}$,
 after marginalizing over
 $\tilde{b}_{1}$, $\tilde{b}_{2}$, $\tilde{C}$, $C_{v}$, $C_{vv}$,
 $\ln(D_{A})$, $\ln(H)$, and the amplitude [${\rm ln}(A)$].
 The last six rows show the 
 1-$\sigma$  constraints on the distance parameters,
 $\ln(D_{A})$ and $\ln(H)$, marginalized over the remaining model
 parameters.}
\label{tab:fiducialbispectrum}
\end{table*}

\begin{table*}
\begin{tabular}{@{}lccccc}
\hline

Priors on $C_{v}$ & Parameter & Model & R BS & PS + BS & PS + R BS \\
& &  & 1-$\sigma$ (per cent) & 1-$\sigma$ (per cent) & 1-$\sigma$ (per cent)\\
\hline

No Priors & $f$ & LAE effects included  & 0.2104 (21.64) & 0.2039 (20.97)& 0.2014 (20.72)\\
Perfect knowledge & $f$ & LAE effects included  & 0.2089 (21.49)& 0.0632 (6.50)& 0.0672 (6.91)\\
0.01 & $f$ & LAE effects included  & 0.2089 (21.49)& 0.0645 (6.64)& 0.0685 (7.05)\\
0.1 & $f$ & LAE effects included  & 0.2090 (21.50)& 0.1262 (12.98)& 0.1270 (13.06)\\
0.5 & $f$ & LAE effects included  & 0.2099 (21.59)& 0.1965 (20.21)& 0.1943 (19.99)\\
\hline
No Priors & ln($D_{A}$) & LAE effects included  & 0.0378 (3.78)& 0.0099 (0.99)& 0.0120 (1.20) \\
0.01 & ln($D_{A}$) & LAE effects included  & 0.0376 (3.76) & 0.0098 (0.98)& 0.0119 (1.19) \\
0.1 & ln($D_{A}$) & LAE effects included  & 0.0376 (3.76) & 0.0099 (0.99) & 0.0119 (1.19)\\
\hline
No Priors & ln($H$) & LAE effects included  & 0.0305 (3.05)& 0.0107 (1.07)& 0.0133 (1.33)\\
0.01 & ln($H$) & LAE effects included  & 0.0302 (3.02)& 0.0106 (1.06)& 0.0131 (1.31)\\
0.1 & ln($H$) & LAE effects included  & 0.0303 (3.03)& 0.0107 (1.07)& 0.0132  (1.32)\\
\hline
\end{tabular}
\caption{Same as Table~\ref{tab:fiducialbispectrum}, but for the
 first-order \lya{} radiative transfer effects given by $C_{\Gamma} = 0.05$, $C_{\rho} = -0.39$, and $C_{v} = 0.11$.}
\label{tab:bispectrum}
\end{table*}

We first consider our fiducial model where we set all \lya{} radiative
transfer coefficients to zero, but marginalize over the \lya{}
effects. With the addition of the bispectrum, the number of
parameters in our model has increased to eight. These include three
cosmological parameters: $f$, $\ln(D_{A})$, and $\ln(H)$; three
radiative transfer parameters: $C_{v}$, $C_{vv}$, and $\tilde{C}$; and
the linear and non-linear galaxy biases: $\tilde{b}_{1}$ and
$\tilde{b}_{2}$. 

We find that the constraints generated from the reduced
bispectrum contain no strong degeneracies between $f$ and the radiative
transfer parameters (see Figure~\ref{fig:2DRBS}). The reduced bispectrum does
however exhibit some degeneracies between $f$, and the galaxy bias
parameters, $\tilde{b}_{1}$ and $\tilde{b}_{2}$. 

To understand this result, let us write the reduced bispectrum given in
Equation~\ref{eq:reducedbispectrumdef} as
\begin{eqnarray}
\nonumber
& & Q_{{\rm Ly\alpha,s}}(\bmath{k}_{1},\bmath{k}_{2},\bmath{k}_{3}) \\
&=&
\frac{2\hat{K}^{(s)}_1(\bmath{k}_{1})\hat{K}^{(s)}_1(\bmath{k}_{2})\hat{K}^{(s)}_2(\bmath{k}_{1},\bmath{k}_{2}){P_{L}(k_1)P_{L}(k_2)
+ \rm{ 2
\,cyc.}}}{{[\hat{K}^{(s)}_1(\bmath{k}_{1})]^{2}[\hat{K}^{(s)}_1(\bmath{k}_{2})]^{2}P_{L}(k_1)P_{L}(k_2)
+ \rm{ 2 \,cyc.}}}, \nonumber \\
\label{eq:reducedbispectrumdef2}
\end{eqnarray}
where
\begin{eqnarray}
\label{eq:K1shat}
 \hat{K}^{(s)}_1(\bmath{k})&\equiv& \frac1{\tilde{b}_1}K^{(s)}_1(\bmath{k}) = 
1 + \tilde{\beta}\mu^{2}(1- C_{v}),\\
\nonumber
\hat{K}^{(s)}_2(\bmath{k}_1,\bmath{k}_2)&\equiv& \frac1{\tilde{b}_1^2}K^{(s)}_2(\bmath{k}_1,\bmath{k}_2) \\
& = & \frac1{\tilde{b}_1}\left[\frac{1}{2}\frac{\tilde{b}_{2}}{\tilde{b}_1} - \frac{1}{2}\tilde{\beta}(\mu^{2}_{1} + \mu^{2}_{2})\tilde{C} + F^{(s)}_{2}(\bmath{k}_{1},\bmath{k}_{2})\right. \nonumber \\
& & \left.\qquad +\tilde{\beta}\mu_{12}^{2}(1-C_{v})G^{(s)}_{2}(\bmath{k}_{1},\bmath{k}_{2})\right]  \nonumber \\
& & + \frac{1}{2}\tilde{\beta}
 (k_{12}\mu_{12})\left[\frac{k_{1z}}{k^{2}_{1}}+\frac{k_{2z}}{k^{2}_{2}}\right]
 +{\cal O}(\mu^4).
\label{eq:K2shat}
\end{eqnarray}
Here, ${\cal O}(\mu^4)$ are the terms that contain four powers of
cosines (see Equation~\ref{eq:2ndredkernel} for the full expression of
$K_2^{(s)}$). 
These terms contribute less, as their contributions are
important only near the line-of-sight direction, for which the number
of available modes is limited. 

Equations~\ref{eq:reducedbispectrumdef2}, \ref{eq:K1shat}, and
\ref{eq:K2shat} show that the reduced bispectrum determines the following
parameter combinations: 
\begin{itemize}
 \item $\tilde{b}_1$ from the overall amplitude of the first four terms
       in $\hat{K}_2^{(s)}$,
 \item $\tilde{b}_2/\tilde{b}_1$ from a constant, $k$-independent term
       in $\hat{K}_2^{(s)}$,
 \item $\tilde{\beta}(1-C_v)$ from $\hat{K}_1^{(s)}$ and
the term proportional to $G_2^{(s)}$ in $\hat{K}_2^{(s)}$,
 \item $\tilde{\beta}\tilde{C}$ from the second term in
       $\hat{K}_2^{(s)}$, and
 \item $\tilde{\beta}$ from the last term before ${\cal O}(\mu^4)$ in
       $\hat{K}_2^{(s)}$.
\end{itemize}
Recalling $\tilde{\beta}=f/\tilde{b}_1$, there are five unknown
variables ($\tilde{b}_1$, $\tilde{b}_2$, $f$, $\tilde{C}_v$, and
$\tilde{C}$), and the reduced bispectrum yields five combinations of
these variables. 

From the Fisher matrix calculations, we find that the reduced bispectrum
primarily yields $\tilde{b}_2/\tilde{b}_1$ and $\tilde{\beta}$. The
information on $\tilde{b}_1$ coming from the first four terms in
$\hat{K}_2^{(s)}$ breaks a complete degeneracy between $\tilde{b}_2$ and
$\tilde{b}_1$ and $f$, but correlations between these parameters
still remain. One can see this in the first two panels in
Figure~\ref{fig:2DRBS}. On the other hand, we do not find much correlation
between $f$ and the radiative transfer parameters, $\tilde{C}$,
$\tilde{C}_v$, and $\tilde{C}_{vv}$ (see the third to fifth panels of
Figure~\ref{fig:2DRBS}).

While the reduced bispectrum does break the degeneracy between $f$ and
$C_v$ seen in our 
power spectrum analysis, it cannot provide a strong constraint on $f$. In
the fourth column of Table \ref{tab:fiducialbispectrum} we
provide the 1-$\sigma$ constraints 
generated from the one-dimensional likelihood distribution for $f$. In
the fiducial case with no additional priors, we find the 1-$\sigma$
constraint on the growth rate of structure, $f$, to be $0.16$
($17$ per cent). It is important to note that, while the
constraints are relatively weak, they are on $f$ as opposed to
$\tilde{\beta}$. Also, $\tilde{\beta}$ and $C_v$ are 
totally degenerate in the LAE power spectrum, and thus the error bar on
$\tilde{\beta}$ is infinite unless we put a prior on $C_v$. Therefore, the
reduced bispectrum provides a massive improvement on the constraint on
$f$: the error bar shrinks from infinity to 17 per cent.

In the left panel of Figure \ref{fig:fprior_fiducial}, we show the
one-dimensional likelihood distributions for the growth rate of
structure, 
$f$, for various priors on $C_{v}$. The addition of priors to $C_{v}$
does not improve the constraints on $f$ from the reduced
bispectrum alone, as the reduced bispectrum contains no degeneracy
between $f$ and $C_{v}$.
 
In the fourth column of Table \ref{tab:fiducialbispectrum}
we also provide the 1-$\sigma$ constraints from the one dimensional
likelihoods for $\ln(D_{A})$ and $\ln(H)$ given various priors on
$C_{v}$. We find that (independent of priors on $C_{v}$) the fiducial
LAE galaxy reduced bispectrum can recover the angular diameter distance
scale at $3$ per cent and the Hubble rate at $2.5$ per
cent. This should be contrasted with the 1.1 per cent and 1.3 per cent
errors on $D_{A}$ and $H$ expected from the fiducial LAE galaxy power
spectrum. Clearly the reduced bispectrum alone
provides weaker distance constraints. This is not
surprising, as the distance information is contained in the shape of the
power spectrum (e.g., baryon acoustic oscillation (BAO) and
Alcock-Paczynski (AP) test), which is largely divided out in the reduced
bispectrum. 

\subsection{LAE reduced bispectrum}

We now consider the inclusion of \lya{} radiative transfer effects by
adding the linear \lya{} radiative transfer model parameters,
$C_{\Gamma} = 0.05$, $C_{\rho} = -0.39$, and $C_{v} = 0.11$ from
\citet{Wyithe:2011p12569}. The inclusion of these parameters modifies
the effective bias parameters, $\tilde{b}_{1}$ and $\tilde{b}_{2}$, and
the \lya{} radiative transfer effects associated with $\tilde{C}$. We
still set the fiducial values of the second-order \lya{}
radiative transfer coefficients to vanish. Although we set $C_{vv} = 0$,
we still marginalize over $C_{vv}$ in our models. 

In the fourth column of Table \ref{tab:bispectrum}, the
1-$\sigma$ constraints on $f$, $\ln(D_{A})$, and $\ln(H)$ are generated
from the likelihood distributions for various priors on $C_{v}$ as per
the previous section. With the inclusion of the \lya{} effects, the
precision with which we can constrain the growth rate of structure $f$
has been reduced to an error of $0.21$ (22 per cent) compared to $0.16$
(17 per cent) for the fiducial model. Once again, this is
due to the reduced effective linear galaxy bias, which reduces the
signal-to-noise ratio of the LAE power spectrum relative to the shot
noise.

\begin{figure*} 
	\begin{center}
		\includegraphics[trim = 8cm 0cm 0.2cm 2cm, scale = 0.165]{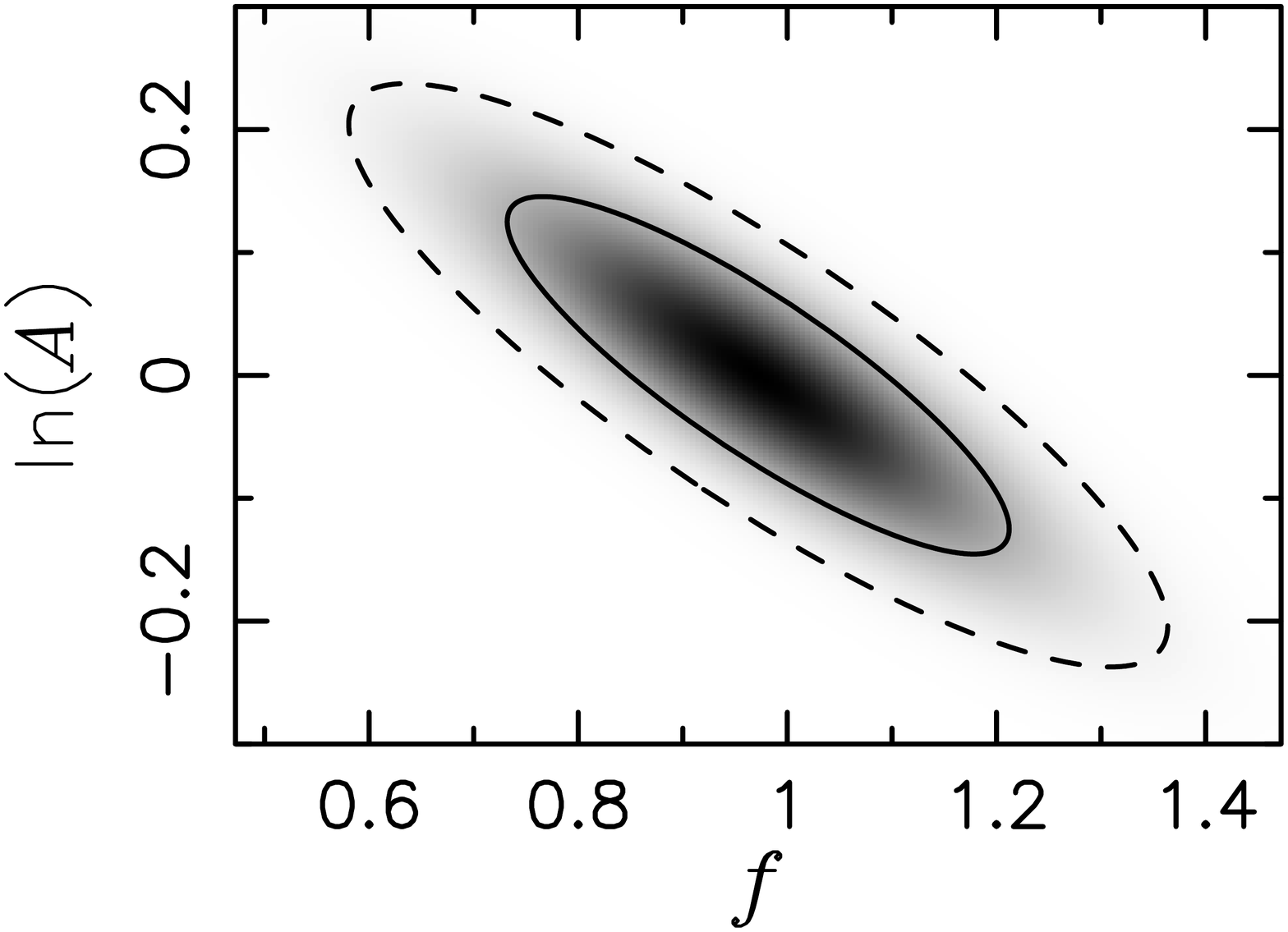}
		\includegraphics[trim = 1.5cm 0cm 0.2cm 2cm, scale = 0.165]{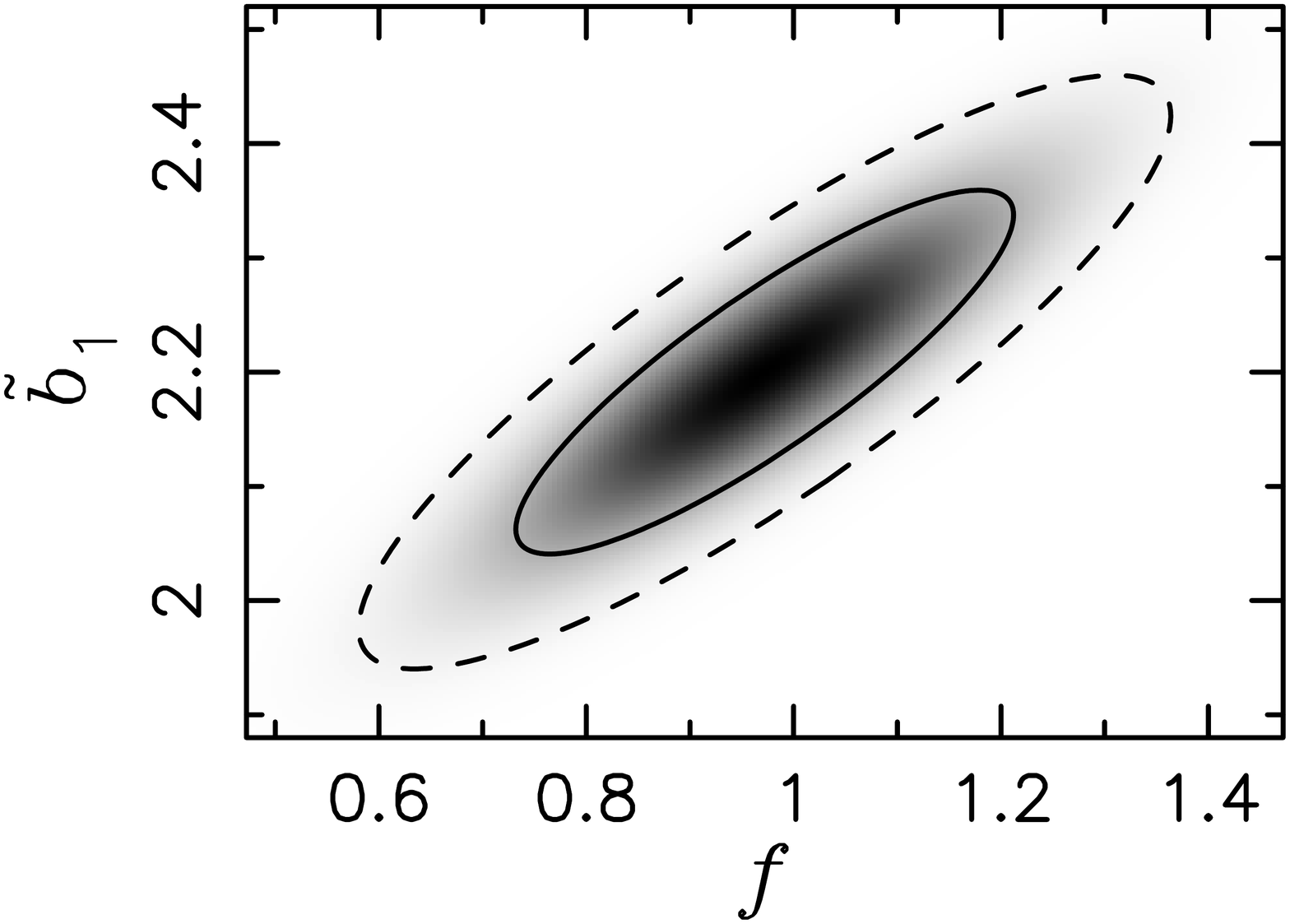}
		\includegraphics[trim = 1.5cm 0cm 0.2cm 2cm, scale = 0.165]{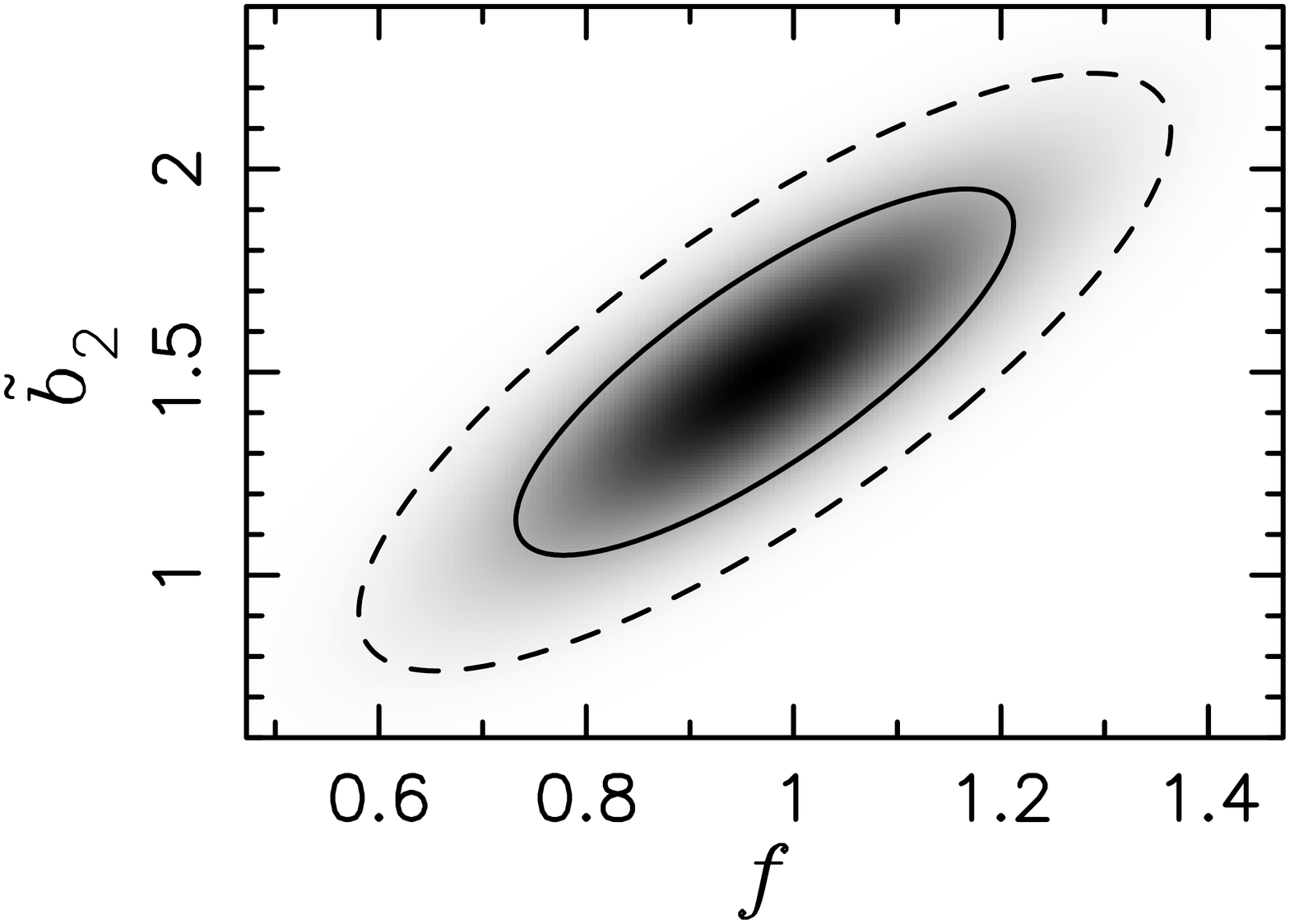}
		\includegraphics[trim = 1.5cm 0cm 6.2cm 2cm, scale = 0.165]{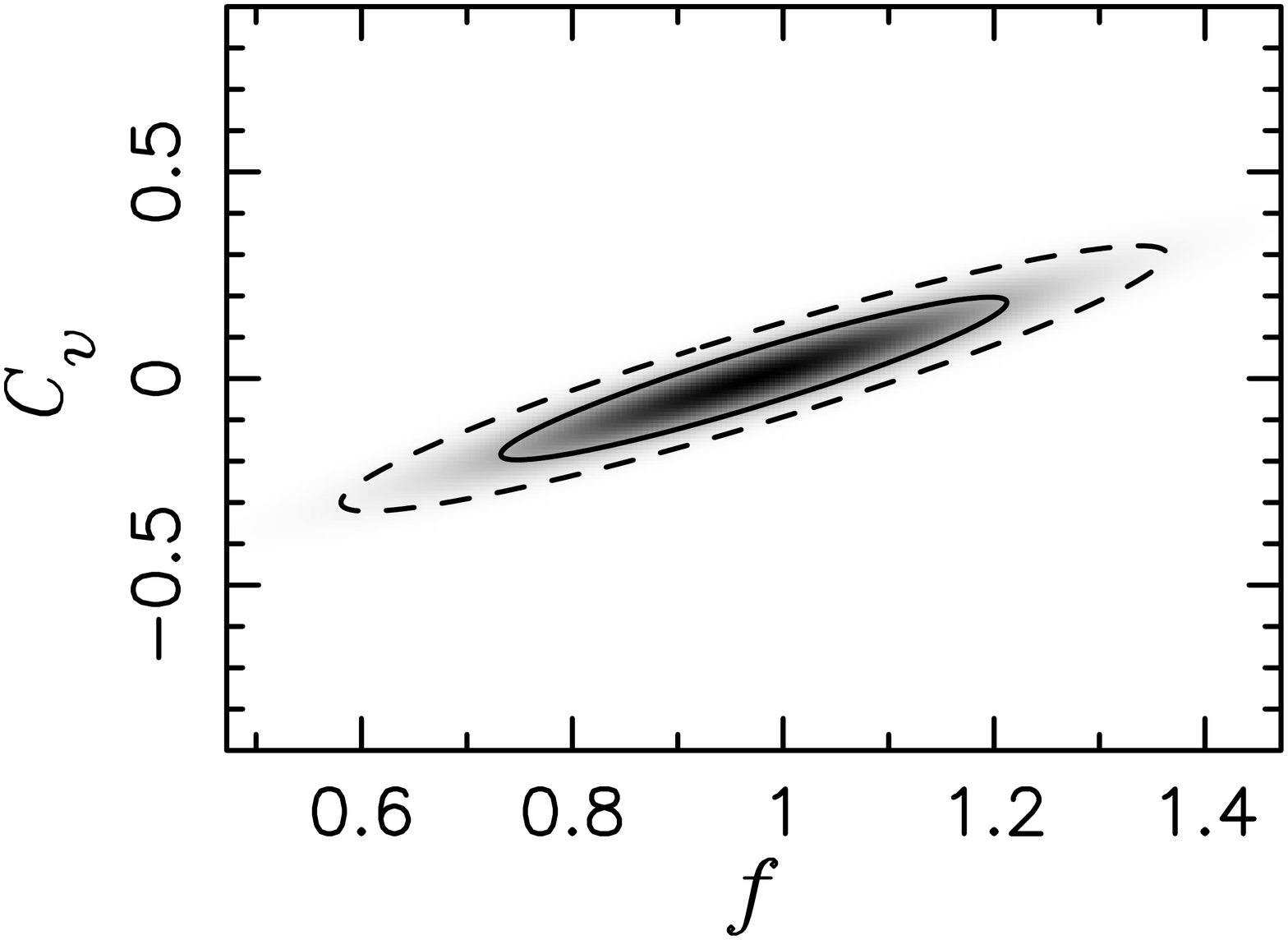}
		\includegraphics[trim = 8cm 3cm 0.2cm 2cm, scale = 0.165]{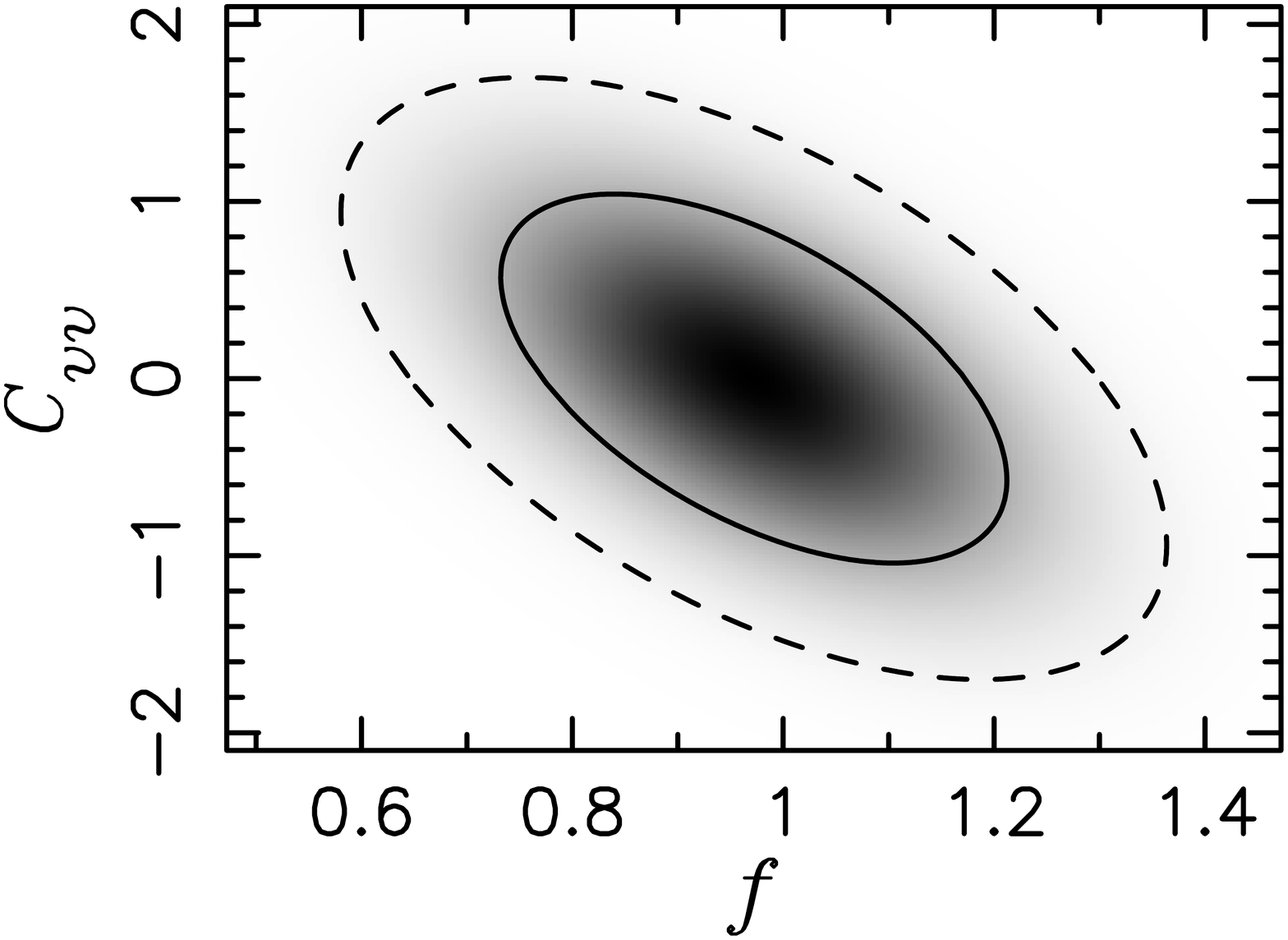}
		\includegraphics[trim = 1.5cm 3cm 0.2cm 2cm, scale = 0.165]{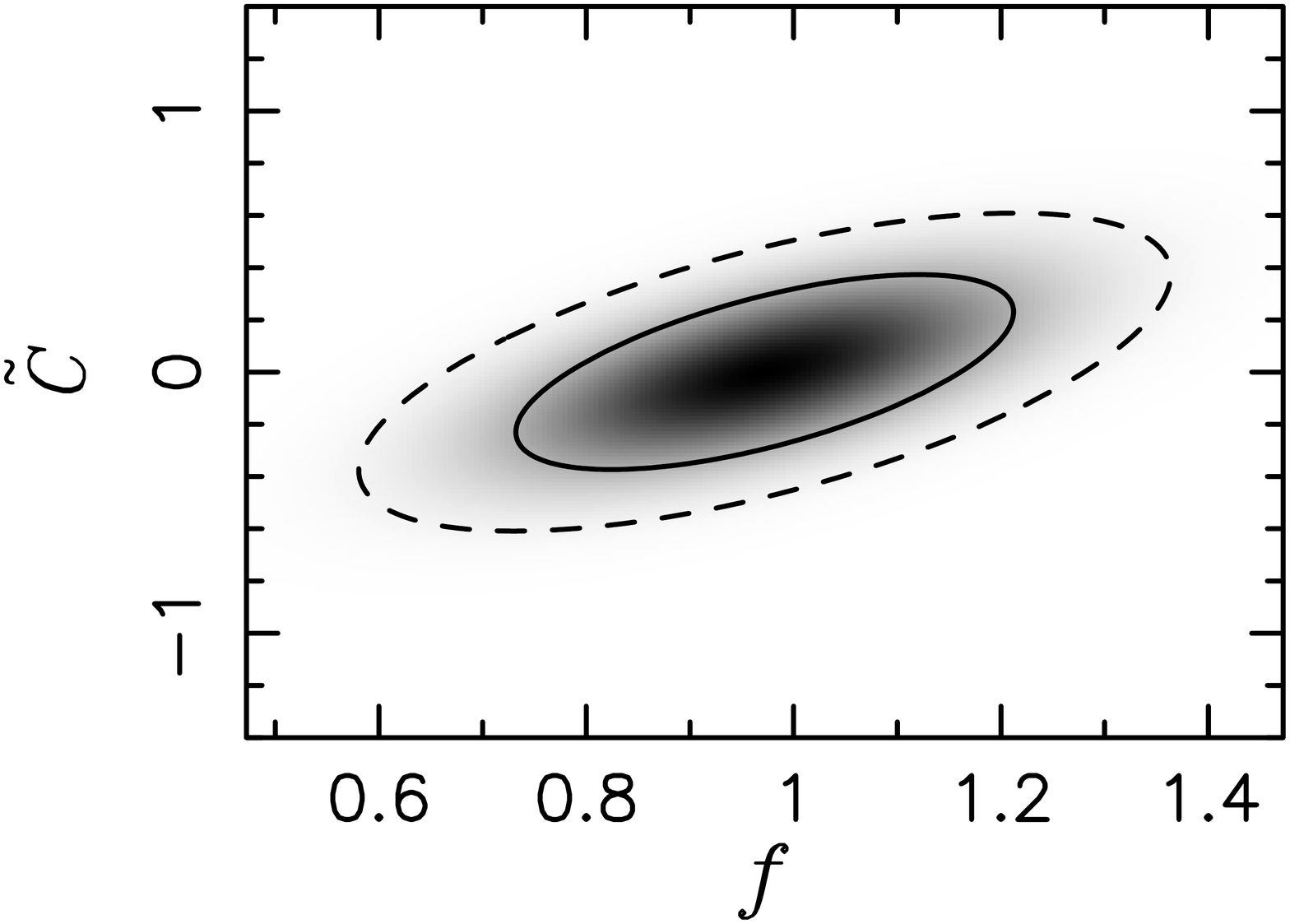}
		\includegraphics[trim = 1.5cm 3cm 0.2cm 2cm, scale = 0.165]{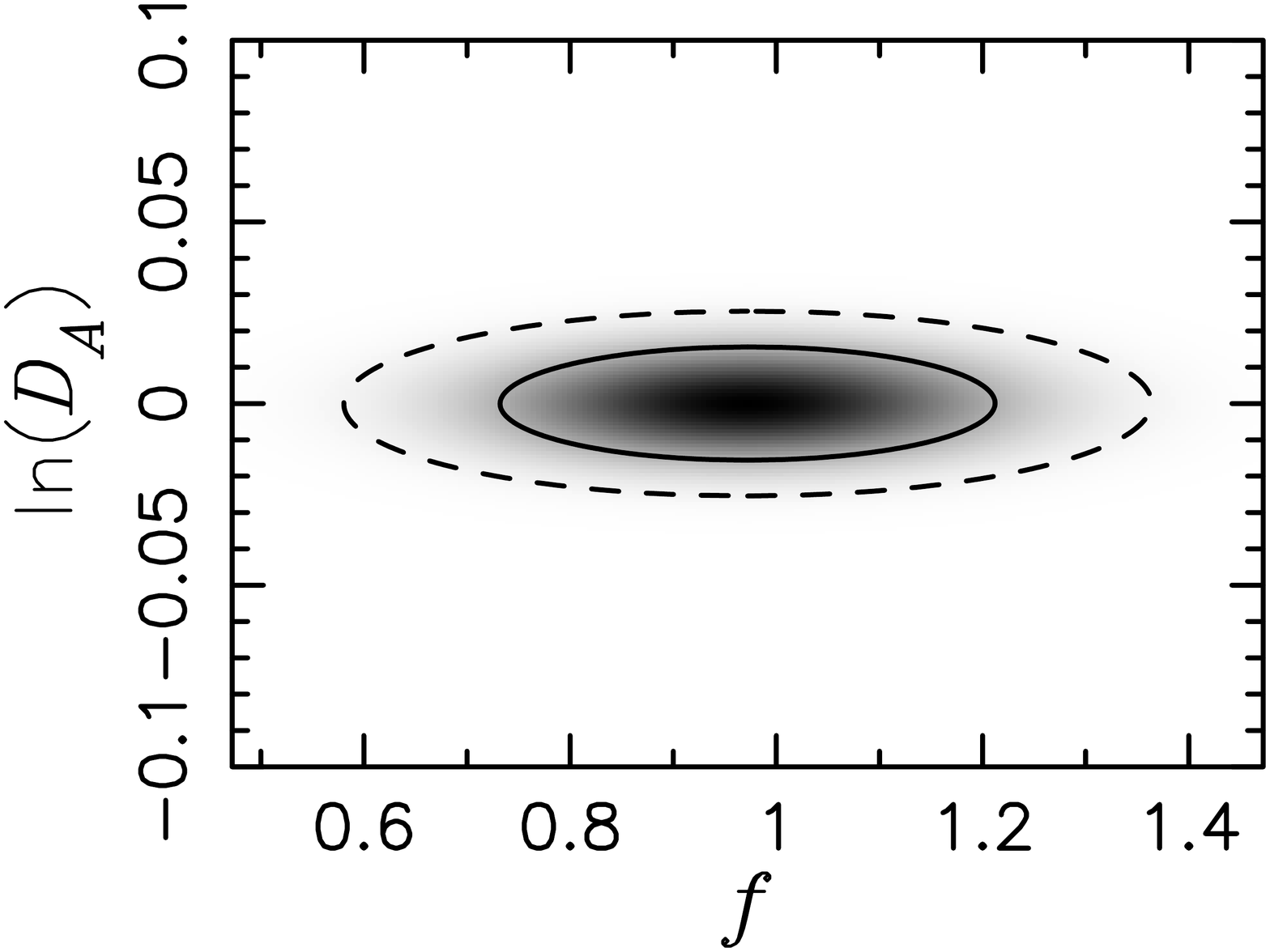}
		\includegraphics[trim = 1.5cm 3cm 6.2cm 2cm, scale = 0.165]{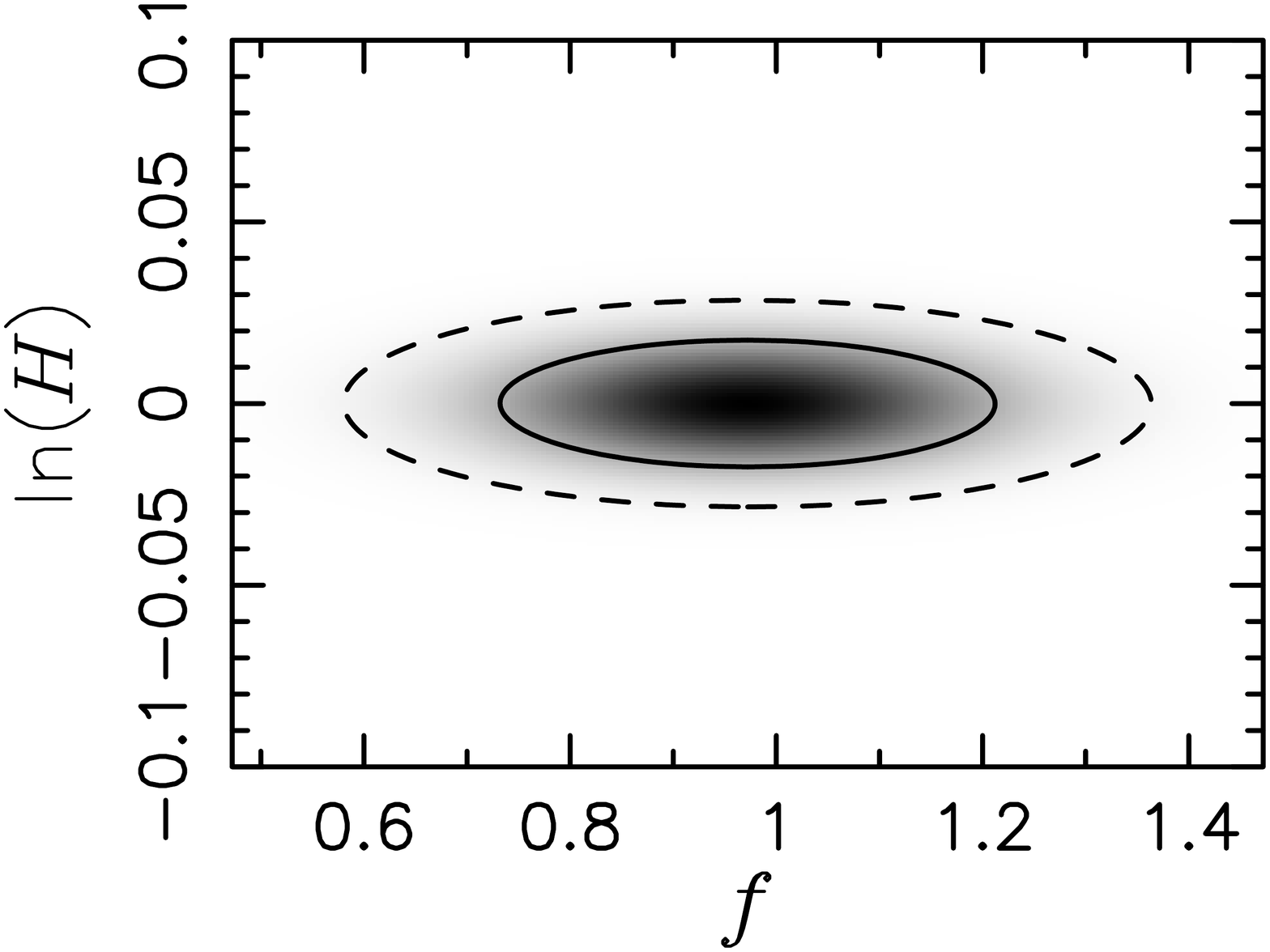}
	\end{center}
\caption{Two-dimensional joint marginalized likelihood distributions
 computed from the fiducial LAE galaxy power spectrum
 combined with the fiducial LAE galaxy {\it reduced} bispectrum (no
 \lya{} radiative 
 transfer effects added, but including marginalization over $C_{v}$,
 $C_{vv}$, and $\tilde{C}$). We show the correlations
 between the growth rate of structure, $f$, and various
 parameters including (clockwise from top left): the amplitude,
 $\ln(A)$, linear bias, $\tilde{b}_{1}$, non-linear bias,
 $\tilde{b}_{2}$, linear peculiar velocity \lya{} effect, $C_{v}$,
 non-linear peculiar velocity \lya{} effect, $C_{vv}$, the non-linear
 combination of other radiative transfer effect, $\tilde{C}$, angular
 diameter distance, $\ln(D_{A})$, and the Hubble rate, $\ln(H)$. The
 solid and dashed curves show the 1- and
 2-$\sigma$ joint marginalized constraints, respectively.}
\label{fig:2DRBSandPS}
\end{figure*}

\section[Cosmological constraints from combining the power spectrum and bispectrum]{Cosmological constraints from combining the power spectrum and bispectrum} \label{sec:higherorder}

We next discuss the improvements on the cosmological constraints
available when we combine the LAE power spectrum with either the LAE
reduced bispectrum or the bispectrum. When combining the reduced
bispectrum (and the bispectrum) to the information from the power spectrum, we
assume that there is no covariance between the power spectrum and the
 reduced bispectrum (or the bispectrum), which is incorrect. 
Therefore, the numerical values of the 1-$\sigma$ constraints
on various parameters reported here should be considered as lower bounds.

\subsection{Fiducial LAE model for power spectrum and bispectrum} \label{sec:fiducial_higherorder}

We first consider our fiducial model in which all \lya{} radiative
transfer coefficients are set to zero. The number of parameters in this
model is nine. While the LAE galaxy reduced bispectrum is insensitive to
the amplitude of the matter power spectrum, we must marginalize over the
amplitude information in the LAE galaxy power spectrum. The parameters include
four cosmological parameters: the amplitude [$\ln(A)$], $f$, $\ln(D_{A})$, and $\ln(H)$; three radiative transfer parameters: $C_{v}$, $C_{vv}$, and $\tilde{C}$; and the linear and non-linear galaxy biases: $\tilde{b}_{1}$ and $\tilde{b}_{2}$. 

\subsubsection{Combined power spectrum and reduced bispectrum} \label{sec:dist}

\begin{figure}
	\begin{center}
		\includegraphics[trim = 2.5cm 2.5cm 2.2cm 3cm, scale = 0.35]{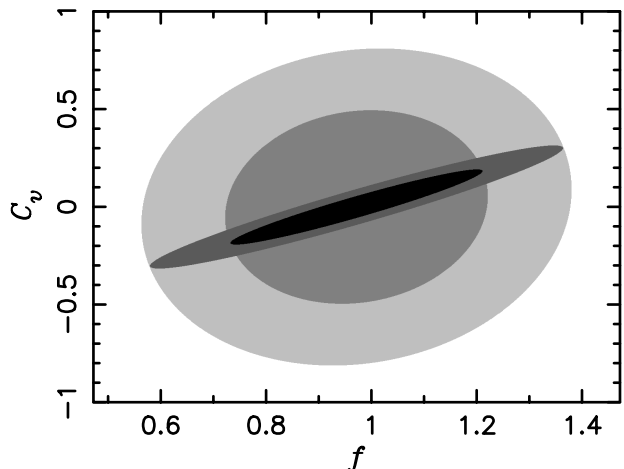}
	\end{center}
\caption{Comparison of the joint two-dimensional constraints on  $f$ and $C_{v}$. Outer two ellipses correspond to the 1- and 2-$\sigma$ constraints generated from the fiducial LAE galaxy reduced bispectrum only. Two narrower ellipses correspond to the 1- and 2-$\sigma$ constraints generated from the fiducial LAE galaxy power spectrum combined with the fiducial LAE galaxy reduced bispectrum.  }
\label{fig:PSimprovement}
\end{figure}

Figure~\ref{fig:2DRBSandPS} shows the expected constraints from a joint
analysis of the reduced bispectrum and the power spectrum on various
pairs of parameters involving $f$. Comparing this figure with
Figure~\ref{fig:2DRBS}, we find that adding the power spectrum does not
improve the constraints on $f$ and the bias parameters very much, but
improves the constraints on all the other
parameters. Figure~\ref{fig:PSimprovement} shows this more clearly:
adding the power spectrum information does not improve the constraint on
$f$, but it substantially improves the constraint on $C_v$. 

What does this imply? This implies that the uncertainty in $f$ is now
dominated by the correlation between $f$ and the bias parameters - the
correlation that we have discussed in Section~\ref{sec:fiducialRB}.
Comparing the fourth and sixth
columns of Table~\ref{tab:fiducialbispectrum} shows this quantitatively.

Comparing the fourth and sixth
columns of Table~\ref{tab:fiducialbispectrum} also shows that 
adding the power spectrum does improve the constraints on $D_A$ and $H$
substantially, as the power 
spectrum contains features such as BAO and AP test, whereas such
information is largely cancelled out in the reduced bispectrum. 

Nevertheless, as the reduced
bispectrum still has some sensitivity to these features (i.e.,
cancellation is not exact), the constraints on $D_A$ and $H$ from the
power spectrum and the reduced bispectrum are slightly better than those
from the power spectrum alone. Comparing the third column of
Table~\ref{tab:betapriorsfiducial} and the sixth
columns of Table~\ref{tab:fiducialbispectrum}, we find that the expected
constraints improve from 1.1 to 1.0 per cent for $D_A$ and 1.3 to 1.2
per cent for $H$.

\begin{figure*} 
	\begin{center}
		\includegraphics[trim = 2.5cm 5cm 2.2cm 2cm, scale = 0.27]{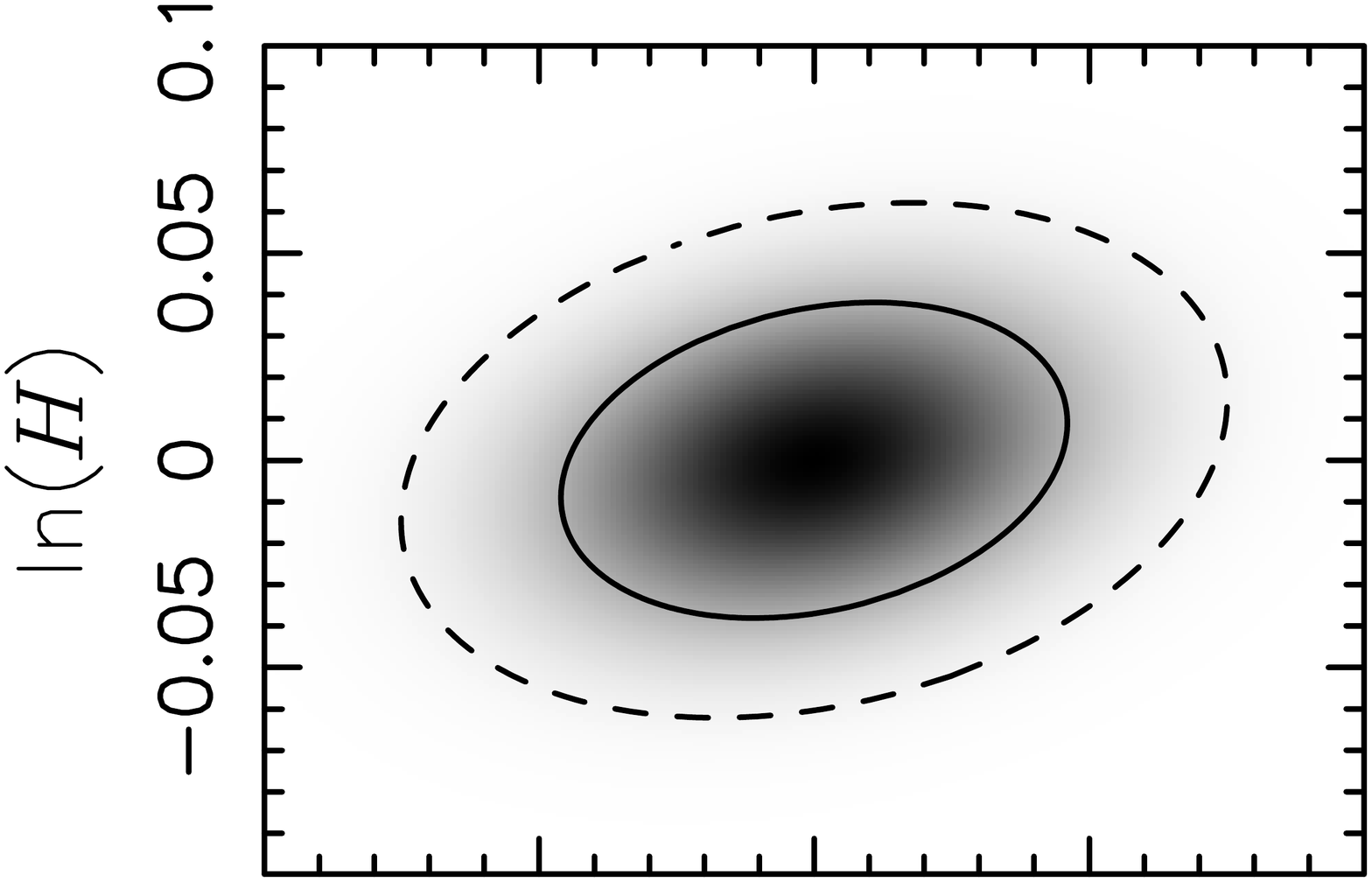}
		\includegraphics[trim = 6.5cm 5cm 2.2cm 2cm, scale = 0.27]{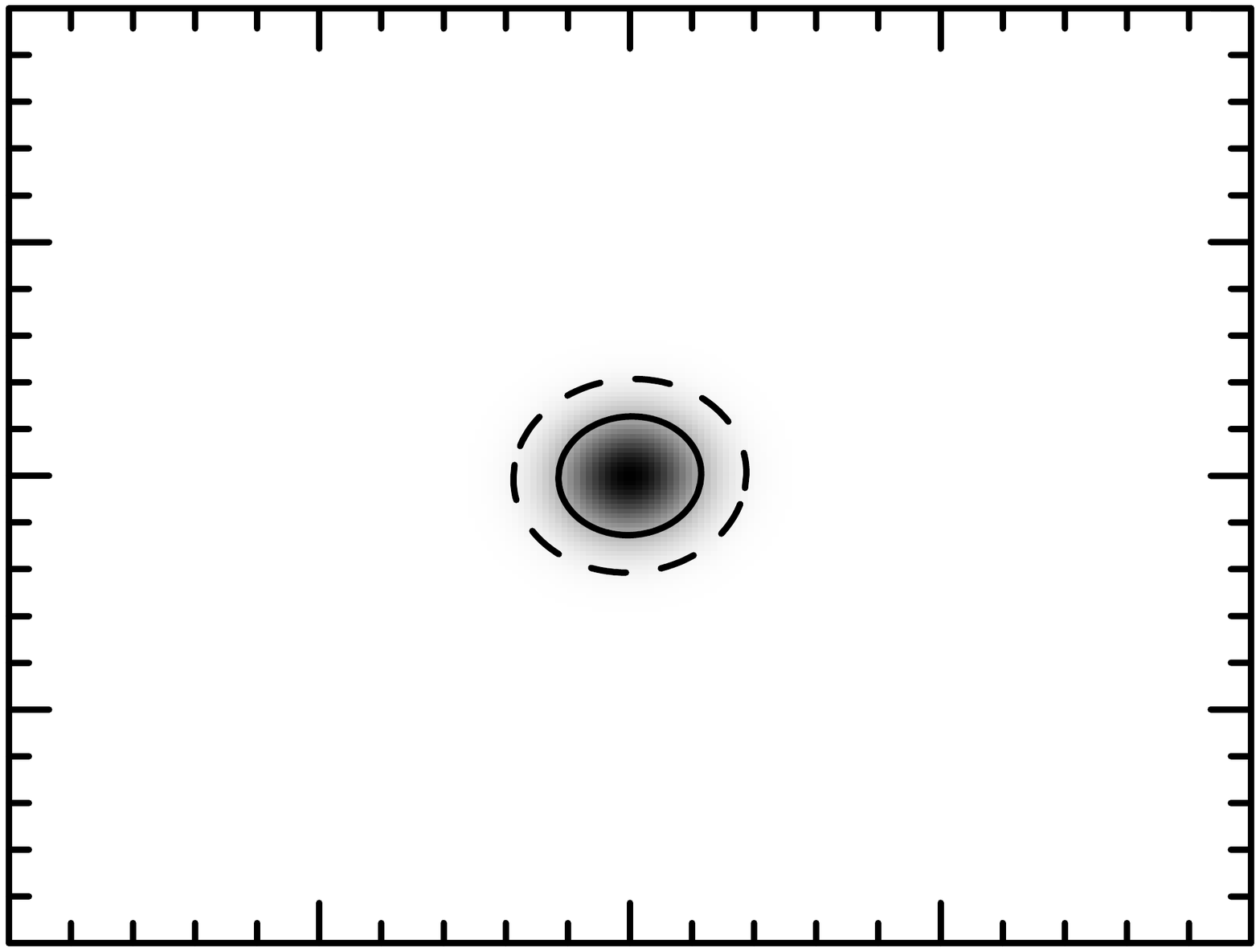}
		\includegraphics[trim = 6.5cm 5cm 2.2cm 2cm, scale = 0.27]{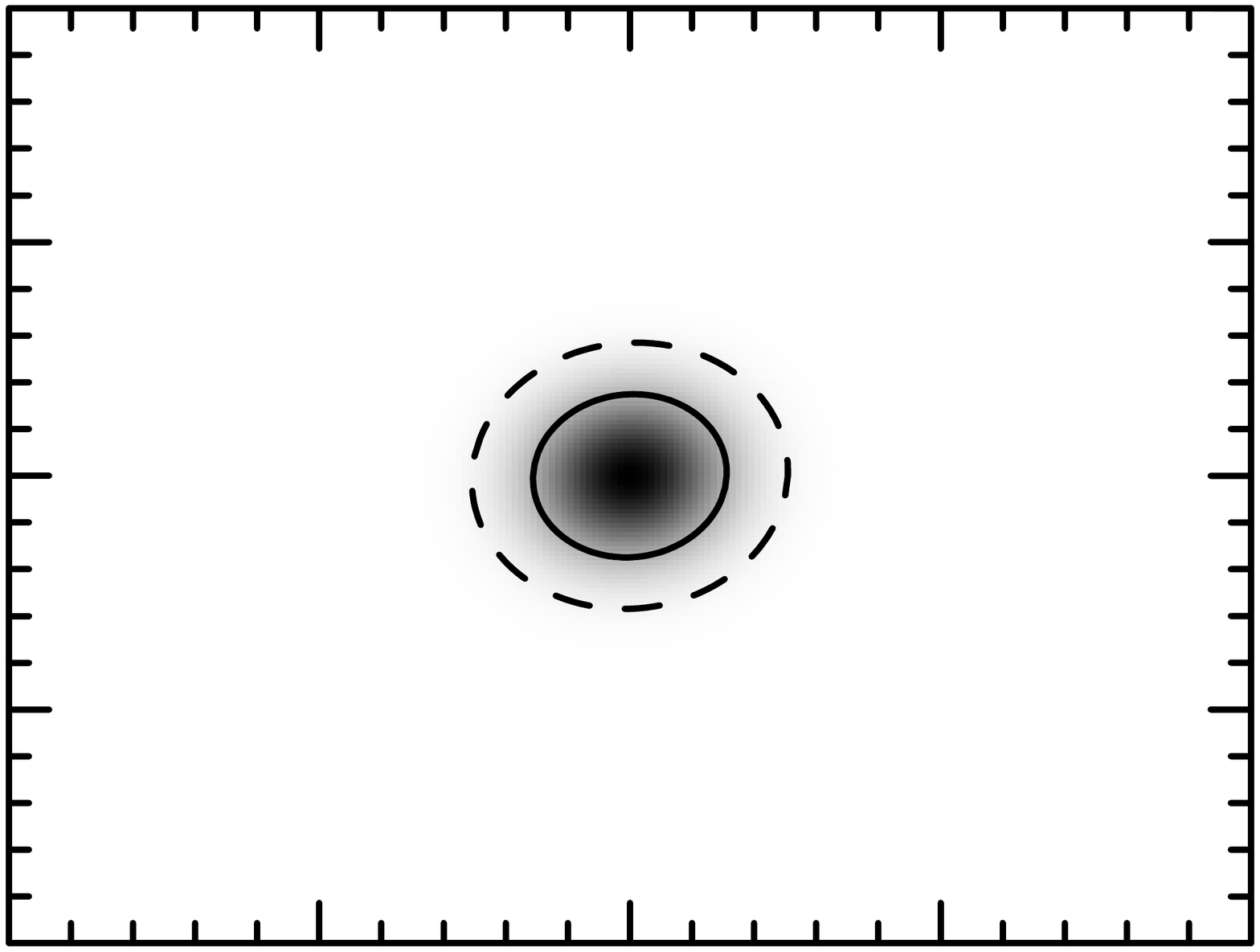}
		\includegraphics[trim = 2.5cm 2cm 2.2cm 2cm, scale = 0.27]{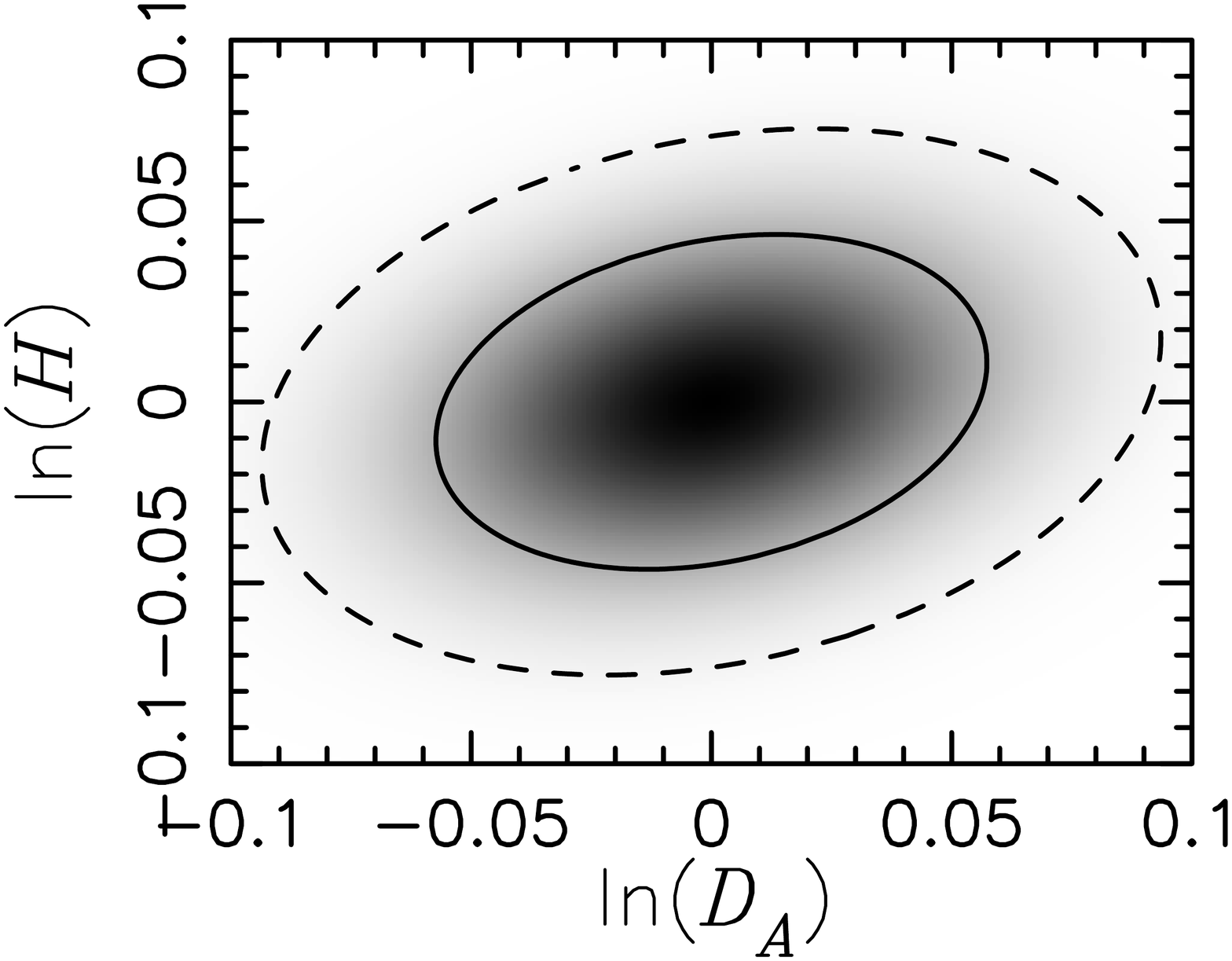}
		\includegraphics[trim = 6.5cm 2cm 2.2cm 2cm, scale = 0.27]{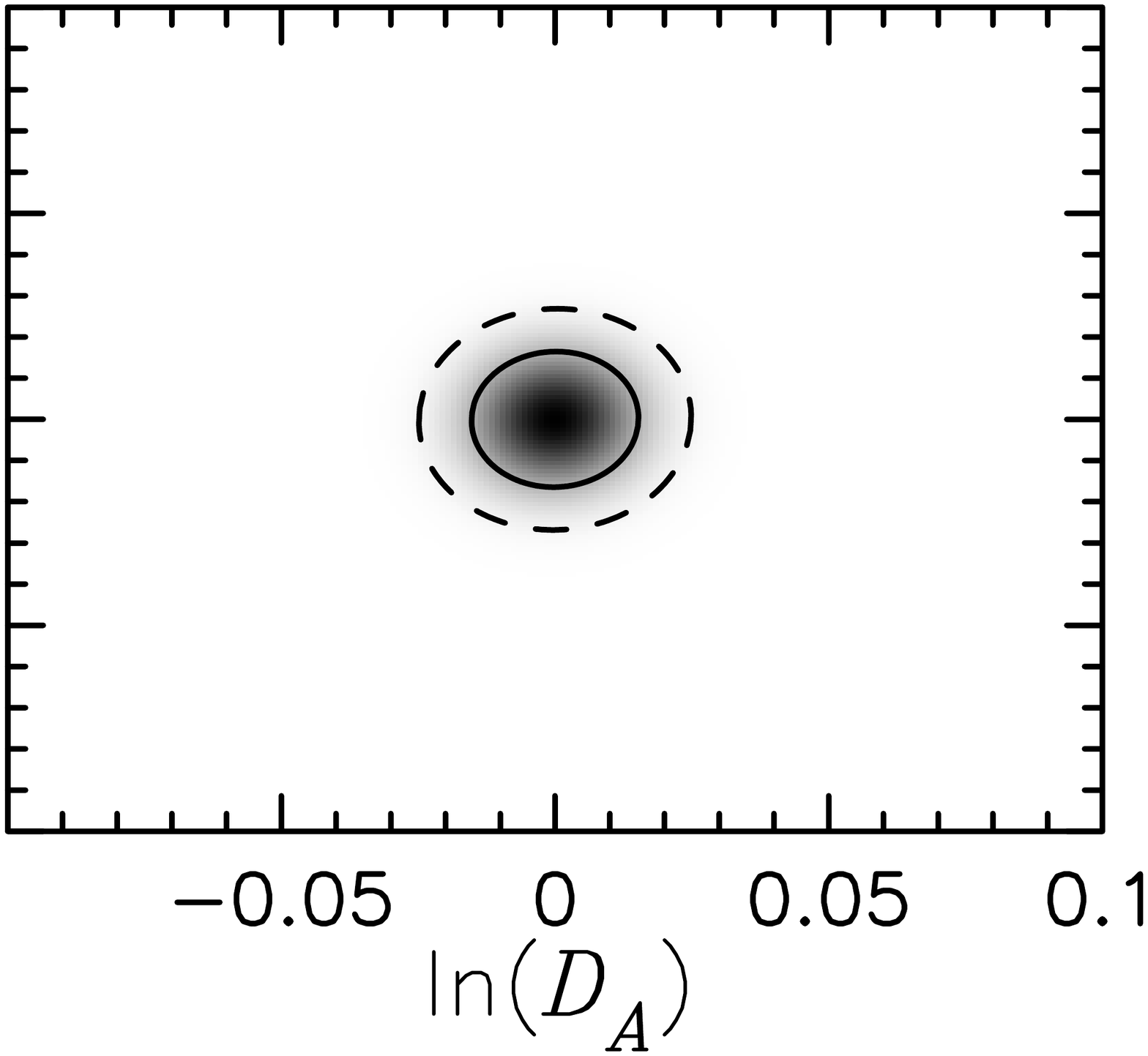}
		\includegraphics[trim = 6.5cm 2cm 2.2cm 2cm, scale = 0.27]{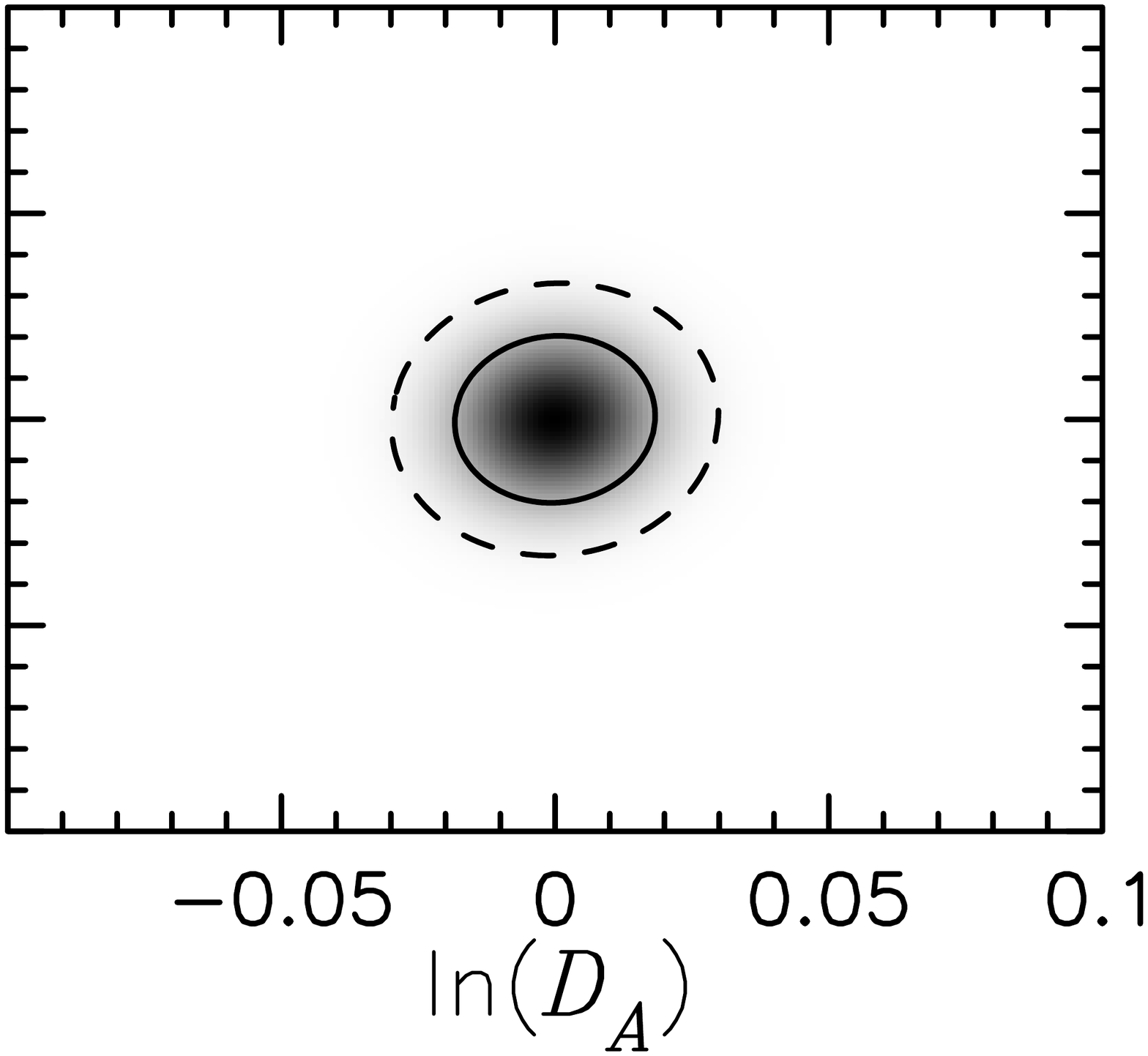}
	\end{center}
\caption{Two-dimensional joint marginalized likelihood distributions for
the angular diameter distance,
 $D_{A}$, and the Hubble rate, $H$. Shown also are the 1- (\textit{solid})
 and 2-$\sigma$ (\textit{dashed}) joint likelihood contours. 
From \textit{left} to \textit{right} show the constraints
 generated from the LAE galaxy reduced bispectrum only, the LAE galaxy
 power spectrum and bispectrum combined, and the LAE galaxy power
 spectrum and reduced bispectrum combined.
\textit{Top panels:}
 Fiducial case, with no \lya{} radiative transfer effects added to the
 fiducial parameters, but marginalizing over $C_{v}$, $C_{vv}$, and
 $\tilde{C}$. \textit{Bottom panels:} The inclusion of the first-order
 \lya{} radiative transfer effects, $C_{v} = 0.11$, $C_{\Gamma} = 0.05$,
 and $C_{\rho} = -0.39$, and marginalizing over $C_{v}$, $C_{vv}$, and
 $\tilde{C}$. } 
\label{fig:DA_Hcomplete}
\end{figure*}

\subsubsection{Combined power spectrum and bispectrum}

Next, we combine the power spectrum with the bispectrum (rather than the
reduced bispectrum). As far as $f$ is concerned, we have the same story:
adding the power spectrum does not improve the expected error bar on
$f$ (see the fifth column of Table~\ref{tab:fiducialbispectrum}).

On the other hand, a joint analysis of the power spectrum and the bispectrum
yields a significant improvement on the angular diameter distance and
the Hubble rate. This is because the bispectrum also contains the BAO
features and the AP test in its wavenumber dependence. However, this
could be due 
to our ignoring a covariance between the power spectrum and the
bispectrum: a correlation between them would degrade the constraints in
a joint analysis. This point requires a further investigation.

\subsection{LAE model for power spectrum combined with bispectrum}

Finally, we consider the inclusion of \lya{} radiative transfer effects
on the recovery on $f$, $\ln(D_A)$, and $\ln(H)$, by adding the linear
\lya{} radiative transfer parameters, $C_{\Gamma} = 0.05$, $C_{\rho} =
-0.39$ and $C_{v} = 0.11$ from
\citet{Wyithe:2011p12569}. 
Table~\ref{tab:bispectrum} shows the results:
the expected constraints are slightly weaker than those from the
fiducial case, which is again due to a smaller effective bias,
$\tilde{b}_1$, reducing the amplitude of the signal relative to the shot noise.

\section[Summary and conclusion]{Summary and conclusion} \label{sec:conclusion}

In this paper we have studied how the radiative transfer
effects alter the power spectrum and bispectrum of LAE galaxies, and how
we can use these properties to separate  the radiative
transfer effects and the cosmological effects, so that we can improve the
cosmological constraints derived from them.

First, as a follow up to \citet{Wyithe:2011p12569}, we
show that the growth rate of structure ($f$) and the parameter
$C_{v}$ describing the radiative transfer effects of velocity gradients
are completely degenerate in the linear power spectrum. Next, by
performing a perturbation theory expansion of the \lya{} radiative
transfer effects, we derive the next-to-leading order
corrections to the density fields of LAE galaxies. This
allows us to derive the leading-order expression for the bispectrum of
LAE galaxies. We then show that the reduced bispectrum alone can
determine $f$ and $C_v$ separately, leaving no degeneracy between
them. Adding the power spectrum information to the reduced bispectrum
does not improve the precision of $f$ further, as the precision of $f$ is now
limited by remaining correlations between $f$ and the galaxy bias parameters,
$b_1$ and $b_2$.

We find that HETDEX-like surveys of LAE galaxies can
determine $f$ to about 20 per cent accuracy, if we do not assume any
prior information on $C_v$. Including the prior on $C_v$, the uncertainty on
$f$ can be reduced down to 7 per cent. Note that this is the uncertainty
on $f$, rather than on $\beta=f/b_1$.

We find that the constraints on the angular diameter
distance and the Hubble expansion rate are not directly affected by the
radiative transfer parameters (with the caveat that we
have assumed that the effect of the UV ionizing background fluctuation
is not scale dependant). The only indirect effect is a
slight reduction of the effective linear galaxy bias, which reduces the
amplitude of the LAE power spectrum with respect to the shot noise, thus
slightly increasing the  uncertainties in the angular diameter
distance and the Hubble rate.
Comparison between the top and bottom panels of Figure
\ref{fig:DA_Hcomplete} shows this graphically.

Finally, to summarize the results of this work, we provide Table
\ref{tab:summary} detailing the constraints on $\beta$, $f$,
$D_{A}$, and $H$ expected from HETDEX-like surveys. This
table shows how 
powerful such surveys are in terms of measuring the distance, the expansion
rate, as well as the growth rate of the structure in a high-redshift
universe, and the determination of these quantities are not
significantly compromised by the \lya{} radiative transfer effect.

\begin{table*}
\begin{tabular}{@{}lcccccccccc}
\hline

Priors on $C_{v}$ & Parameter & Model & PS & &  R BS &  & PS + BS & &  PS + R BS  & \\
& &  & 1-$\sigma$ & per cent & 1-$\sigma$ & per cent & 1-$\sigma$ & per cent & 1-$\sigma$ & per cent \\
\hline
No Priors & $\beta$ & Galaxy  & 0.0213 & 4.8 & - & - & - & - & - & - \\
No Priors & $f$ & Fiducial  & - & - & 0.1645 & 16.9 & 0.1602 & 16.5 & 0.1579 & 16.2\\
0.01 & $f$ & Fiducial  & 0.0218 ($\tilde{\beta}$) & 4.9 &0.1635 & 16.8 & 0.0520 & 5.3 & 0.0576 & 5.9\\
0.1 & $f$ & Fiducial  & 0.0491 ($\tilde{\beta}$) & 11.1 & 0.1636 & 16.8 & 0.1055 & 10.9 & 0.1063 & 10.9\\
\hline
No Priors & $f$ & LAE effects & - & - & 0.2104  & 21.6 & 0.2039 & 21.0 & 0.2014 & 20.7\\
0.01 & $f$ & LAE effects  & 0.0283 ($\tilde{\beta}$) & 5.6 & 0.2089& 21.5 & 0.0645 & 6.7& 0.0685& 7.1\\
0.1 & $f$ & LAE effects  & 0.0633 ($\tilde{\beta}$) & 12.5 & 0.2090& 21.5 & 0.1262 & 13.0 &  0.1270& 13.1\\
\hline
No Priors & ln($D_{A}$) & Fiducial  & 0.0110 & 1.10 & 0.0303 & 3.03 & 0.0076 &0.76 &  0.0103 & 1.03\\
No Priors & ln($D_{A}$) & LAE effects & 0.0128 & 1.28 & 0.0378 & 3.78 & 0.0099 & 0.99 & 0.0120 & 1.20 \\
No Priors & ln($H$) & Fiducial  & 0.0132 & 1.32 & 0.0251  & 2.51 &   0.0084 & 0.84 & 0.0115 & 1.15\\
No Priors & ln($H$) & LAE effects  & 0.0151 & 1.51 & 0.0305 & 3.05 & 0.0107 & 1.07 & 0.0133 & 1.33\\
\hline
\end{tabular}
\caption{Summary table of the relevant 1-$\sigma$ constraints (and
 fractional errors) generated from the one-dimensional likelihood
 distributions for $\beta$, $f$, $\ln(D_{A})$, and $\ln(A)$. Models
 considered are `Galaxy' (no \lya{} radiative transfer effects, and no
 marginalization over the radiative transfer parameters), `Fiducial'
 (no \lya{} radiative transfer effects, but marginalized over the
 radiative transfer parameters), and `LAE effects' (non-zero \lya{} radiative
 transfer parameters are included and marginalize over). We consider a
 power spectrum only model (PS), reduced bispectrum only (R BS), power
 spectrum combined with the  bispectrum (PS + BS), and the power spectrum
 combined with the reduced bispectrum (PS + R BS). We explore three
 priors on the  \lya{} radiative parameter associated with the peculiar
 velocity, $C_{v}$: no prior at all, prior of 0.01, and prior of 0.1.} 
\label{tab:summary}
\end{table*}

\section*{Acknowledgments}

BG acknowledges the support of the Australian Postgraduate Award.  The
Centre for All-sky Astrophysics is an Australian Research Council
Centre of Excellence, funded by grant CE110001020. BG would also like to acknowledge 
the partial travel support to the USA provided by the Astronomical Society of Australia (ASA). 
BG would like to thank the Texas Cosmology Center at the University of Texas
at Austin for their hospitality during this work, and  the Kavli Institute
for the Physics and Mathematics of the Universe for their hospitality
during the completion of this work.  
We would like to thank Donghui Jeong for helpful discussions
contributing to the completion of this work as well as to the derivation
provided in Appendix C. 
EK is supported in part by NSF grant AST-0807649 and
NASA grant NNX08AL43G.

\bibliography{Paper2}

\begin{thebibliography}{36}
\expandafter\ifx\csname natexlab\endcsname\relax\def\natexlab#1{#1}\fi

\bibitem[{Bernardeau {et~al.}(2002)Bernardeau, Colombi, Gazta{\~n}aga, \&
  Scoccimarro}]{Bernardeau:2002p8437}
Bernardeau F., Colombi S., Gazta{\~n}aga E., Scoccimarro R., 2002, Physics
  Reports, 367, 1

\bibitem[{{Blake} {et~al.}(2012){Blake}, {Brough}, {Colless}, {Contreras},
  {Couch}, {Croom}, {Croton}, {Davis}, {Drinkwater}, {Forster}, {Gilbank},
  {Gladders}, {Glazebrook}, {Jelliffe}, {Jurek}, {Li}, {Madore}, {Martin},
  {Pimbblet}, {Poole}, {Pracy}, {Sharp}, {Wisnioski}, {Woods}, {Wyder}, \&
  {Yee}}]{blake/etal:2012}
{Blake} C., {Brough} S., {Colless} M., {Contreras} C., {Couch} W., {Croom} S.,
  {Croton} D., {Davis} T.~M., {Drinkwater} M.~J., {Forster} K., {Gilbank} D.,
  {Gladders} M., {Glazebrook} K., {Jelliffe} B., {Jurek} R.~J., {Li} I.-h.,
  {Madore} B., {Martin} D.~C., {Pimbblet} K., {Poole} G.~B., {Pracy} M.,
  {Sharp} R., {Wisnioski} E., {Woods} D., {Wyder} T.~K., {Yee} H.~K.~C., 2012,
  Monthly Notices of the Royal Astronomical Society, 425, 405

\bibitem[{{Blake} {et~al.}(2011{\natexlab{a}}){Blake}, {Brough}, {Colless},
  {Contreras}, {Couch}, {Croom}, {Davis}, {Drinkwater}, {Forster}, {Gilbank},
  {Gladders}, {Glazebrook}, {Jelliffe}, {Jurek}, {Li}, {Madore}, {Martin},
  {Pimbblet}, {Poole}, {Pracy}, {Sharp}, {Wisnioski}, {Woods}, {Wyder}, \&
  {Yee}}]{blake/etal:2011c}
{Blake} C., {Brough} S., {Colless} M., {Contreras} C., {Couch} W., {Croom} S.,
  {Davis} T., {Drinkwater} M.~J., {Forster} K., {Gilbank} D., {Gladders} M.,
  {Glazebrook} K., {Jelliffe} B., {Jurek} R.~J., {Li} I.-H., {Madore} B.,
  {Martin} D.~C., {Pimbblet} K., {Poole} G.~B., {Pracy} M., {Sharp} R.,
  {Wisnioski} E., {Woods} D., {Wyder} T.~K., {Yee} H.~K.~C.,
  2011{\natexlab{a}}, Monthly Notices of the Royal Astronomical Society, 415,
  2876

\bibitem[{{Blake} {et~al.}(2011{\natexlab{b}}){Blake}, {Davis}, {Poole},
  {Parkinson}, {Brough}, {Colless}, {Contreras}, {Couch}, {Croom},
  {Drinkwater}, {Forster}, {Gilbank}, {Gladders}, {Glazebrook}, {Jelliffe},
  {Jurek}, {Li}, {Madore}, {Martin}, {Pimbblet}, {Pracy}, {Sharp}, {Wisnioski},
  {Woods}, {Wyder}, \& {Yee}}]{blake/etal:2011b}
{Blake} C., {Davis} T., {Poole} G.~B., {Parkinson} D., {Brough} S., {Colless}
  M., {Contreras} C., {Couch} W., {Croom} S., {Drinkwater} M.~J., {Forster} K.,
  {Gilbank} D., {Gladders} M., {Glazebrook} K., {Jelliffe} B., {Jurek} R.~J.,
  {Li} I.-H., {Madore} B., {Martin} D.~C., {Pimbblet} K., {Pracy} M., {Sharp}
  R., {Wisnioski} E., {Woods} D., {Wyder} T.~K., {Yee} H.~K.~C.,
  2011{\natexlab{b}}, Monthly Notices of the Royal Astronomical Society, 415,
  2892

\bibitem[{{Blake} {et~al.}(2011{\natexlab{c}}){Blake}, {Kazin}, {Beutler},
  {Davis}, {Parkinson}, {Brough}, {Colless}, {Contreras}, {Couch}, {Croom},
  {Croton}, {Drinkwater}, {Forster}, {Gilbank}, {Gladders}, {Glazebrook},
  {Jelliffe}, {Jurek}, {Li}, {Madore}, {Martin}, {Pimbblet}, {Poole}, {Pracy},
  {Sharp}, {Wisnioski}, {Woods}, {Wyder}, \& {Yee}}]{blake/etal:2011a}
{Blake} C., {Kazin} E.~A., {Beutler} F., {Davis} T.~M., {Parkinson} D.,
  {Brough} S., {Colless} M., {Contreras} C., {Couch} W., {Croom} S., {Croton}
  D., {Drinkwater} M.~J., {Forster} K., {Gilbank} D., {Gladders} M.,
  {Glazebrook} K., {Jelliffe} B., {Jurek} R.~J., {Li} I.-H., {Madore} B.,
  {Martin} D.~C., {Pimbblet} K., {Poole} G.~B., {Pracy} M., {Sharp} R.,
  {Wisnioski} E., {Woods} D., {Wyder} T.~K., {Yee} H.~K.~C.,
  2011{\natexlab{c}}, Monthly Notices of the Royal Astronomical Society, 418,
  1707

\bibitem[{Fry \& Gaztanaga(1993)}]{Fry:1993p8713}
Fry J.~N., Gaztanaga E., 1993, Astrophysical Journal, 413, 447

\bibitem[{Gawiser {et~al.}(2007)Gawiser, Francke, Lai, Schawinski, Gronwall,
  Ciardullo, Quadri, Orsi, Barrientos, Blanc, Fazio, Feldmeier, sheng Huang,
  Infante, Lira, Padilla, Taylor, Treister, Urry, van Dokkum, \&
  Virani}]{Gawiser:2007p13156}
Gawiser E., Francke H., Lai K., Schawinski K., Gronwall C., Ciardullo R.,
  Quadri R., Orsi A., Barrientos L.~F., Blanc G.~A., Fazio G., Feldmeier J.~J.,
  sheng Huang J., Infante L., Lira P., Padilla N., Taylor E.~N., Treister E.,
  Urry C.~M., van Dokkum P.~G., Virani S.~N., 2007, The Astrophysical Journal,
  671, 278

\bibitem[{Guaita {et~al.}(2010)Guaita, Gawiser, Padilla, Francke, Bond,
  Gronwall, Ciardullo, Feldmeier, Sinawa, Blanc, \& Virani}]{Guaita:2010p13210}
Guaita L., Gawiser E., Padilla N., Francke H., Bond N.~A., Gronwall C.,
  Ciardullo R., Feldmeier J.~J., Sinawa S., Blanc G.~A., Virani S., 2010, The
  Astrophysical Journal, 714, 255

\bibitem[{Hill {et~al.}(2008)Hill, Gebhardt, Komatsu, Drory, MacQueen, Adams,
  Blanc, Koehler, Rafal, Roth, Kelz, Grupp, Murphy, Palunas, Gronwall,
  Ciardullo, Bender, Hopp, \& Schneider}]{Hill:2008p12595}
Hill G.~J., Gebhardt K., Komatsu E., Drory N., MacQueen P.~J., Adams J., Blanc
  G.~A., Koehler R., Rafal M., Roth M.~M., Kelz A., Grupp F., Murphy J.,
  Palunas P., Gronwall C., Ciardullo R., Bender R., Hopp U., Schneider D.~P.,
  2008, Panoramic Views of the Universe, ASP Conf. Series, 399, 115

\bibitem[{Hill {et~al.}(2004)Hill, Gebhardt, Komatsu, \&
  MacQueen}]{Hill:2004p12593}
Hill G.~J., Gebhardt K., Komatsu E., MacQueen P.~J., 2004, The New Cosmology:
  Conference on Strings and Cosmology; The Mitchell Symposium on Observational
  Cosmology, AIP Conference Proc., 743, 224

\bibitem[{Hui \& Gnedin(1997)}]{Hui:1997p339}
Hui L., Gnedin N.~Y., 1997, Royal Astronomical Society, 292, 27

\bibitem[{Iye {et~al.}(2006)Iye, Ota, Kashikawa, Furusawa, Hashimoto, Hattori,
  Matsuda, Morokuma, Ouchi, \& Shimasaku}]{Iye:2006p13032}
Iye M., Ota K., Kashikawa N., Furusawa H., Hashimoto T., Hattori T., Matsuda
  Y., Morokuma T., Ouchi M., Shimasaku K., 2006, Nature, 443, 186

\bibitem[{Jain \& Bertschinger(1994)}]{Jain:1994p8584}
Jain B., Bertschinger E., 1994, The Astrophysical Journal, 431, 495

\bibitem[{Jeong \& Komatsu(2006)}]{Jeong:2006p5318}
Jeong D., Komatsu E., 2006, The Astrophysical Journal, 651, 619

\bibitem[{Kaiser(1984)}]{Kaiser:1984p8759}
Kaiser N., 1984, Astrophysical Journal, 284, L9

\bibitem[{Kaiser(1987)}]{Kaiser:1987p8786}
---, 1987, Royal Astronomical Society, 227, 1

\bibitem[{Kashikawa {et~al.}(2006)Kashikawa, Shimasaku, Malkan, Doi, Matsuda,
  Ouchi, Taniguchi, Ly, Nagao, Iye, Motohara, Murayama, Murozono, Nariai, Ohta,
  Okamura, Sasaki, Shioya, \& Umemura}]{Kashikawa:2006p13047}
Kashikawa N., Shimasaku K., Malkan M.~A., Doi M., Matsuda Y., Ouchi M.,
  Taniguchi Y., Ly C., Nagao T., Iye M., Motohara K., Murayama T., Murozono K.,
  Nariai K., Ohta K., Okamura S., Sasaki T., Shioya Y., Umemura M., 2006, The
  Astrophysical Journal, 648, 7

\bibitem[{Komatsu {et~al.}(2011)Komatsu, Smith, Dunkley, Bennett, Gold,
  Hinshaw, Jarosik, Larson, Nolta, Page, Spergel, Halpern, Hill, Kogut, Limon,
  Meyer, Odegard, Tucker, Weiland, Wollack, \& Wright}]{Komatsu:2011p12557}
Komatsu E., Smith K.~M., Dunkley J., Bennett C.~L., Gold B., Hinshaw G.,
  Jarosik N., Larson D., Nolta M.~R., Page L., Spergel D.~N., Halpern M., Hill
  R.~S., Kogut A., Limon M., Meyer S.~S., Odegard N., Tucker G.~S., Weiland
  J.~L., Wollack E., Wright E.~L., 2011, The Astrophysical Journal Supplement,
  192, 18

\bibitem[{Kova{\v c} {et~al.}(2007)Kova{\v c}, Somerville, Rhoads, Malhotra, \&
  Wang}]{Kovac:2007p13181}
Kova{\v c} K., Somerville R.~S., Rhoads J.~E., Malhotra S., Wang J., 2007, The
  Astrophysical Journal, 668, 15

\bibitem[{Laursen {et~al.}(2011)Laursen, Sommer-Larsen, \&
  Razoumov}]{Laursen:2011p8323}
Laursen P., Sommer-Larsen J., Razoumov A.~O., 2011, The Astrophysical Journal,
  728, 52

\bibitem[{Lehnert {et~al.}(2010)Lehnert, Nesvadba, Cuby, Swinbank, Morris,
  Cl{\'e}ment, Evans, Bremer, \& Basa}]{Lehnert:2010p13102}
Lehnert M.~D., Nesvadba N. P.~H., Cuby J.-G., Swinbank A.~M., Morris S.,
  Cl{\'e}ment B., Evans C.~J., Bremer M.~N., Basa S., 2010, Nature, 467, 940,
  (c) 2010: Nature

\bibitem[{Linder(2005)}]{Linder:2005p13557}
Linder E.~V., 2005, Physical Review D, 72, 43529

\bibitem[{McDonald(2006)}]{McDonald:2006p8376}
McDonald P., 2006, Physical Review D, 74, 103512

\bibitem[{Morales \& Wyithe(2010)}]{Morales:2010p10051}
Morales M.~F., Wyithe J. S.~B., 2010, Annual Review of Astronomy and
  Astrophysics, 48, 127

\bibitem[{Orsi {et~al.}(2008)Orsi, Lacey, Baugh, \& Infante}]{Orsi:2008p13553}
Orsi A., Lacey C.~G., Baugh C.~M., Infante L., 2008, Monthly Notices of the
  Royal Astronomical Society, 391, 1589

\bibitem[{Ouchi {et~al.}(2010)Ouchi, Shimasaku, Furusawa, Saito, Yoshida,
  Akiyama, Ono, Yamada, Ota, Kashikawa, Iye, Kodama, Okamura, Simpson, \&
  Yoshida}]{Ouchi:2010p13150}
Ouchi M., Shimasaku K., Furusawa H., Saito T., Yoshida M., Akiyama M., Ono Y.,
  Yamada T., Ota K., Kashikawa N., Iye M., Kodama T., Okamura S., Simpson C.,
  Yoshida M., 2010, The Astrophysical Journal, 723, 869

\bibitem[{Scoccimarro {et~al.}(1999)Scoccimarro, Couchman, \&
  Frieman}]{Scoccimarro:1999p9850}
Scoccimarro R., Couchman H. M.~P., Frieman J.~A., 1999, The Astrophysical
  Journal, 517, 531

\bibitem[{Sefusatti {et~al.}(2006)Sefusatti, Crocce, Pueblas, \&
  Scoccimarro}]{Sefusatti:2006p752}
Sefusatti E., Crocce M., Pueblas S., Scoccimarro R., 2006, Physical Review D,
  74, 23522

\bibitem[{Sefusatti \& Komatsu(2007)}]{Sefusatti:2007p9477}
Sefusatti E., Komatsu E., 2007, Physical Review D, 76, 83004

\bibitem[{Seo \& Eisenstein(2003)}]{Seo:2003p11154}
Seo H.-J., Eisenstein D.~J., 2003, The Astrophysical Journal, 598, 720

\bibitem[{Shoji {et~al.}(2009)Shoji, Jeong, \& Komatsu}]{Shoji:2009p8219}
Shoji M., Jeong D., Komatsu E., 2009, The Astrophysical Journal, 693, 1404

\bibitem[{Smith {et~al.}(2008)Smith, Sheth, \& Scoccimarro}]{Smith:2008p8996}
Smith R.~E., Sheth R.~K., Scoccimarro R., 2008, Physical Review D, 78, 23523

\bibitem[{Wyithe \& Dijkstra(2011)}]{Wyithe:2011p12569}
Wyithe J. S.~B., Dijkstra M., 2011, Monthly Notices of the Royal Astronomical
  Society, 415, 3929

\bibitem[{Yoshikawa {et~al.}(2001)Yoshikawa, Taruya, Jing, \&
  Suto}]{Yoshikawa:2001p8880}
Yoshikawa K., Taruya A., Jing Y.~P., Suto Y., 2001, The Astrophysical Journal,
  558, 520

\bibitem[{Zheng {et~al.}(2010)Zheng, Cen, Trac, \&
  Miralda-Escud{\'e}}]{Zheng:2010p8285}
Zheng Z., Cen R., Trac H., Miralda-Escud{\'e} J., 2010, The Astrophysical
  Journal, 716, 574

\bibitem[{Zheng {et~al.}(2011)Zheng, Cen, Trac, \&
  Miralda-Escud{\'e}}]{Zheng:2011p8289}
---, 2011, The Astrophysical Journal, 726, 38

\end{thebibliography}

\appendix
\section[Ly-alpha radiative coefficients]{$\bmath{\lya{}}$ radiative transfer coefficients} \label{app:Lyaeffects}
Throughout this work, we denote the \lya{} radiative transfer effects as
constants, which encompass the derivatives of the transmission function
with respect to the \lya{} radiative transfer effect. Here we outline
the derivations for the first-order \lya{} radiative transfer
coefficients, from which the second-order constants can be easily
calculated. In the following derivations, each of the
\lya{} radiative transfer effects is expressed as a function of an
arbitrary transmission function, $\mathcal{T}$
\citep{Wyithe:2011p12569}. 

First, let us consider the fluctuations in the ionizing background,
which are taken from the expression in Equation \ref{eq:Taylorex1}, 
\begin{eqnarray} 
\nonumber
& &\frac{(\Gamma - \Gamma_{0})}{\bar{n}^{(0)}_{\rm
 Ly\alpha}}\frac{\partial\bar{n}_{\rm
 Ly\alpha}}{\partial\Gamma}\line(0,1){15}_{\line(0,1){15}F_{0},\Gamma_{0}}\\
\nonumber
 &=&  \left(\frac{\Gamma - \Gamma_{0}}{\bar{n}^{(0)}_{\rm
       Ly\alpha}\Gamma_{0}}\right)\frac{\partial
 \rm{log}(\mathcal{T})}{\partial\rm
 log(\Gamma)}\line(0,1){15}_{\line(0,1){15}\mathcal{T}_{0},\Gamma_{0}}\frac{\partial\bar{n}_{\rm
 Ly\alpha}}{\partial
 \rm{log}(\mathcal{T})}\line(0,1){15}_{\line(0,1){15}F_{0},\mathcal{T}_{0}}\\
&\equiv& 
\label{eq:example1}
\delta_{\Gamma}C_{\Gamma},
\end{eqnarray}
where we have defined $\delta_{\Gamma} \equiv
\left(\frac{\Gamma - \Gamma_{0}}{\Gamma_{0}}\right)$ and 
\begin{eqnarray}
C_{\Gamma} \equiv  \frac{1}{\bar{n}^{(0)}_{\rm Ly\alpha}}\frac{\partial
 \rm{log}(\mathcal{T})}{\partial\rm
 log(\Gamma)}\line(0,1){15}_{\line(0,1){15}\mathcal{T}_{0},\Gamma_{0}}\frac{\partial\bar{n}_{\rm
 Ly\alpha}}{\partial
 \rm{log}(\mathcal{T})}\line(0,1){15}_{\line(0,1){15}F_{0},\mathcal{T}_{0}}.
\end{eqnarray}

Following the same idea, we redefine the other two terms in Equation \ref{eq:Taylorex1} as fluctuations in the density field, and in the line-of-sight velocity field. For the density field, we find 
\begin{eqnarray}
\frac{1}{\bar{n}^{(0)}_{\rm Ly\alpha}}(\rho - \rho_{0})\frac{\partial\bar{n}_{\rm Ly\alpha}}{\partial\rho}\line(0,1){15}_{\line(0,1){15}F_{0},\rho_{0}} \equiv \delta_{\rho}C_{\rho},
\end{eqnarray}
where
\begin{eqnarray}
C_{\rho} \equiv \frac{1}{\bar{n}^{(0)}_{\rm Ly\alpha}}\frac{\partial \rm{log}(\mathcal{T})}{\partial\rm log(\rho)}\line(0,1){15}_{\line(0,1){15}\mathcal{T}_{0},\rho_{0}}\frac{\partial\bar{n}_{\rm Ly\alpha}}{\partial \rm{log}(\mathcal{T})}\line(0,1){15}_{\line(0,1){15}F_{0},\mathcal{T}_{0}}.
\end{eqnarray}
However for the velocity gradient, we rewrite the derivative as 
\begin{eqnarray}
& &\frac{1}{\bar{n}^{(0)}_{\rm Ly\alpha}}\left(\frac{{\rm d}v_{z}}{{\rm d}(ar_{\rm com})}- H\right)\frac{\partial\bar{n}_{\rm Ly\alpha}}{\partial\left(\frac{{\rm d}v_{z}}{{\rm d}(ar_{\rm com})}\right)}\line(0,1){15}_{\line(0,1){15}F_{0},\rho_{0}} \nonumber \\
&=& \frac{1}{\bar{n}^{(0)}_{\rm Ly\alpha}}\frac{1}{a}\left(\frac{{\rm d}v_{z}}{{\rm d}(r_{\rm com})}- Ha\right)\frac{\partial\bar{n}_{\rm Ly\alpha}}{(\frac{1}{a})\partial\left(\frac{{\rm d}v_{z}}{{\rm d}r_{\rm com}}\right)}\line(0,1){15}_{\line(0,1){15}F_{0},\rho_{0}} \nonumber \\
&=& \frac{{\rm d}v_{\rm{pec}}}{\rm d(r_{com})}\frac{\partial\bar{n}_{\rm Ly\alpha}}{\partial\left(\frac{{\rm d}v_{z}}{\rm dr_{com}}\right)}\line(0,1){15}_{\line(0,1){15}F_{0},\rho_{0}}, 
\end{eqnarray}
where we have expressed the total velocity as $v$ = $H(a r_{\rm com})$ + $v_{\rm pec}$, yielding the velocity gradient
\begin{eqnarray}
\frac{{\rm d}v}{{\rm d}(r_{\rm com})} = Ha + \frac{{\rm d}v_{\rm pec}}{{\rm d}(r_{\rm com})}.
\end{eqnarray}
Following the case for the ionizing background, we rewrite the partial
derivative as
\begin{eqnarray}
& &\frac{\partial\bar{n}_{\rm Ly\alpha}}{\partial\frac{{\rm d}v_{z}}{{\rm d}r_{\rm com}}}\line(0,1){15}_{\line(0,1){15}F_{0},\rho_{0}} \nonumber \\
&=& \frac{\partial\rm{log}(\mathcal{T})}{\partial{\rm log}(\frac{{\rm d}v_{z}}{{\rm d}r_{\rm com}})}\line(0,1){15}_{\line(0,1){15}\mathcal{T}_{0},\rho_{0}}\frac{\partial{\rm log}(\frac{{\rm d}v_{z}}{{\rm d}r_{\rm com}})}{\partial\frac{{\rm d}v_{z}}{{\rm d}r_{\rm com}}}\line(0,1){15}_{\line(0,1){15}\,0}\frac{\partial\bar{n}_{\rm Ly\alpha}}{\partial\rm{log}(\mathcal{T})}\line(0,1){15}_{\line(0,1){15}F_{0},\mathcal{T}_{0}}  .\nonumber \\
\end{eqnarray}
Using
\begin{eqnarray}
\left(\frac{{\rm d}v_{z}}{{\rm d}r_{\rm com}}\right)_{0} &=& Ha,\\
\frac{\partial{\rm log}(\frac{{\rm d}v_{z}}{{\rm d}r_{\rm com}})}{\partial\frac{{\rm d}v_{z}}{{\rm d}r_{\rm com}}}\line(0,1){15}_{\line(0,1){15}\,0}&=&\frac{1}{\textit{Ha}},
\end{eqnarray}
we find
\begin{eqnarray}
\nonumber
& &
\frac{1}{\bar{n}^{(0)}_{\rm Ly\alpha}}\left(\frac{{\rm d}v_{z}}{{\rm
				       d}(ar_{\rm com})}-
				       H\right)\frac{\partial\bar{n}_{\rm
Ly\alpha}}{\partial\frac{{\rm d}v_{z}}{{\rm d}(ar_{\rm
com})}}\line(0,1){15}_{\line(0,1){15}F_{0},\rho_{0}} \\
\nonumber
&=& \frac{1}{\bar{n}^{(0)}_{\rm Ly\alpha}}\frac{1}{Ha}\frac{{\rm
d}v_{\rm pec}}{{\rm d}(r_{\rm com})}\frac{\partial{\rm
log}(\mathcal{T})}{\partial{\rm log}(\frac{{\rm d}v_{z}}{{\rm d}r_{\rm
com}})}\line(0,1){15}_{\line(0,1){15}\mathcal{T}_{0},\rho_{0}}\frac{\partial\bar{n}_{\rm
Ly\alpha}}{\partial{\rm
log}(\mathcal{T})}\line(0,1){15}_{\line(0,1){15}F_{0},\mathcal{T}_{0}}\\
&\equiv&  C_v\delta_v,
\end{eqnarray}
where $\delta_{v} \equiv \frac{1}{Ha}\frac{{\rm d}v_{\rm pec}}{{\rm d}(r_{\rm com})}$, and
\begin{eqnarray}
C_{v} \equiv \frac{1}{\bar{n}^{(0)}_{\rm Ly\alpha}}\frac{\partial{\rm log}(\mathcal{T})}{\partial{\rm log}(\frac{{\rm d}v_{z}}{{\rm d}r_{\rm com}})}\line(0,1){15}_{\line(0,1){15}\mathcal{T}_{0},\rho_{0}}\frac{\partial\bar{n}_{\rm Ly\alpha}}{\partial{\rm log}(\mathcal{T})}\line(0,1){15}_{\line(0,1){15}F_{0},\mathcal{T}_{0}}.
\end{eqnarray}

\section[Lyman-alpha radiative transfer kernels]{$\bmath{\lya{}}$ radiative transfer kernels} \label{app:full}
Here we outline the derivation of the higher-order real- and redshift-space kernels for the fluctuations in the number density of LAEs. 
\subsection[Real-space Lyman-alpha kernels]{Real-space $\bmath{\lya{}}$ kernels} 
In this subsection, we derive the symmetrized kernels,
$Z^{(s)}_n$, which give the LAE galaxies in real space as
\begin{eqnarray}
\delta_{\lya{}}(\bmath{k},z) &= &\sum^{\infty}_{n=1} D^{n}(z)\int \frac{\rm d^3\bmath{q}_{1}}{(2\pi)^3} \int \frac{\rm d^3\bmath{q}_{n-1}}{(2\pi)^3}\nonumber \\
& & \times \int d^3\bmath{q}_{n}\delta^{D}(\bmath{k}-\sum^{n}_{i=1} \bmath{q}_{i})Z^{(s)}_{n} (\bmath{q}_{1},\bmath{q}_{2},...,\bmath{q}_{n})  \nonumber \\
& & \times\delta_{1}(\bmath{q}_{1})\delta_{1}(\bmath{q}_{2})...\delta_{1}(\bmath{q}_{n}).
\end{eqnarray}

Expanding out Equation \ref{eq:Lyafluct2ndOrder} and keeping terms only to
second order in fluctuations,
\begin{eqnarray} \label{eq:expansion1}
\delta_{\lya{}}(\bmath{x}) = \delta_{g}(\bmath{x}) + \bar{n}^{(1)}_{\lya{}} + \bar{n}^{(2)}_{\lya{}} + \delta_{g}(\bmath{x})\bar{n}^{(1)}_{\lya{}}.
\end{eqnarray}
Now, substituting in Equations \ref{eq:galaxybias}, \ref{eq:1stOrder}
and \ref{eq:2ndOrder} into Equation \ref{eq:expansion1} and expanding,
again only keeping terms up to the second order in fluctuations,
\begin{eqnarray} \label{eq:expansion2}
\nonumber
& &
\delta_{\lya{}}(\bmath{x})\\
 &=& b_{1}\delta(\bmath{x}) + \frac{1}{2}b_{2}\delta(\bmath{x})^{2} + \delta_{\Gamma}(\bmath{x})C_{\Gamma} + \delta_{\rho}(\bmath{x}) C_{\rho} + \delta_{v}(\bmath{x})C_{v} \nonumber \\
& & + \frac{1}{2}\left[C_{\Gamma \Gamma}\delta_{\Gamma}^{2}(\bmath{x}) + C_{\rho \rho}\delta_{\rho}^{2}(\bmath{x}) + C_{v v}\delta_{v}^{2}(\bmath{x})\right]  \nonumber \\
& & + C_{\Gamma \rho}\delta_{\Gamma}(\bmath{x})\delta_{\rho}(\bmath{x}) + C_{\Gamma v}\delta_{\Gamma}(\bmath{x})\delta_{v}(\bmath{x}) + C_{\rho v}\delta_{\rho}(\bmath{x}) \delta_{v}(\bmath{x})\nonumber \\
& & +  b_{1}\delta(\bmath{x})\left[\delta_{\Gamma}(\bmath{x})C_{\Gamma} + \delta_{\rho}(\bmath{x}) C_{\rho} + \delta_{v}(\bmath{x})C_{v}\right].
\end{eqnarray}
To model the effect of the fluctuating ionizing background, we convolve
the overdensity of sources (in this case the galaxies),
$\delta_{g}(\bmath{x})$, with a function of the form $\propto {\rm
exp}[-(\bmath{x} - \bmath{x_{o}})/\lambda]/[(\bmath{x} -
\bmath{x_{o}})^{2}]$ \citep{Morales:2010p10051}, where $\lambda$ is the
mean free path of the ionizing photons. Hence we find the Fourier
transform of the ionizing fluctuations as
\begin{eqnarray}
 \delta_{\Gamma}(\bmath{k}) = \delta_{g}(\bmath{k})\frac{{\rm arctan}(|\bmath{k}|\lambda)}{|\bmath{k}|\lambda},
\end{eqnarray}
which, to the second order, yields
\begin{eqnarray} \label{eq:Gammaexp}
 \delta_{\Gamma}(\bmath{k}) = \left[b_{1}\delta(\bmath{k}) + \frac{1}{2}b_{2}\delta(\bmath{k})^{2}\right]\frac{{\rm arctan}(|\bmath{k}|\lambda)}{|\bmath{k}|\lambda}.
\end{eqnarray}

Taking the Fourier transform of Equation \ref{eq:expansion2}, and using
Equation \ref{eq:PTdens}, \ref{eq:PTvel}, and \ref{eq:Gammaexp} for the
density field, peculiar-velocity field, and the ionizing background,
respectively, and again keeping terms up to the second order only, one
finally obtains 
\begin{eqnarray}
\delta_{\lya{}}(\bmath{k}) &=& \left[b_{1} + C_{\Gamma}b_{1}A(\bmath{k}) + C_{\rho}\right]\delta^{(1)}(\bmath{k}) +  C_{v}\delta_{v}^{(1)}(\bmath{k}) \nonumber \\
& &+ \left[b_{1} + C_{\Gamma}b_{1}A(\bmath{q}_{1},\bmath{q}_{2}) + C_{\rho}\right]\delta^{(2)}(\bmath{q}_{1},\bmath{q}_{2})\nonumber \\
& & + C_{v}\delta_{v}^{(2)}(\bmath{q}_{1},\bmath{q}_{2}) + \left[\frac{1}{2}b_{2} + \frac{1}{2}C_{\Gamma}b_{2}A(\bmath{q}_{1},\bmath{q}_{2})\right. \nonumber \\
& & \left.+ \frac{1}{2}C_{\Gamma\Gamma}b^{2}_{1}A(\bmath{q}_{1})A(\bmath{q}_{2}) + \frac{1}{2}C_{\rho\rho} + C_{\Gamma\rho}b_{1}A(\bmath{q}_{1}) \right. \nonumber \\
& & \left.  +  b^{2}_{1}C_{\Gamma}A(\bmath{q}_{1}) + b_{1}C_{\rho} \right]\delta^{(1)}(\bmath{q}_{1})\delta^{(1)}(\bmath{q}_{2}) \nonumber \\
& & +  \left[C_{\Gamma v}b_{1} A(\bmath{q}_{1}) + C_{\rho v} + b_{1}C_{v}\right]\delta^{(1)}(\bmath{q}_{1})\delta_{v}^{(1)}(\bmath{q}_{2})  \nonumber \\
& &  + \frac{1}{2}C_{vv}\delta_{v}^{(1)}(\bmath{q}_{1})\delta_{v}^{(1)}(\bmath{q}_{2}),
\end{eqnarray}
where $\bmath{k} = \bmath{q}_{1} + \bmath{q}_{2}$, $A(\bmath{q}_{1}) \equiv
\frac{{\rm arctan}(|\bmath{q}_{1}|\lambda)}{|\bmath{q}_{1}|\lambda}$, and
$A(\bmath{q}_{1},\bmath{q}_{2}) \equiv \frac{{\rm arctan}(|\bmath{q}_{1}
+ \bmath{q}_{2}|\lambda)}{|\bmath{q}_{1} +
\bmath{q}_{2}|\lambda}$. Where also, $\delta^{(1)}(\bmath{k})$ is read
as the $n=1$ term  and $\delta^{(2)}(\bmath{q}_{1},\bmath{q}_{2})$ is
read as the $n=2$ term of Equation \ref{eq:PTdens}. In the
peculiar-velocity effects, $\delta_{v}^{(1)}(\bmath{k}) =
-f\mu^{2}\eta^{(1)}(\bmath{k})$ and
$\delta_{v}^{(2)}(\bmath{q}_{1},\bmath{q}_{2}) =
-f\mu^{2}_{12}\eta^{(2)}(\bmath{k})$, the terms $\eta^{(1)}(\bmath{k})$ and
$\eta^{(2)}(\bmath{k})$ are the $n=1$ and $n=2$ terms of Equation
\ref{eq:PTvel}, respectively.  In the above expression, $\mu$ is the
cosine of the angle between the line-of-sight vector, $\hat{\bmath{z}}$, and the
wavevector, $\bmath{k}$, i.e., $\mu \equiv
\bmath{k}\cdot\hat{\bmath{z}}/k$, and $f = 
\frac{{\rm d ln}D(a)}{{\rm d ln}a}$ is the growth rate of structure
whose derivative is a function of the scale factor, $a$. 

After we
symmetrize the above expression and collect into first- and second-order
expressions, we obtain the first- and second-order real-space kernels as
\begin{eqnarray}
Z^{(s)}_{1}(\bmath{k}) =  b_{1}\left(1 + C_{\Gamma}\frac{\rm{arctan}(\bmath{|k|}\lambda)}{\bmath{|k|}\lambda}\right) + C_{\rho} - f\mu^{2}C_{v}
\end{eqnarray}
\begin{eqnarray}
Z^{(s)}_{2} (\bmath{q}_{1},\bmath{q}_{2}) & = & \frac{1}{2}b^{2}_{1}C_{\Gamma \Gamma}A(\bmath{q}_{1})A(\bmath{q}_{2}) + \frac{1}{2}C_{\rho \rho} + b_{1}C_{\rho} \nonumber \\
& &+ \frac{b_{2}}{2}\left[1 + C_{\Gamma}A(\bmath{q}_{1},\bmath{q}_{2})\right] -\frac{1}{2}fC_{\rho v}(\mu^{2}_{1}+\mu^{2}_{2}) \nonumber \\
& & + \frac{1}{2}b_{1}\left(C_{\Gamma\rho} + b_{1}C_{\Gamma}\right)\left[A(\bmath{q}_{1}) + A(\bmath{q}_{2})\right] \nonumber \\
& &  - \frac{1}{2}b_{1}C_{v}f(\mu^{2}_{1} + \mu^{2}_{2}) -f\mu_{12}^{2}C_{v}G^{(s)}_{2}(\bmath{q}_{1},\bmath{q}_{2}) \nonumber \\
& & + \left\{b_{1}\left[1 + C_{\Gamma}A(\bmath{q}_{1},\bmath{q}_{2})\right] + C_{\rho}\right\}F^{(s)}_{2}(\bmath{q}_{1},\bmath{q}_{2}) \nonumber \\
& & -\frac{1}{2} fb_{1}C_{\Gamma v}\left[\mu^{2}_{2}A(\bmath{q}_{1}) + \mu^{2}_{1}A(\bmath{q}_{2})\right] \nonumber \\
& & + \frac{1}{2}f^{2}\mu^{2}_{1}\mu^{2}_{2}C_{v v}.
\end{eqnarray}

\subsection[Redshift-space Lyman-alpha kernels]{Redshift-space $\bmath{\lya{}}$ kernels}
{In this subsection, we derive the symmetrized kernels,
$K^{(s)}_n$, which give the LAE galaxies in redshift space as}
\begin{eqnarray} 
\delta_{\lya{},s}(\bmath{k},z) &= &\sum^{\infty}_{n=1} D^{n}(z)\int \frac{\rm d^3\bmath{q}_{1}}{(2\pi)^3} \int \frac{\rm d^3\bmath{q}_{n-1}}{(2\pi)^3}  \nonumber \\
& & \times \int d^3\bmath{q}_{n}\delta^{D}(\bmath{k}-\sum^{n}_{i=1} \bmath{q}_{i})K^{(s)}_{n} (\bmath{q}_{1},\bmath{q}_{2},...,\bmath{q}_{n})  \nonumber \\
& & \times\delta_{1}(\bmath{q}_{1})\delta_{1}(\bmath{q}_{2})...\delta_{1}(\bmath{q}_{n}),
\end{eqnarray}

In galaxy redshift surveys we measure positions of galaxies
in redshift space as opposed to real space. To generate the
redshift-space expressions for the \lya{} kernels we perform the
following coordinate transform:
\begin{eqnarray}\label{eq:red_trans}
\bmath{s} = \bmath{x} + (1+z)\frac{v(\bmath{x})\cdot\hat{\bmath{z}}}{H(z)},
\end{eqnarray}
where $\bmath{s}$ denotes redshift space and $v(\bmath{x})$ is the
line-of-sight peculiar velocity. We assume the standard `plane-parallel
approximation,' in which the peculiar velocity vector and the
line-of-sight direction are parallel (or anti-parallel), and are chosen to be
along the $\hat{\bmath{z}}$-direction. In other words, we ignore curvature
of the sky.
We can rewrite Equation \ref{eq:red_trans} as
\begin{eqnarray}\label{eq:red_trans}
\bmath{s} = \bmath{x} + fu_{z}(\bmath{x})\hat{\bmath{z}},
\end{eqnarray}
where $\bmath{u}(\bmath{x}) \equiv (1+z)\frac{v(\bmath{x})}{fH(z)}$. 
We can then relate the fluctuations in real space and redshift space
using the mass conservation:
\begin{eqnarray}
[1 + \delta_{\lya{},s}(\bmath{s})]d^{3}\bmath{s} = [1 + \delta_{\lya{}}(\bmath{x})]d^{3}\bmath{x}.
\end{eqnarray}
Taking the Fourier transform of both sides, yields the exact expression relating the real and redshift space quantities:
\begin{eqnarray}
\delta_{\lya{},s}(\bmath{k}) &=& \delta_{\lya{}}(\bmath{k}) \\
& & + \int d^{3}x\,{\rm e}^{-i\bmath{k}\cdot\bmath{x}}\left({\rm e}^{-ik_{z}fu_{z} }- 1\right)[1 + \delta_{\lya{}}(\bmath{x})] \nonumber,
\end{eqnarray}
where $u_{z}(\bmath{k},z) \equiv -\frac{i\mu}{k}\eta(\bmath{k},z)$, and
$\eta(\bmath{k},z)$ is given by Equation \ref{eq:PTvel}. 
Expanding the exponential as a power series and keeping terms up to the second order only,
\begin{eqnarray} \label{eq:fullredshift}
\delta_{\rm{Ly}\alpha,s}(\bmath{k}) &=& \delta_{\rm{Ly}\alpha}(\bmath{k}) - \delta_{v}(\bmath{k}) - \int d^{3}xe^{-i\bmath{k.x}} \nonumber \\
& & \times\left[ik_{z}fu_{z}(\bmath{x})\delta_{\rm{Ly\alpha}}(\bmath{x})+\frac{1}{2}k_{z}^{2}f^{2}u_{z}^{2}(\bmath{x})\right].\nonumber \\
\end{eqnarray}

Now, inserting Equations \ref{eq:PTvel} and \ref{eq:realspace} into
\ref{eq:fullredshift}, and symmetrizing the arguments, we obtain the
first- and the second-order redshift-space kernels as
\begin{eqnarray}
K^{(s)}_{1} (\bmath{k}) & = & b_{1}\left[1 + C_{\Gamma}A(\bmath{k})\right] + C_{\rho} + f\mu^{2}(1- C_{v}),
\end{eqnarray}
\begin{eqnarray}
K^{(s)}_{2} (\bmath{q}_{1},\mathbf{q}_{2}) & = & Z^{(s)}_{2} (\bmath{q}_{1},\bmath{q}_{2}) + f\mu_{12}^{2}G^{(s)}_{2} (\bmath{q}_{1},\bmath{q}_{2}) \nonumber \\
& & + \frac{1}{2}(q_{12}\mu_{12}
 f)^{2}\left[\frac{q_{1z}q_{2z}}{q^{2}_{1}q^{2}_{2}}\right] \nonumber \\
& & + \frac{1}{2}(q_{12}\mu_{12} f)\left[\frac{q_{1z}}{q^{2}_{1}}Z^{(s)}_{1}(\bmath{q_{2}}) + \frac{q_{2z}}{q^{2}_{2}}Z^{(s)}_{1}(\bmath{q_{1}})\right],\nonumber\\
\end{eqnarray}
where 
\begin{eqnarray}
\mu_{12}&\equiv& (\bmath{q}_{1} +
 \bmath{q}_{2})\cdot\hat{\bmath{z}}/q_{12},\\ 
q_{12} &\equiv& |\bmath{q}_{1} + \bmath{q}_{2}|,\\
q_{1z} &\equiv& \bmath{q}_{1}\cdot\hat{\bmath{z}} = q_{1}\mu_{1}.
\end{eqnarray}

\section[Calculation of bispectrum estimator volume]{Calculation of $\bmath{V_{B}}$} \label{app:VB}
To calculate the sample variance of the bispectrum (and reduced
bispectrum), we need the volume of the bispectrum estimator, $V_{B}$,
used in Equation \ref{eq:num_triangles}, which determines the number of
triangular configurations sampled at each position. We begin
with the definition of $V_B$:
\begin{eqnarray}
 V_{B}  =  \int_{\bmath{k}_{1}}d^{3}q_{1}\int_{\bmath{k}_{2}}d^{3}q_{2}\int_{\bmath{k}_{3}}d^{3}q_{3}\,\delta^{D}(\bmath{q}_{123}).
\end{eqnarray}
Rewriting $\delta^{D}(\bmath{q}_{123})$, we find 
\begin{eqnarray}
 V_{B}  =  \int\frac{d^{3}x}{(2\pi)^{3}}\int_{\bmath{k}_{1}}d^{3}q_{1} \,e^{i\bmath{x}\cdot\bmath{q}_{1}}\int_{\bmath{k}_{2}}d^{3}q_{2}\, e^{i\bmath{x}\cdot\bmath{q}_{2}}\int_{\bmath{k}_{3}}d^{3}q_{3}\, e^{i\bmath{x}\cdot\bmath{q}_{3}}. \nonumber
\end{eqnarray}
Now
\begin{eqnarray}
\int_{\bmath{k}_{3}}d^{3}q_{3}\, e^{i\bmath{x}\cdot\bmath{q}_{3}} & \simeq &  4\pi\,[\, k_{3}\frac{{\rm{sin}}(k_{3} x)}{x}\Delta k + O(\Delta k^{3}\,) \,] \nonumber \\
& \simeq &  4\pi\, k_{3}\frac{{\rm{sin}}(k_{3} x)}{x}\Delta k,
\end{eqnarray}
yielding
\begin{eqnarray}
V_{B} & = & 32\pi k_{1}k_{2}k_{3}(\Delta k)^{3}\int^{\infty}_{0}\frac{{\rm{sin}}(k_{1} x){\rm{sin}}(k_{2} x){\rm{sin}}(k_{3} x)}{x}dx. \nonumber \\ 
\end{eqnarray}
Evaluating the integral,
\begin{eqnarray}
& &\int^{\infty}_{0}\frac{{\rm{sin}}(k_{1} x){\rm{sin}}(k_{2} x){\rm{sin}}(k_{3} x)}{x}dx   \nonumber \\
& = & \frac{1}{2} \int^{\infty}_{-\infty} \left( \frac{{\rm sin}(k_{1}x)}{x}\right){\rm sin}(k_{2}x){\rm sin}(k_{3}x)d x \nonumber \\
& = & \frac{1}{2}\int^{\infty}_{-\infty} \left(
					  \frac{1}{2}k_{1}\int^{1}_{-1}e^{i
					  x k_{1} \mu}d\mu\right){\rm
sin}(k_{2}x){\rm sin}(k_{3}x)d x \nonumber \\
&=&  -\frac{\pi}{8}k_{1}\int^{1}_{-1}d\mu\int^{\infty}_{-\infty} \frac{dx}{2\pi} \left[e^{i x (k_{1} \mu+k_{2}+k_{3})} -e^{i x (k_{1} \mu-k_{2}+k_{3})}\right.\nonumber\\ 
& & \left. \qquad\qquad\qquad\qquad -  e^{i x (k_{1} \mu+k_{2}-k_{3})} + e^{i x (k_{1} \mu-k_{2}-k_{3})} \right] \nonumber \\
& = & \frac{\pi}{8}\int^{1}_{-1}d\mu\,\left[ \delta^{D}\left(\mu-\frac{k_{2}-k_{3}}{k_{1}}\right) + \delta^{D}\left(\mu+\frac{k_{2}-k_{3}}{k_{1}}\right)- \right.\nonumber \\
& & \left.\qquad\qquad {} \delta^{D}\left(\mu+\frac{k_{2}+k_{3}}{k_{1}}\right)-\delta^{D}\left(\mu-\frac{k_{2}+k_{3}}{k_{1}}\right)\right].\nonumber \\
& & 
\end{eqnarray}
Here, we have used
\begin{eqnarray}
{\rm sin}(k_{2}x){\rm sin}(k_{3}x) & = & -\frac{1}{4}\left(e^{i x k_{2}}e^{i x k_{3}} - e^{i x k_{3}}e^{-i x k_{2}}\right. \nonumber \\
& & \left. - e^{i x k_{2}}e^{-i x k_{3}} + e^{-i x k_{2}}e^{-i x k_{3}} \right).\nonumber \\
\end{eqnarray}

Once we consider the triangular condition, and that $k_{1} \geq k_{2} \geq k_{3}$, we obtain
\begin{eqnarray}
V_{B} = 8\pi^{2}k_{1}k_{2}k_{3}(\Delta k)^{3} \times \left\{ \begin{array}{cl} 
      1  &  {\rm normal \,\, triangles}\\ 
       1-\Theta_{0} & {\rm if} \,\, k_{1} = k_{2} + k_{3} \\
       \Theta_{0} & {\rm if} \,\, k_{2} = k_{1} + k_{3} \,\,{\rm or}\,\, \\
       & k_{3} = k_{1} + k_{2}
\end{array} \right.
\end{eqnarray}
where $\Theta$ is the Heaviside theta function, and $\Theta_{0} = \frac{1}{2}$.

\section[Derivation of bispectrum distance derivatives]{Derivation of the bispectrum distance derivates}

To calculate the Fisher matrix for the angular
diameter distance and the Hubble rate, we need to calculate derivatives
of the bispectrum with respect to the distances, namely,
$\frac{\partial B(k_{1},k_{2},k_{3},\mu_{1},\mu_{2},\mu_{3})}{\partial
{\rm ln}(D_{A})}$ and
$\frac{\partial B(k_{1},k_{2},k_{3},\mu_{1},\mu_{2},\mu_{3})}{\partial
{\rm ln}(H)}$.
However, we are dealing with the variables
$k_{1},k_{2},k_{3},\mu_{1},\mu_{2}$ and $\mu_{3}$, which are not all
independent of one another. Hence we perform the chain rule with respect to
the independent functions,
$k_{1}$, $k_{2}$, $k_{3}(\theta_{12},\mu_{1},\phi),\mu_{1},\mu_{2}(\mu_{1},\phi)$,
and $\mu_{3}(k_{1},k_{2},\theta_{12},\mu_{1},\phi)$. Here, we have
written the dependent variables, $k_{3}$, $\mu_{2}$ and $\mu_{3}$, as
functions of the independent variables $k_{1}$, $k_{2}$, $\theta_{12}$,
$\mu_{1}$ and $\phi$, which fully describe the shape and orientation of the
bispectrum triangles. 

Performing the chain rule,
\begin{eqnarray} \label{eq:DA}
\frac{\partial B}{\partial {\rm ln}(D_{A})} &=& \frac{\partial B}{\partial k_{1}}\frac{\partial k_{1}}{\partial {\rm ln}k_{1}}\frac{\partial {\rm ln}k_{1}}{\partial {\rm ln}(D_{A})} + \frac{\partial B}{\partial k_{2}}\frac{\partial k_{2}}{\partial {\rm ln}k_{2}}\frac{\partial {\rm ln}k_{2}}{\partial {\rm ln}(D_{A})} \nonumber \\
& & + \frac{\partial B}{\partial k_{3}}\frac{\partial k_{3}}{\partial {\rm ln}k_{3}}\frac{\partial {\rm ln}k_{3}}{\partial {\rm ln}(D_{A})} + \frac{\partial B}{\partial \mu_{1}}\frac{\partial \mu_{1}}{\partial {\rm ln}(D_{A})}\nonumber \\
& & + \frac{\partial B}{\partial \mu_{2}}\frac{\partial \mu_{2}}{\partial {\rm ln}(D_{A})} + \frac{\partial B}{\partial \mu_{3}}\frac{\partial \mu_{3}}{\partial {\rm ln}(D_{A})},
\end{eqnarray}
and
\begin{eqnarray} \label{eq:H}
\frac{\partial B}{\partial {\rm ln}(H)} &=& \frac{\partial B}{\partial k_{1}}\frac{\partial k_{1}}{\partial {\rm ln}k_{1}}\frac{\partial {\rm ln}k_{1}}{\partial {\rm ln}(H)} + \frac{\partial B}{\partial k_{2}}\frac{\partial k_{2}}{\partial {\rm ln}k_{2}}\frac{\partial {\rm ln}k_{2}}{\partial {\rm ln}(H)} \nonumber \\
& & + \frac{\partial B}{\partial k_{3}}\frac{\partial k_{3}}{\partial {\rm ln}k_{3}}\frac{\partial {\rm ln}k_{3}}{\partial {\rm ln}(H)} + \frac{\partial B}{\partial \mu_{1}}\frac{\partial \mu_{1}}{\partial {\rm ln}(H)} \nonumber \\
& & + \frac{\partial B}{\partial \mu_{2}}\frac{\partial \mu_{2}}{\partial {\rm ln}(H)} + \frac{\partial B}{\partial \mu_{3}}\frac{\partial \mu_{3}}{\partial {\rm ln}(H)},
\end{eqnarray}
where for notational convenience we have dropped the dependant variables on all functions. For the determination of the actual derivatives, in the evaluation of the second-order kernels, we write $k_{12} = k_{3}$, $k_{13} = k_{2}$, and $k_{23} = k_{1}$; and $\mu_{12} = -\mu_{3}$, $\mu_{13} = -\mu_{2}$, and $\mu_{23} = -\mu_{1}$. Following this we find, for example
\begin{eqnarray}
\frac{\partial k_{12}}{\partial k_{1}} = \frac{\partial k_{3}}{\partial k_{1}} = 0, \, \frac{\partial k_{13}}{\partial k_{1}} = 0 \,\,{\rm and} \, \,\frac{\partial k_{23}}{\partial k_{1}} = 1.
\end{eqnarray}
We apply similar logic when calculating the derivatives with respect to $k_{2}$ and $k_{3}$ as well as when performing the derivatives with respect to $\mu_{1},\mu_{2}$, and $\mu_{3}$. 

When performing the derivatives of the bispectrum with respect to, e.g.,
$k_{1}$, we hold $k_{2}$, $k_{3}$, $\mu_{1}$, $\mu_{2}$ and $\mu_{3}$
fixed, despite $k_{3}$ and $\mu_{3}$ depending on $k_{1}$ as per the
definition of the chain rule. We have chosen to write Equations
\ref{eq:DA} and \ref{eq:H} in the form above, to then insert the
following derivatives of $k_{i}$ and $\mu_{i}$ from
\citet{Shoji:2009p8219}, 
\begin{eqnarray}
\frac{\partial {\rm ln}k_{i}}{\partial {\rm ln}(D_{A})}  = 1 - \mu^{2}_{i},
\end{eqnarray}
\begin{eqnarray}
\frac{\partial {\rm ln}k_{i}}{\partial {\rm ln}(H)}  = - \mu^{2}_{i},
\end{eqnarray}
\begin{eqnarray}
\frac{\partial \mu_{i}}{\partial {\rm ln}(D_{A})}  = -\mu_{i}(1 - \mu^{2}_{i}),
\end{eqnarray}
and
\begin{eqnarray}
\frac{\partial \mu_{i}}{\partial {\rm ln}(H)}  = -\mu_{i}(1 - \mu^{2}_{i}).
\end{eqnarray} 

\section[Generation of the N-dimensional likelihoods]{Generation of the $N$-dimensional likelihoods}
In Section \ref{sec:PS} we use the the Fisher matrix to compute the one
and two-dimensional likelihoods marginalizing over the remaining
parameters in a model of size $n$. In a similar way one can compute
$N$-dimensional likelihoods marginalizing over $n-N$ parameters: 
\begin{eqnarray}
& & \mathscr{L}(\bmath{x}_{1},\bmath{x}_{2},...,\bmath{x}_{N}) \\
& = & {\rm exp}\left\{-\frac{1}{2}\left[\sum^{N}_{i=1} \bar{x}^{2}_{i} \left( F_{ii} - \sum^{n-N}_{k,l\neq i}F_{ik}(\bar{F}_{kl})^{-1}F_{li} \right) \right. \right. \nonumber \\
& & \left. \left. + 2\sum^{N}_{j>i} \bar{x}_{i}\bar{x}_{j}\left(F_{ij} - \sum^{n-N}_{k,l\neq i,j}F_{ik}(\bar{F}_{kl})^{-1}F_{lj}\right)\right]\right\}. \nonumber \\
\end{eqnarray}

\end{document}